%% file: ms.tex
\documentclass[iop]{emulateapj}

\usepackage{etoolbox}
\usepackage{color}

\newtoggle{figures}
\togglefalse{figures} 

\toggletrue{figures}

\input{mynewcommands.tex}

\newcommand{\nsys}{19} 
\newcommand{\nconst}{3} 
\newcommand{\nsystot}{22} 

\newcommand{\nrvs}{218}  

\shorttitle{RV monitoring of \ik\ HB stars}

\shortauthors{Shporer et al.}

\begin{document}

\title{Radial Velocity Monitoring of Kepler Heartbeat Stars$^\dagger$} \thanks{$\dagger$ The data presented herein were obtained at the W.M.~Keck Observatory, which is operated as a scientific partnership among the California Institute of Technology, the University of California and the National Aeronautics and Space Administration. The Observatory was made possible by the generous financial support of the W.M.~Keck Foundation.}

\author{
Avi Shporer\altaffilmark{1, 2}, 
Jim Fuller\altaffilmark{3, 4}, 
Howard Isaacson\altaffilmark{5},
Kelly Hambleton\altaffilmark{6, 7},
Susan E.~Thompson\altaffilmark{8, 9},
Andrej Pr{\v s}a\altaffilmark{6},
Donald W.~Kurtz\altaffilmark{7},
Andrew W.~Howard\altaffilmark{10},
Ryan M.~O'Leary\altaffilmark{11}
} 
\altaffiltext{1}{Jet Propulsion Laboratory, California Institute of Technology, 4800 Oak Grove Drive, Pasadena, CA 91109, USA}
\altaffiltext{2}{NASA Sagan Fellow}
\altaffiltext{3}{TAPIR, Walter Burke Institute for Theoretical Physics, Mailcode 350-17, Caltech, Pasadena, CA 91125, USA}
\altaffiltext{4}{Kavli Institute for Theoretical Physics, Kohn Hall, University of California, Santa Barbara, CA 93106, USA}
\altaffiltext{5}{Department of Astronomy, University of California, Berkeley CA 94720, USA}
\altaffiltext{6}{Department of Astrophysics and Planetary Science, Villanova University, 800 East Lancaster Avenue, Villanova, PA 19085, USA}
\altaffiltext{7}{Jeremiah Horrocks Institute, University of Central Lancashire, Preston, PR1 2HE, UK}
\altaffiltext{8}{NASA Ames Research Center, Moffett Field, CA 94035, USA}
\altaffiltext{9}{SETI Institute, 189 Bernardo Avenue Suite 100, Mountain View, CA 94043, USA}
\altaffiltext{10}{Institute for Astronomy, University of Hawaii, 2680 Woodlawn Drive, Honolulu, HI 96822, USA}
\altaffiltext{11}{JILA, University of Colorado and NIST, 440 UCB, Boulder, 80309-0440, USA}

\begin{abstract}

\openup 0.5em

Heartbeat stars (HB stars) are a class of eccentric binary stars with close periastron passages. 
The characteristic photometric HB signal evident in their light curves is produced by a combination of tidal distortion, heating, and Doppler boosting near orbital periastron. Many HB stars continue to oscillate after periastron and along the entire orbit, indicative of the tidal excitation of oscillation modes within one or both stars. 
These systems are among the most eccentric binaries known, and they constitute astrophysical laboratories for the study of tidal effects. 
We have undertaken a radial velocity (RV) monitoring campaign of \ik\ HB stars in order to measure their orbits. We present our first results here, including a sample of \nsystot\ \ik\ HB systems, where for \nsys\ of them we obtained the Keplerian orbit and for \nconst\ other systems we did not detect a statistically significant RV variability.
Results presented here are based on \nrvs\ spectra obtained with the Keck/HIRES spectrograph during the 2015 \ik\ observing season, and they have allowed us to obtain the largest sample of HB stars with orbits measured using a single instrument, which roughly doubles the number of HB stars with an RV measured orbit. 
The \nsys\ systems measured here have orbital periods from 7 to 90~d and eccentricities from 0.2 to 0.9. We show that HB stars draw the upper envelope of the eccentricity -- period distribution. Therefore, HB stars likely represent a population of stars currently undergoing high eccentricity migration via tidal orbital circularization, and they will allow for new tests of high eccentricity migration theories.

\end{abstract}


\section{Introduction}
\label{sec:intro}

Heartbeat stars are an exciting class of stellar binaries which have been discovered in large numbers only recently by the \ik\ photometric survey \citep[e.g.,][]{thompson12, beck14}. Their name originates from the characteristic light curve signal seen once per orbital period, induced by the close periastron passage of a highly eccentric binary star system, a signal whose shape resembles that of a heartbeat in an electrocardiogram. 

The photometric signal seen at periastron results from a combination of several processes, including tidal distortion, heating, and Doppler boosting. In addition, many of the systems exhibit tidally excited stellar pulsations that maintain constant amplitude throughout the orbit. They result from near-resonances between the multiples of the orbital frequency and stellar oscillation modes \citep{cowling41, zahn75, kumar95, fuller12, burkart12}. Another attractive quality of heartbeat stars is that they show a photometric signal whether or not the system shows eclipses. Therefore, heartbeat stars (hereafter HB stars) are astrophysical laboratories for the study of tidal interactions in stellar binaries.

Since the first discoveries of HB stars in \ik\ data \citep{welsh11, thompson12} the number of known HB stars has substantially increased and is currently at 173 \citep{kirk16}. However, the \ik\ light curves alone are not sufficient for taking advantage of the scientific opportunities HB stars offer. We have undertaken a radial velocity (RV) monitoring campaign of \ik\ HB stars using Keck/HIRES \citep{vogt94} in order to measure their orbits. Our observations were performed during the 2015 \ik\ observing season and we report our first results here.

We use our results to study the eccentricity -- period diagram, which is important for testing tidal circularization theory. Such a study can only be done with a large sample of HB stars, as we have characterized here, as opposed to a few individual systems. We show that heartbeat stars generally lie near the upper extreme of the eccentricity distribution as a function of orbital period. We infer that the lack of systems with higher eccentricity is a result of prior orbital circularization, and that heartbeat stars represent systems that are likely undergoing slow tidal orbital circularization. Therefore, circularization timescales for heartbeat stars are likely to be comparable to their ages, and testing this supposition with detailed analyses of these systems will yield valuable constraints on tidal and orbital evolution theories. 

The paper is arranged as follows. We describe the Keck/HIRES observations, data analysis, and orbit fitting in \secr{dataanal}, and in \secr{res} we describe the results. In \secr{dis} we discuss our results, our attempts to constrain the companion mass, the study of the eccentricity -- period relation, and some future prospects. We conclude with a summary in \secr{sum}.

\section{Observations and Data Analysis}
\label{sec:dataanal}

\subsection{Target selection}
\label{sec:targets}

We have selected our targets from the list of 173 known \ik\ HB stars, all flagged with the ``HB" flag in the \ik\ eclipsing binary (EB) online catalog\footnote{\url{http://keplerEBs.villanova.edu}} \citep{prsa11, slawson11, kirk16}. Although HB systems do not necessarily show eclipses they do show a drop and/or rise in flux during periastron passage which is reminiscent of an eclipse, or an inverted eclipse. Therefore, many of the methods designed to detect stellar eclipses detect also a periastron HB signal, in addition to identification by visual inspection, leading to their inclusion in the EB catalog.

We chose systems with an orbital period shorter than 90~d to allow orbital phase coverage within one observing season, and with a brightness of \kp\ $\leq$ 14.0 mag (where \kp\ is the \ik\ magnitude) to keep the exposure time short. In addition, using the stellar effective temperature and surface gravity \citep{huber14} we have tried to include only main sequence stars and avoid giant stars, as the latter were already the focus of the work of \cite{beck14}. Next we prioritized the systems according to a combination of several criteria, including (1) the stellar effective temperature, as hotter stars are more challenging for RV measurements, (2) the presence of tidal pulsations or rotational modulation (due to stellar activity) in the \ik\ light curve, making the system more interesting scientifically, and (3) target brightness. The list of targets we observed is given in \tabr{targets} along with the stellar parameters from \cite{huber14}, including a total of \nsystot\ targets. Although we did not analyze these light curves here we show them in \appr{rvlc} for completeness. As can be seen in \appr{rvlc} figures, the relative flux variation during periastron passage has a typical full amplitude from 10$^{-4}$ to 10$^{-3}$. Some of the systems show visually identifiable tidal pulsations, with an amplitude of up to several 10$^{-4}$ in relative flux (e.g.~KID 8164262). In addition, a few of the systems show eclipses (e.g.~KID 5790807). As described in detail below, for \nsys\ targets we measured the RV orbit and for \nconst\ targets we did not detect a statistically significant RV variability.

\input{targets_table.tex}

\subsection{Keck/HIRES observations and data analysis}
\label{sec:hires}

The Keck/HIRES data analyzed and presented here includes \nrvs\ exposures obtained during 43 nights from May to October 2015. The access to a relatively large number of nights while using a small amount of telescope time per night was critical for the success of this program. It allowed us to sample the entire orbital phase of our targets, and sample the periastron phase more intensely since it is that phase where most of the RV variability takes place for eccentric binary systems. For each of the systems presented here we have obtained at least 7 RV measurements, in order to fit a Keplerian orbital model that in our case includes 5 fitted parameters, since the orbital period is already precisely known from \ik\ photometry (see more details in \secr{fit}).
  
We used the Keck/HIRES instrumental setup of the California Planet Search as described in \cite{howard09}. At the beginning of each observing night we used a Thorium-Argon lamp to align the spectral format to within one-half pixel of the historical position, where one pixel represents 1.3 \kms. This careful setup is the first step is calculating the RVs presented here. Each spectrum was acquired with the C2 decker (angular size of 0.87~arcsec $\times$ 14.0~arcsec), allowing for background sky and scattered light to be removed and resulting in a resolving power of $R\approx60,000$. The exposure time was in the range of $0.5-5.0$~min, depending on target brightness, and the spectra we obtained have a signal-to-noise ratio of $10-20$ per pixel.

As a first step of the spectral data analysis we obtained the wavelength solution (assigning a wavelength value for every pixel) with a precision of 0.1 pixels using a Thorium-Argon calibration lamp spectrum taken at the beginning of each night's observing. To derive the RV measurements we used the method described in \cite{chubak12}, using the telluric A and B absorption bands ($7,594-7,621$~\AA\ and $6,867-6,884$~\AA\ respectively) due to absorption by molecular Oxygen in the Earth’s atmosphere.  
 
We chose the reference B-type star HD~79439 to serve as the telluric lines wavelength zero-point, and measure the position of the target stars' telluric lines relative to those of the reference star. This zero-point offset corrects for drift in the CCD position throughout the night and observing variables such as non-uniform illumination of the spectral slit. We subtract this offset from any measured shift in the position of the target stars' spectral lines in order to determine their true Doppler shift.
 
We measure the position of the target stars' spectral lines using four wavelength segments rich in stellar absorption lines. Those four segments are located at $6,795-6,867$~\AA, $7,067-7,146$~\AA, $7,398-7,489$~\AA, and $7,518-7,593$~\AA, which are adjacent to but not overlapping with the telluric A and B bands. The four wavelength segments of the target star are cross-correlated with a HIRES spectrum of Vesta, which serves as a solar proxy reference spectrum. The mean and RMS of the four RV measurements serve as the RV value and uncertainty respectively. This raw RV measurement between the target star and Vesta is then corrected for barycentric motion of the reference star and the target, determined by the JPL Solar System ephemeris\footnote{\url{http://ssd.jpl.nasa.gov}}. This way, any contributions to the RV measurement that are not due to the radial motion of the target star relative to the reference star have been accounted for by the telluric lines and the barycentric corrections. 
 
Finally, each RV measurement is set to the RV scale of \cite{nidever02} and \cite{latham02}, by using an offset determined by the observations of 110 stars in the overlap of the samples of \cite{chubak12} and \cite{nidever02}. All \nrvs\ RV measurements are listed in Appendix A. The method we used calculates the RVs in an absolute scale, and the typical errors for slowly rotating stars (with rotation periods at the level of 10~d or longer) are at the 0.1 \kms\ level. As the majority of the heartbeat stars observed here have an increased rotation rate (with rotation periods at the level of 1~d) the resulting RV errors are typically at the range of 0.1--1.0 \kms\ (for six systems the RV precision is at the level of a few \kms, see Appendix A), sufficient to measure the RV variations of the stellar components' orbital motion.

\subsection{Keplerian orbit fitting}
\label{sec:fit}

We have fitted a Keplerian orbit model to the RV measurements using the adaptive MCMC approach described in \cite{shporer09}. In this approach the width of the distribution from which the step sizes are drawn is adjusted every $10^4$ steps, which we refer to as a minichain, in order to keep the step's acceptance rate at 25\% \citep{gregory05, holman06}. This adaptive approach eliminates a possible dependence of the fitted parameters on the width of the distribution from which the step sizes are drawn. Each chain consists of 100 minichains, or 10$^6$ steps total, and we ran 5 chains for each system. We then generated the posterior probability distribution of each parameter by combining the 5 chains while ignoring the initial 20\% steps of each chain. We took the distribution median to be the best-fit value and the values at the 84.13 and 15.87 percentiles to be the +\sig{1} and -\sig{1} confidence uncertainties, respectively.

The Keplerian orbital model includes six parameters, the period $P$, periastron time $T_0$, RV semi-amplitude $K$, system's center of mass RV $\gamma$ (commonly referred to as RV zero point), orbital eccentricity $e$, and argument of periastron $\omega$. Since the \ik\ photometric data constrain the orbital period significantly better than the RV measurements we adopted the photometric orbital period value, $P_{\rm phot}$, and its uncertainty from the \ik\ EB catalog, and used them as the mean and width, respectively, of a Gaussian prior distribution on $P$. We implemented that prior by drawing a value at random from the Gaussian prior distribution in each step of the MCMC analysis. Therefore, our fitted model included 5 free parameters. Those 5 parameters were fitted by 7 -- 12 RVs per system.

When stepping through the five-dimensional parameter space we used the parameters \ecosw\ and \esinw\ instead of $e$ and $\omega$, following, e.g., \cite{eastman13}. We made the parameter conversion at each step and when the chain reached a position where $e \geq 1$ we set $\chi^2$ to infinity to make sure the step is not accepted. 

We set $\chi^2$ to infinity also when $T_0$ reached a position where it is more than $P_{\rm phot}$/2 away (in absolute value) from the initial $T_0$ position, so $T_0$ was allowed to vary within a span of $P_{\rm phot}$, or the full orbital phase. The initial $T_0$ position was arbitrarily set, taken to be the predicted time of eclipse, based on the eclipse ephemeris listed in the \ik\ EB catalog \citep{kirk16}, which was closest to the middle of the time period covered by the RV measurements. For non-eclipsing systems the \ik\ EB catalog reports the time of minimum flux, although that distinction is not important since we are using this time only as an arbitrary starting point for $T_0$. Therefore, we used a uniform prior on $T_0$ and allowed it to vary throughout the entire orbital phase, and we did not use the eclipse time (or time of minimum flux) to constrain the periastron time. 

For some of the systems we have analyzed we noticed that the best fit $\chi^2$ value is significantly larger than the expectation value (or the mean) of the $\chi^2$ distribution for the given number of degrees of freedom $\nu$ (which equals $n - 5$, for $n$ RV measurements and 5 fitted parameters). As this could be the result of underestimated RV errors we have added a mechanism to our analysis to correct for that. Once the analysis was done we checked the distance between the best fit $\chi^2$ and the expectation value of the $\chi^2$ distribution for $\nu$ degrees of freedom. If that distance was larger than $2\sqrt{2\nu}$, which is twice the $\chi^2$ standard deviation for $\nu$ degrees of freedom, we repeated the analysis while adding in quadrature a systematic uncertainty to the RV measurements uncertainties. We refer to that systematic uncertainty as jitter, and it was set to make $\chi^2$ equal $\nu$. Therefore, the analysis was iterated until the best fit $\chi^2$ was close enough to the expectation value.

Our analysis included also a component that tests whether a target shows no RV variability. This was done by calculating $\chi^2_{\rm null}$, the $\chi^2$ value of a constant RV model where for each system that constant was the RVs weighted mean. We declared a system to have no statically significant RV variability if $\chi^2_{\rm null}$ was smaller than the 99.9 percentile of a $\chi^2$ distribution with $\nu$ equal to the number of RVs minus one.

\section{Results}
\label{sec:res}

We have obtained the Keplerian orbital solution for \nsys\ systems. \tabr{orbits} lists the fitted orbital parameters, including the orbital period from the \ik\ EB catalog, taken as a prior in the MCMC analysis, and the companion's mass function, \fms, defined as:
\begin{equation}
\label{eq:fm}
f(m) \equiv \frac{P (K\sqrt{1 - e^2})^3}{2\pi G} = \frac{M_2^3 \sin^3 i}{(M_1 + M_2)^2},
\end{equation}
where $i$ is the orbital plane inclination angle, and \mo\ and \mt\ are the masses of the primary star and secondary star, respectively.

\tabr{stats} lists a few statistics describing the fitted model, including the fitted model $\chi^2$, number of degrees of freedom (which is simply the number of RV measurements minus five, for the five fitted parameters), number of analysis iterations (see \secr{fit}), the RV systematic uncertainty by which the RV errors were increased in quadrature (RV jitter; defined to be zero when only a single analysis iteration was performed), and the RV residuals scatter. Here and throughout this paper the scatter is estimated in a robust way, using the median absolute deviation (MAD) where the standard deviation is calculated as 1.4826$\times$MAD (\citealt{hoaglin83, beers90}; see also \citealt{shporer14}). As shown in \tabr{stats}, a second analysis iteration, where a non-zero jitter was introduced was done for only 9 of the \nsys\ systems. The RV residual scatter is at the 1.0 \kms\ level, similarly to the typical jitter value (for systems where it was introduced). This matches well the typical RV errors and the expected precision for RVs derived using the telluric bands method for this population of stars that tend to rotate faster than Sun-like stars (see \secr{hires}).

In Figures~1 through 5 we present the RV curves for all \nsys\ systems with a measured orbit. Each system is presented with two panels. In the top panel we show the RVs vs.~time (black) along with the best fit model (solid red line). In the bottom panel we show the phase-folded RV curve (black) along with a continuum of orbits that correspond to a \sig{3} marginalization (red). For completeness we present in \appr{rvlc} figures of the phase-folded RV curves overplotted by the phase-folded \ik\ light curves for all \nsystot\ systems studied here.

We have tested our results in several ways:
\begin{itemize}
\item We changed the stopping condition of the MCMC iterations. Instead of requiring $\chi^2$ to be within $2\sqrt{2\nu}$ from the expectation value of the $\chi^2$ distribution for $\nu$ degrees of freedom, we required it to be within $\sqrt{2\nu}$ and $4\sqrt{2\nu}$ from it, in two separate applications of our analysis.
\item For each of the \nsys\ systems we used the RV times and injected an RV orbit identical to the fitted model and applied the same analysis.
\item We repeated the entire analysis, to test the repeatability of our results given the random component of the MCMC analysis.
\end{itemize}
In all tests above the results were fully consistent with the original results.

\input{orbits_table.tex}

\input{stats_table.tex}

\iftoggle{figures}{
\clearpage
\input{figures_orbits1.tex}

\clearpage
\input{figures_orbits2.tex}

\clearpage
\input{figures_orbits3.tex}

\clearpage
\input{figures_orbits4.tex}

\clearpage
\input{figures_orbits5.tex}

\clearpage
}

For three other targets we monitored we did not detect any appreciable RV variability. All three were identified by measuring a $\chi^2_{\rm null}$ smaller than the 99.9 percentile of the $\chi^2$ distribution (see \secr{fit}). In fact, all three had $\chi^2_{\rm null}$ smaller than the 99.0 percentile, while all other \nsys\ systems were above the 99.999 ($= 100 - 10^{-3}$) percentile. The three non-variable systems are listed in \tabr{const}, their RV measurements are included in Appendix~A, and their RV curves are plotted in Appendix~B. The RV scatter of these systems is in the range of 0.47 -- 1.42 \kms, which compares well with the residuals RV scatter of the \nsys\ systems with a fitted Keplerian orbit (see \tabr{stats}).

\input{const_table.tex}

\section{Discussion}
\label{sec:dis}

The \nsys\ HB stars whose RV orbits were measured here constitute the largest sample to date where the RVs were measured with a single instrument. Considering the orbital period range covered here, $P \leq 90$~d, this new sample roughly doubles the number of HB stars with RV-measured orbits \citep{maceroni09, welsh11, hambleton13, beck14, hareter14, schmid15, smullen15, hambleton16} listed in \tabr{hblit}.

The stellar parameters listed in \tabr{targets} show that our sample contains stars hotter than the Sun, of spectral types F and A, and effective temperatures in the range of 6,200~--~8,100 K. A minor caveat is that these stellar parameters are taken from \cite{huber14}, who revised the \ik\ input catalog (KIC; \citealt{brown11}) stellar properties of stars observed by \ik. Therefore, these parameters might be less precise than spectroscopically-derived parameters, and could be affected by light from the binary stellar companion and/or other stars on the same line of sight (whether or not they are gravitationally bounded to the HB system). Nonetheless, this characteristic of our sample is not likely to significantly change with more precise parameters. 

The fact that HB stars tend to have relatively hot primaries (hotter than the Sun, as noted above) is at least partially an observational bias. Since we focus here on main sequence stars, hotter stars are larger in radius and have lower surface gravity. Hence they have a larger tidal distortion for the same tidal force, leading to a larger photometric signal observed by \ik\ during periastron. This is supported by the correlation between stellar \teff\ and the photometric amplitude of the HB signal at periastron, identified by \cite{thompson12}. It is therefore easier to detect HB systems with hot primaries. In addition, hot stars typically have a lower level of stellar activity than cool stars, making it easier to detect the HB photometric signal for hot stars. However, the sample of heartbeat stars studied here has been shaped by several subjective selection criteria (see \secr{targets}). Hence, this sample on its own is not appropriate for investigating the temperature distribution or circularization time scales of eccentric binaries.

As can be seen in Figures~1--5 and Tables~2--3, some of the orbital solutions are of better quality than others, where the quality is quantified by how well the fitted parameters are constrained, the residual scatter, and the fitted model $\chi^2$. In general, the quality of the orbital solutions worsens with decreasing number of RVs per target and with increasing eccentricity. The high eccentricity systems tend to have longer periods with only 1--2 observable periastron events during the observing season, making the observations more time critical and difficult to schedule. Those orbits can be refined in the future with additional RVs, especially during periastron passage.

\subsection{RV Non-variable Stars}
\label{sec:const}

As already noted in \secr{res}, three of the targets we observed show no RV variability at the 1 \kms\ level (see \tabr{const} and Appendix~B). The reason for the RV non-variability is unclear. These three targets have brightness and stellar parameters similar to the \nsys\ systems with measured orbits (see \tabr{targets}). The non-variability could be the result of a few possible scenarios, some of them similar to the false positive scenarios of eclipsing and transiting systems \citep{bryson13, coughlin14, abdul16}. 

One possible scenario is a triple or higher multiplicity system where the spectrum is dominated by lines from a bright star with no RV variability, which may or may not be bound to the binary heartbeat system and is located on the same line of sight. A more detailed study of all spectra obtained here, including searching for additional sets of lines, is ongoing and will be reported in a future publication.

In a similar scenario, the \ik\ photometric heartbeat signal does not originate from the star whose RVs were monitored but from another nearby star that is at least partially blended with the target on \ik's pixels that are 4 arcsec wide. High angular resolution imaging is needed to further study this scenario.

In a third scenario the target is a single variable star with variability mimicking that of a HB signal. The mechanism inducing the photometric variability can be for example stellar pulsations, or rotation and stellar spots. High quality spectra combined with detailed \ik\ light curve analysis are required to further study this scenario. 

In fact, we have identified KID~11409673 as a rapidly oscillating peculiar A star (roAp). These strongly magnetic stars are oblique pulsators, pulsating in high radial overtone p-modes with their pulsation axis inclined to the rotation axis and closely aligned to their magnetic axis \citep[e.g.,][]{kurtz82, holdsworth16}. We believe the 12.3 d photometric periodicity is the rotation period, and the photometric variability arises from a combination of stellar rotation and persistent stellar spots. A full study of this roAp star will be presented in a future publication.

Finally, it is possible that our RV measurements are not sensitive enough to detect the binary companion. This could occur if the system has a low orbital inclination or the companion mass is small. For a system to have a stellar-mass companion and an RV amplitude at or below the 1 \kms\ level, the orbital inclination needs to be exceptionally small, with $i \lesssim 1$ deg, meaning a completely face-on configuration, which is possible but is statistically unlikely for our sample size. On the other hand, a low-mass companion is also unlikely because it will need to be below the $\sim$40 \mjup\ level to avoid RV detection, and at that mass level it is not expected to generate a photometric HB signal at the observed amplitudes. In addition, if the binary companion induces an RV amplitude at the level of the RV scatter of the three non-variable systems then we would expect that scatter to be close to the low end of the RV amplitude distribution of the \nsys\ systems with a fitted orbit. However, the latter has a range of 13 -- 65 \kms\ (see \tabr{orbits}), which seems to be distinct than the RV scatter of the non-variable systems, of 0.5 -- 1.4 \kms\ (see \tabr{const}). Therefore, this scenario is considered unlikely.

\subsection{Companion Mass}
\label{sec:m2}

For the \nsys\ systems with a measured orbit, we calculated the companion mass, \mt, using \eqr{fm} that can be rearranged into a cubic polynomial in \mt:
\begin{equation}
\label{eq:m2}
\sin^3i M_2^3 - f(m) M_2^2 - 2 f(m) M_1 M_2 - f(m) M_1^2 = 0,
\end{equation}
which has only one real root. The polynomial coefficients in \eqr{m2} are composed of \fms, which we measured directly from the orbital solution (\tabr{orbits}), \mo, for which we use the values of \citet[][see \tabr{targets}]{huber14}, and $\sin^3i$. To derive the companion mass values and uncertainties we generated an \mt\ distribution by solving for the polynomial roots for a distribution of \fms\ and \mo. 

To address our lack of knowledge of the orbital plane inclination angle, and in turn of the $\sin^3i$ coefficient in \eqr{m2}, we have chosen three approaches. 

In the first approach, we assumed that $\sin^3i=1$ which corresponds to an edge-on system, providing the companion's minimum mass. 

In the second approach, we used the median value of the $\sin^3i$ distribution, which equals 0.6495\footnote{Given the highly asymmetric nature of this distribution we chose to use the median instead of the distribution expectation value (the mean), which equals 0.5890.}. Therefore, the \mt\ estimate derived in this approach reflects our current knowledge of \fms\ and \mo\ and shows the likely value of \mt, and the uncertainty we can hope to achieve once the inclination angle is estimated in the future, for example from modeling the \ik\ light curve. 

In the third approach we used the entire $\sin^3i$ distribution, so the results are an accurate reflection of our current knowledge of \mt.

In the second and third approaches above, we have assumed $i$ is distributed uniformly in $\sin i$ since the orbital angular momentum axis has no preferred direction. One subtlety here is that systems with high inclination angles have larger RV amplitudes, hence are detected more efficiently in RV surveys of binary stars and star-planet systems. However, HB stars analyzed here were identified photometrically and the photometric signal does not have the same dependence on inclination angle as the RV amplitude. For example, the KOI-54 system has a large photometric amplitude of $\sim$0.6 \% despite a face-on configuration with $i = 5.50 \pm 0.10$ deg \citep{welsh11}.

The results of the three approaches are listed in \tabr{m2}, where for completeness we list also \fms\ and \mo. Examining the values of \mo\ and \mt\ shows that for the majority of systems it is likely that \mt\ \textless\ \mo. However, for a few systems the companion mass may be comparable to or larger than the primary mass, although the current uncertainties are large (e.g., KID 4659476, KID 9016693). This raises the possibility that in those systems the secondary is not a main sequence star because in that case we would expect it to dominate the spectrum. Therefore, those systems are candidates for a compact object companion and are interesting targets for further study. Although we should note that those systems could have a large orbital inclination angle, where the companion's mass is close to the minimum mass (first approach above), making the companion a main sequence star with mass close but smaller than the primary mass. This is supported by the detection of HB systems with binary mass ratios close to one \citep{smullen15}. In such systems it might be possible to identify in the spectrum the spectral lines of the secondary. A detailed study of the spectra collected here, including a systematic search for spectral lines of the secondary, is beyond the scope of this work (and will be a subject of a future publication) as here we focus on RV measurements of the primary and measurement of the orbit for a large sample of HB stars.

\input{m2_table.tex}

\subsection{The Eccentricity-Period Relation}
\label{sec:ep}

\begin{figure*}
\begin{center}
\includegraphics[scale=0.6]{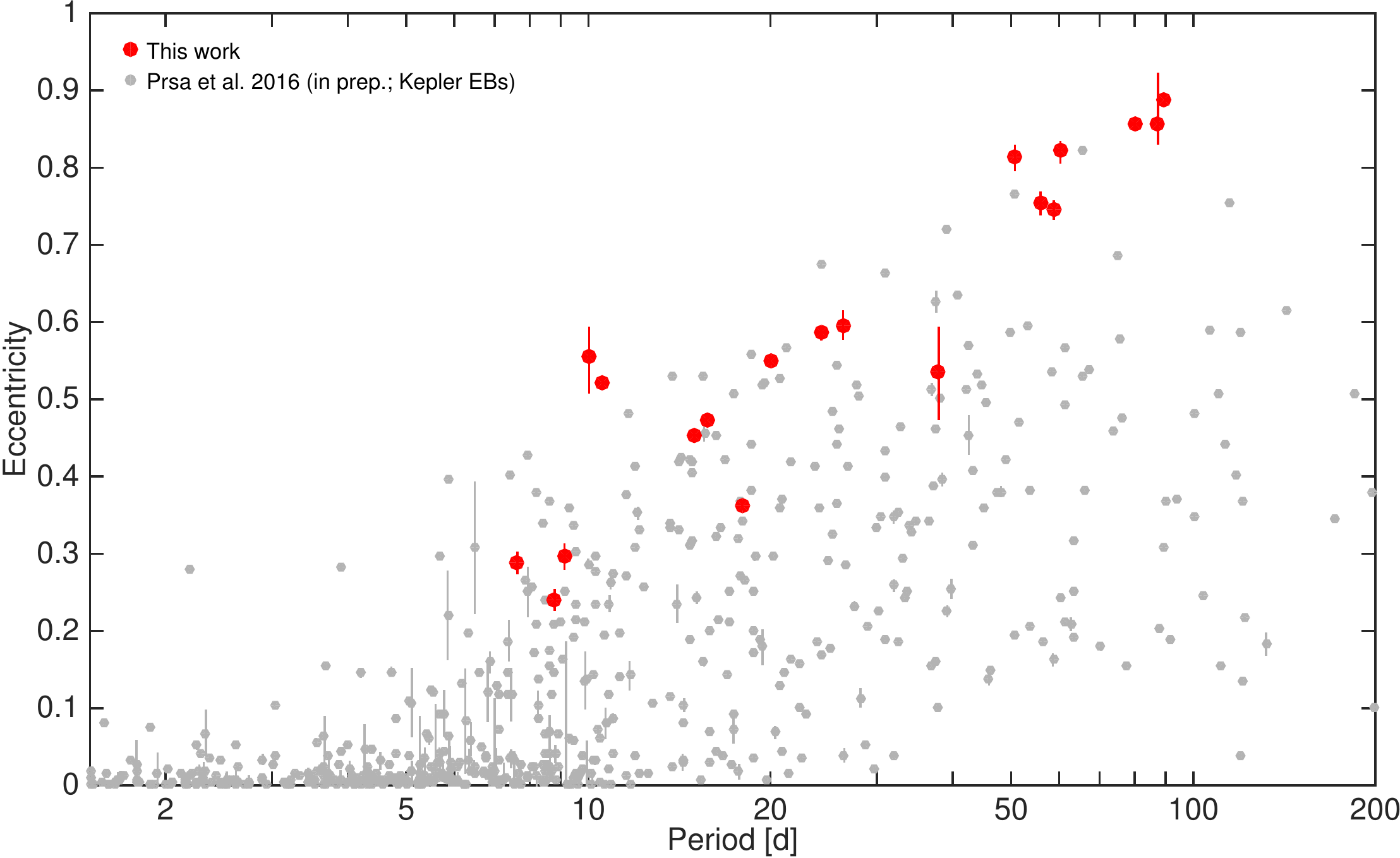}\\
\vspace{6mm}
\includegraphics[scale=0.6]{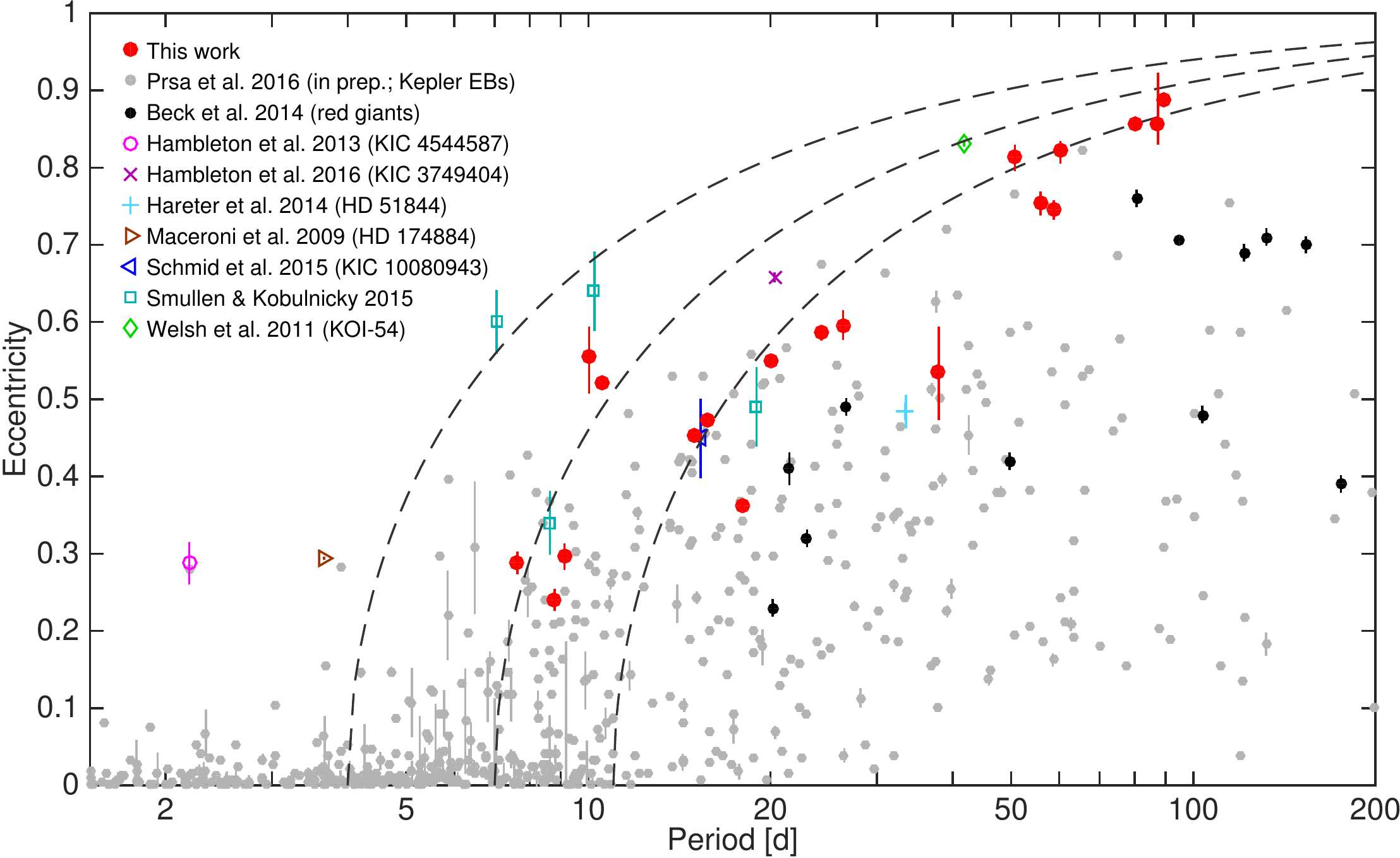}
\caption{\label{fig:ep}
Orbital eccentricity vs.~orbital period. In both panels the \nsys\ HB systems with orbits measured here are shown in red, and in gray we mark \ik\ EBs where the eccentricity was derived through analysis of the eclipse light curves (from Pr{\v s}a et al.~in prep.). The top panel shows how the HB stars are typically positioned at the top envelope of the eccentricity-period distribution. In the bottom panel we add all known HB stars with orbits measured using RVs and $P < 200$ d (see legend and \tabr{hblit}). The dashed gray lines mark an eccentricity-period relation of $e = \sqrt{1 - (P_0/P)^{(2/3)}}$, which is the expected functional form assuming conservation of angular momentum. The three curves use $P_0$ of 4, 7, and 11 d, showing that it is difficult to use a single curve to match the upper envelope of the distribution throughout the entire period range. See \secr{ep} for further discussion.
}
\end{center}
\end{figure*}

\figr{ep} shows the eccentricity -- period ($e-P$) diagram. In the top panel we show the \nsys\ systems whose orbit was measured here marked in red, and for comparison we show in gray \ik\ EBs where the eccentricity was derived through analysis of their eclipse light curve (Pr{\v s}a et al., in prep.). A visual examination of the figure shows that the eccentricity of most HB systems analyzed here is close to the high end of eccentricity range of similar orbital period systems. In other words, {\it HB systems draw the envelope of the  $e-P$ distribution}. The figure also shows that our sample of \nsys\ HB systems encompass most of the period range across which tidal orbital circularization takes place. This is reflected by the wide range of eccentricity of the systems observed here (0.2 -- 0.9), spanning almost the entire eccentricity range.

In the bottom panel of \figr{ep} we added all other HB stars with orbits measured using RVs and with orbital periods within 200 d \citep{maceroni09, welsh11, hambleton13, beck14, hareter14, schmid15, smullen15, hambleton16}. We list those systems in \tabr{hblit}. The dashed gray curves show lines of constant orbital angular momentum with an $e-P$ relation of $e = \sqrt{1 - (P_0/P)^{(2/3)}}$. We plotted lines with circularization period $P_0$ \citep[e.g.,][]{mazeh08} of 4, 7, and 11 d. This was in an attempt to match the envelope of the $e-P$ distribution, although it can be seen that no single curve can match the envelope throughout the entire period range. This suggests that HB systems are born with a range of angular momenta and eventually tidally circularize to a range of periods $P_0$, with $P_0$ typically below 10 d for main sequence binaries. Another possible explanation is that the population of HB systems studied here has a wide age range, since as shown by \cite{meibom05} the circularization period and the shape of the $e-P$ distribution depends on the population age.

\input{hblit_table.tex}

Longer period systems, beyond $P$ = 90 d, were not included in our targets since we wanted to monitor the entire orbit in one observing season. At short periods, the orbital eccentricity grows smaller and the heartbeat signal becomes less concentrated near periastron and instead appears as typical ellipsoidal modulations. Therefore, the occurrence of systems classified as HB systems may decrease at short orbital periods, although the classification becomes somewhat arbitrary. 

Measuring the orbits of additional HB stars, and other high-eccentricity systems, will better shape the $e-P$ upper envelope. Still, a close visual examination of \figr{ep} bottom panel shows that there are several systems with eccentricity well beyond that of similar period systems, raising the possibility they do not belong to the same distribution, which in turn suggests they could be impacted by other physical processes in addition to tidal circularization. Those systems include for example KID 4544587 \citep{hambleton13}, some of the systems studied by \cite{smullen15}, and also two of the systems studied here at $P\sim10$ d and $e\sim0.5$ (KID 6775034 and KID 11649962). Possible mechanisms that can account for the high eccentricity of these systems are interaction with a third body, negligible tidal circularization, or young system age (although the latter is less likely given the typical ages of \ik\ field stars). Therefore, systems with increased eccentricity compared to similar period systems provide an opportunity to study these processes. 

It is also worth noting that the red giant HB systems studied by \citet[][marked in black in \figr{ep} bottom panel]{beck14} show smaller eccentricities than other HB systems at the same period. This is likely the result of the larger radii $R$ of red giant stars, as the circularization time scale is proportional to $R^{-5}$ in tidal theories with a constant lag angle or lag time. The red giant systems also have systematically longer periods than the heartbeat stars we monitored with RVs, which is likely required for these systems to have retained significant eccentricity. However, the substantial eccentricity of these red giant HB systems suggests that their tidal circularization timescales are not extremely short, and we speculate that the occurrence of these eccentric red giant systems is evidence against highly efficient tidal dissipation in sub-giants as has been suggested by \cite{schlaufman13}. A more detailed study examining the stellar radius and semi-major axis distribution of red giant systems is required for a firm conclusion.

\subsection{Tidal Circularization}
\label{sec:tid}

Tidal friction will act to circularize the orbits of HB stars, although the mechanisms and time scales of tidal circularization remain poorly understood. Pioneering works such as \cite{zahn75,zahn77} have suggested that tidal friction is more efficient in stars with convective envelopes where an effective turbulent friction can damp the equilibrium tidal distortion. In stars without convective envelopes, tidal dissipation can still occur via dissipation of dynamical tides (e.g., gravity waves) in the radiative envelope, but may be less efficient. Our sample contains stars on both sides of the convective/radiative transition, and could be used to constrain or revise existing tidal theories. In many HB stars (e.g., KID 9016693 in Fig.~10), tidally excited oscillations are present in the light curve, and can be used to study tidal dissipation via dynamical tides. Although a detailed investigation is beyond the scope of this work, we examine here some basic tidal parameters for our HB systems.

\begin{figure*}
\begin{center}
\includegraphics[scale=0.65]{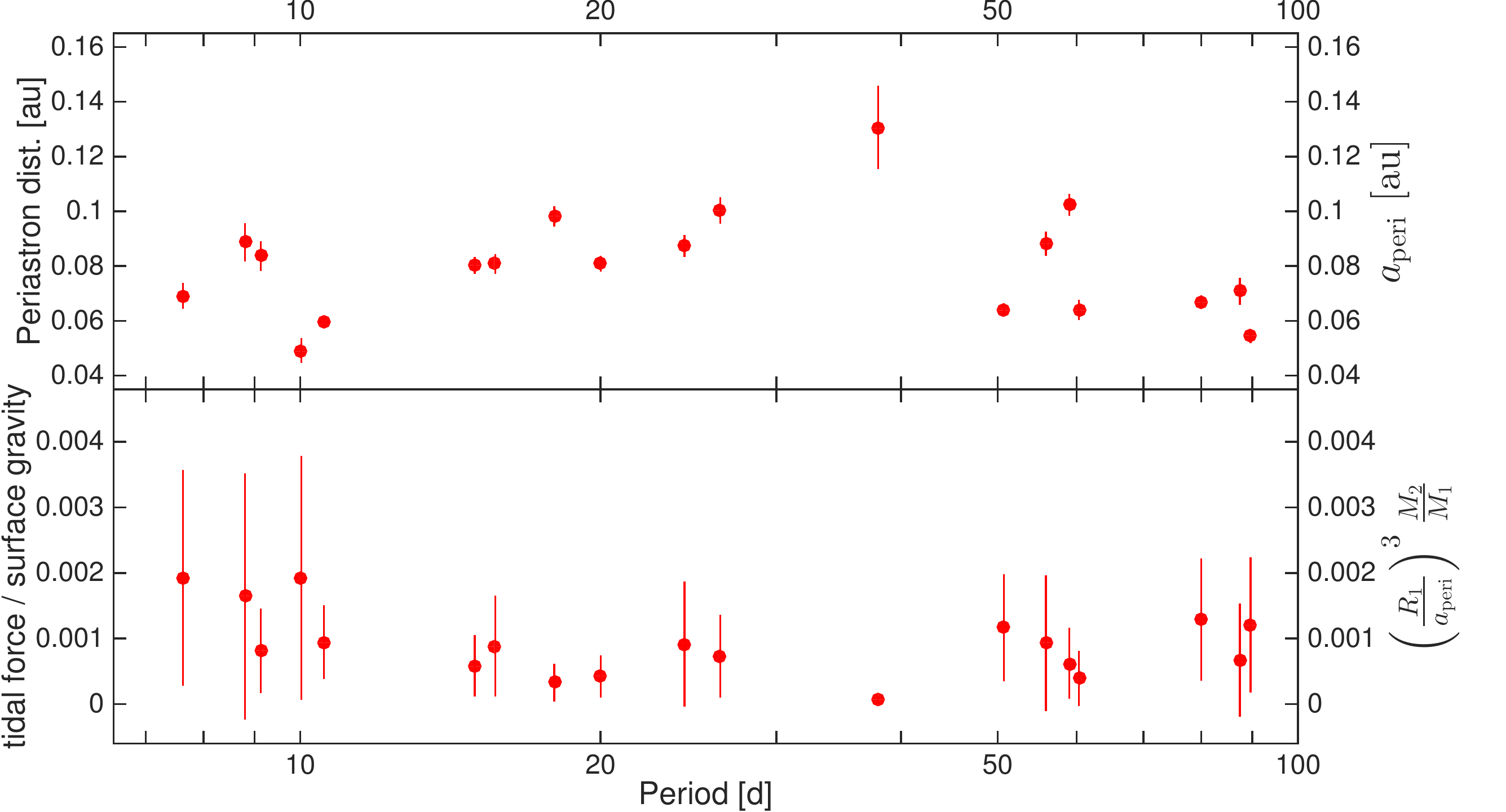}
\caption{\label{fig:peridist}
Periastron distance (top panel; $a_{\rm peri} = a(1-e)$) and the tidal force acting on the primary star at periastron divided by the star's surface gravity (bottom panel; see \eqr{tide}), both as a function of orbital period (x-axis in log scale).
}
\end{center}
\end{figure*}

Since the HB stars' orbital eccentricities show a strong correlation with orbital period, we investigate the period dependence of two other parameters that are closely related to the tidal force acting on the primary star. The first is the periastron distance:
\begin{equation}
\label{eq:peridist}
a_{\rm peri} = a(1-e),
\end{equation}
where $a$ is the orbital semi-major axis. As shown in the top panel of \figr{peridist}, $a_{\rm peri}$ does not show a correlation with period, and has a roughly constant value of 0.080~au with a scatter of 0.021~au. For comparison, the mean and scatter of the semi-major axis (not shown) are 0.26~au and 0.14~au, respectively.

The second parameter we investigate is the tidal force acting on the primary star at periastron divided by the star's surface gravity:
\begin{equation}
\label{eq:tide}
\frac{F_{\rm tide}}{F_{\rm gravity}}=\left(\frac{GR_1M_2}{a_{\rm peri}^3}\right)\left(\frac{GM_1}{R_1^2}\right)^{-1}= \left(\frac{R_1}{a_{\rm peri}}\right)^3\frac{M_2}{M_1},
\end{equation}
where $R_1$ is the primary star's radius (see \tabr{targets}) and $G$ the gravitational constant. Here we used the $M_2$ values derived from the second approach described in \secr{m2}, where in \eqr{m2} $\sin^3(i)$ is replaced by the distribution median of 0.6495 (see \tabr{m2} second column from the right). As can be seen in \figr{peridist} bottom panel, this parameter also does not correlate with period, and the typical error bar is comparable to the scatter. The reason for the latter is that this ratio depends strongly on stellar parameters ($M_1, M_2, R_1$) which are usually less constrained than orbital parameters. The errors on $a_{\rm peri}$ are in comparison much smaller since that parameter depends weakly on the masses of both stars and more strongly on the period and eccentricity.

The relatively small range of $a_{\rm peri}$ and tidal forcing amplitudes for HB stars in \figr{peridist} likely reflects the sensitivity of tidal circularization time scales to the amplitude of tidal forcing (see e.g., \citealt{zahn75,zahn77,hut81}). HB systems with smaller $a_{\rm peri}$ are very rare due to short circularization time scales. Systems with larger $a_{\rm peri}$ are abundant, but exhibit smaller photometric variations and have not been flagged as HB systems. This is also likely to be the cause of the somewhat narrow range in angular momentum per unit mass (which scales as $P_0^{1/3}$) of HB systems in \figr{ep}.

The system with the largest $a_{\rm peri}$ (KID 10334122, $P$=37.95 d, $a_{\rm peri}$ = 0.131$\pm$0.015 au) is not surprisingly also the system with the lowest $F_{\rm tide}/F_{\rm gravity}$ (see \figr{peridist}), since its lower eccentricity (compared to systems with similar period) results in a larger $a_{\rm peri}$ and a weaker tidal force. It is also not surprising to find this system positioned below the envelope in the $e-P$ diagram (see \figr{ep}). That system has a large uncertainty on $a_{\rm peri}$ because of the large uncertainty on its eccentricity ($e=0.534^{+0.060}_{-0.058}$), and it has a small uncertainty on $F_{\rm tide}/F_{\rm gravity}$ resulting from relatively low uncertainties of $M_1$ and $R_1$ (see \tabr{targets}).

\begin{figure*}
\begin{center}
\includegraphics[scale=0.65]{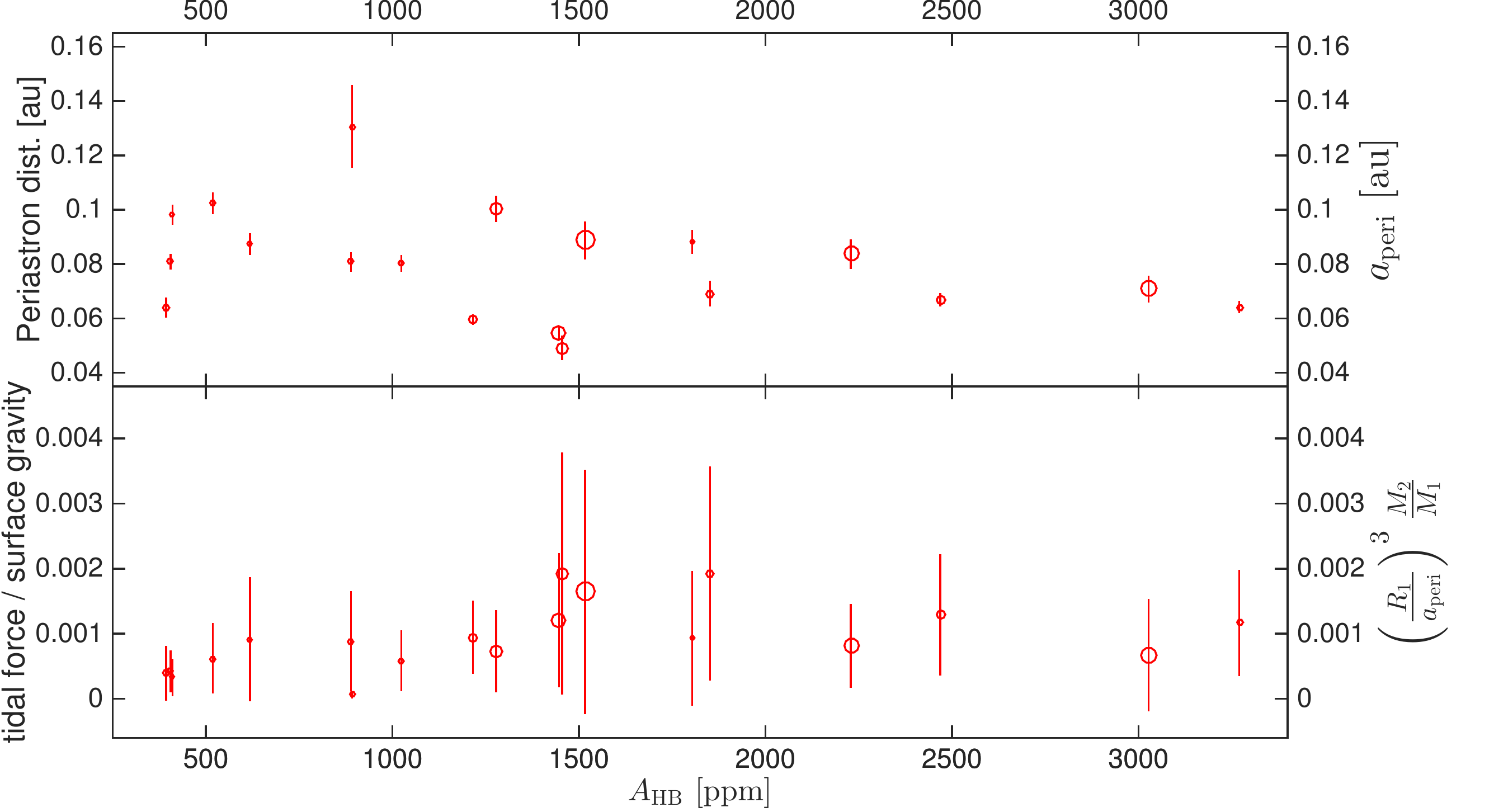}
\caption{\label{fig:photamp}
Periastron distance (top panel; $a_{\rm peri} = a(1-e)$) and the tidal force acting on the primary star at periastron divided by the star's surface gravity (bottom panel; see \eqr{tide}), both as a function of the full amplitude of the photometric HB signal at periastron. In both panels the markers' radius is linear in the primary star's \teff\ (See \tabr{targets}).
}
\end{center}
\end{figure*}

In \figr{photamp} we investigate how the two parameters mentioned above relate to the photometric amplitude of the HB signal during periastron ($A_{\rm HB}$; see \tabr{targets}). The Y-axes of the two panels in \figr{photamp} are the same as in \figr{peridist}, while the X-axis is the photometric amplitude, and the markers' radii are linear in the primary star \teff\ (See \tabr{targets}). The data in both panels do not show a clear correlation, although $A_{\rm HB}$ is expected to increase with decreasing periastron distance and increasing tidal force \citep{kumar95, thompson12}. This suggests that our sample is incomplete and suffers from observational bias, and/or, that our understanding of $A_{\rm HB}$ is incomplete. The data do show, however, that hotter stars have larger $A_{\rm HB}$, perhaps indicating that tidal circularization is less efficient in hotter stars in our sample, although the cause of this correlation is presently unclear.

\subsection{Higher Multiplicity Systems}
\label{sec:multi}

Many of the HB stars examined here may be members of triple or higher multiplicity systems. Several works (e.g., \citealt{duquennoy91,meibom05,tokovinin06,raghavan10}) have found that the fraction of higher multiplicity systems amongst short period and highly eccentric binaries is very high, exceeding 90\% for binaries with $P<3$~d \citep{tokovinin06}. Indeed, tertiary bodies may excite the orbital eccentricity of HB progenitors via Kozai-Lidov oscillations, producing close periastron passages and allowing them to be detected as HB systems. Quadruple systems composed of two binaries may also be common amongst HB systems, and these types of systems have been predicted by \cite{pejcha13}, who found that HB systems formation is greatly enhanced in quadruple systems relative to triple and binary systems.

Despite the probable multiplicity of many HB systems, we find it unlikely that many of our Keplerian RV solutions have been greatly affected by orbital motion induced by a third body. A putative third body exciting the orbital eccentricity would likely have an orbital period, $P_3$, more than several times longer than the HB period (depending on its mass, inclination, and eccentricity) in order to allow long term dynamical stability \citep{kise94}. If a typical HB progenitor system was born with an orbital period $P \gtrsim 30$~d before having its orbital eccentricity excited, we expect $P_3 \gtrsim 1~{\rm yr}$ for most third bodies, with much larger values of $P_3$ in most cases. Although the orbital RV amplitude induced by such a third body can be several \kms, we expect only a small RV change over the $\sim$100~d baseline of our observations, and therefore this motion can be safely included as a constant RV offset. We are currently pursuing a follow-up survey on many of these HB systems that will reveal whether any of them have tertiary companions with orbital periods $P_3 \lesssim 5 \, {\rm yr}$.

\subsection{Future Prospects}
\label{sec:future}

The orbits measured here comprise a relatively large sample obtained with a single instrument, and they can facilitate several follow-up scientific studies in addition to that of tidal circularization and the shape of the $e-P$ distribution (\secr{ep}). Other examples include:

\begin{itemize}

\item Stars in binary systems with highly eccentric orbits are expected to rotate near a ``pseudo-synchronous'' rotation period, where the stellar rotation is synchronized with the orbital motion close to periastron which is faster than the mean orbital motion along the entire orbit. The pseudo-synchronous rotation rate depends on eccentricity and orbital period \citep{hut81}. However, the pseudo-synchronous rotation period also depends on the tidal prescription adopted (e.g., constant tidal $Q$ and constant time lag yield different predictions). The orbital eccentricities measured here, combined with the orbital periods, allow for a calculation of the expected pseudo-synchronous rotation periods for these stars. If the rotation periods can be directly measured, for example in HB systems also showing stellar activity, this will allow for direct tests of tidal theories. We list in \tabr{psynch} the theoretically predicted pseudo-synchronous rotation period $P_{\rm ps}$ derived using Eq.~42 of \cite{hut81}. We also list the orbital period and eccentricity in the same table. 

\input{psynch_table.tex}

\item Combining the orbital RV model parameters constrained here (especially $e$ and $K$) with models of the \ik\ HB light curve, which constrain $i$, results in an improved measurement of \mt\ (see \secr{m2}). That will allow to measure the tidal force acting on the primary star, and in turn the relation between that force and the HB photometric amplitude. 

\item Measurement of the orbital inclination angle (see previous paragraph) and the stellar spin inclination angle (when possible) gives the sky-projection of the stellar obliquity, which is a key parameter in understanding binary stars formation and orbital evolution \citep[e.g.,][]{naoz14}. The stellar spin inclination angle can be constrained by measuring the stellar rotation period (for example from stellar activity), the sky-projected rotation rate (from rotational broadening of spectral lines), and stellar radius (using spectroscopy and light curve modeling).

\item The measurement of the RV orbit paves the way for a detailed analysis of individual systems showing tidal pulsations \citep[e.g.,][]{welsh11,fuller12,burkart12,hambleton13,oleary14,hambleton16}.

\item Continued RV monitoring can unveil extraneous bodies in the system by looking for a long term RV trend. This can be complemented by other data sets, such as spectroscopy, high angular resolution imaging, astrometry, and searching for pulsation phase/frequency modulation in the \ik\ data \citep{shibahashi12, murphy14}. Measuring the occurrence rate of a third object will allow for testing formation and orbital evolution theory \citep[e.g.][]{naoz16}. Such a combined approach of using several different data sets to measure the occurrence rate of a third object has already proven successful in the study of short period gas giant planets \citep{knutson14, ngo15, piskorz15}.

\end{itemize}

\section{Summary}
\label{sec:sum}

We have presented here the first results from our RV monitoring campaign of \ik\ HB stars. Our results include a sample of \nsys\ \ik\ HB stars in the orbital period range of $7-90$~d for which we derived a Keplerian orbital solution from RV monitoring using Keck/HIRES. This is currently the largest sample of HB stars for which an RV orbit was obtained using a single instrument, and it roughly doubles the number of such systems with periods up to 90~d.

We have shown that HB stars populate the upper envelope of the $e-P$ diagram, which is a distinguishing feature for testing tidal circularization theories. This sample will support additional studies that require a sample of HB systems with well measured orbits, for example testing pseudo-synchronization theory and examining the physics of tidally excited stellar pulsations.

We also presented three objects for which we did not detect RV variability, and list a few possible scenarios that can explain that observation. Those objects should be studied in more detail in order to explain the RV non-variability.

We plan to continue our RV monitoring in order to increase the sample size by measuring the orbits of additional HB systems, and to look for long term RV trends indicative of a third object in the system. We will pursue the latter by complementing the RV monitoring with spectroscopy, imaging, astrometry, and examining the \ik\ light curve for modulations in the pulsation phases.

Finally, we note that the ongoing K2 mission \citep{howell14} and the future TESS mission \citep{ricker14} and PLATO mission \citep{rauer14} are expected to detect many more HB systems. For K2 and TESS, the stars are expected to be typically brighter than \ik\ HB stars and therefore more accessible to RV monitoring. They are also expected to have shorter orbital periods due to the shorter temporal coverage, which is useful for studying the transition period below which binary systems are fully circularized.

\acknowledgments

We are grateful to the referee, Maxwell Moe, for his thorough reading of the manuscript and his meticulous comments that have helped improve this paper.
We warmly thank Ben Fulton, Evan Sinukoff, Lauren Weiss, Lea Hirsch, Erik Petigura, and Geoff Marcy for contributions to the Keck/HIRES observations.
This work was performed in part at the Jet Propulsion Laboratory, under contract with the California Institute of Technology (Caltech) funded by NASA through the Sagan Fellowship Program executed by the NASA Exoplanet Science Institute.
JF acknowledges partial support from NSF under grant no.~AST-1205732 and through a Lee DuBridge Fellowship at Caltech.
The authors wish to recognize and acknowledge the very significant cultural role and reverence that the summit of Mauna Kea has always had within the indigenous Hawaiian community.  We are most fortunate to have the opportunity to conduct observations from this mountain.
Some of the data presented in this paper were obtained from the Mikulski Archive for Space Telescopes (MAST). STScI is operated by the Association of Universities for Research in Astronomy, Inc., under NASA contract NAS5-26555. Support for MAST for non-HST data is provided by the NASA Office of Space Science via grant NNX09AF08G and by other grants and contracts.
This research has made use of NASA's Astrophysics Data System Service.
This paper includes data collected by the \ik\ mission. Funding for the \ik\ mission is provided by the NASA Science Mission directorate.
We acknowledge the support of the \ik\ Guest Observer Program.


{\it Facilities: \ik, Keck:I (HIRES)} 



\clearpage
\appendix

\section{\thesection. Radial Velocity Table}
\label{app:rvtab}

\tabr{rvs} lists all \nrvs\ Keck/HIRES RV measurements obtained here. The table columns include KIC ID, mid exposure time (BJD), RV (\kms), and RV error (\kms).

\input{RVs_table.tex}

\section{\thesection. Simultaneous light curve and radial velocity curve plots }
\label{app:rvlc}

Figures 9 through 12 show the phase-folded RV curve (measurements in black, model in red) along with the phase-folded and binned \ik\ light curve (blue). Figures 9 through 11 show the \nsys\ systems for which we fitted a Keplerian orbit and Fig.~12 shows the \nconst\ systems for which we do not detect RV variability. We did not analyze the \ik\ data here and the light curves are shown for completeness.

\iftoggle{figures}{
\input{figures_rvlc1.tex}
\clearpage
\input{figures_rvlc2.tex}
\clearpage
\input{figures_rvlc3.tex}
\input{figures_rvlc_const.tex}
}

\end{document}

%% file: mynewcommands.tex
\newcommand{\mo}{\ensuremath{M_1}}
\newcommand{\mt}{\ensuremath{M_2}}

\newcommand{\fms}{\ensuremath{f(m)}}

\newcommand{\mjup}{\ensuremath{M_{\rm J}}}

\newcommand{\teff}{\ensuremath{T_{\rm eff}}}
\newcommand{\logg}{\ensuremath{\log g}}

\newcommand{\kms}{km~s$^{-1}$}

\newcommand{\ik}{{\it Kepler}}

\newcommand{\kp}{$K_{p}$}

\newcommand{\sig}[1]{\ensuremath{#1\sigma}}

\newcommand{\figr}[1]{Fig.~\ref{fig:#1}}

\newcommand{\secr}[1]{Sec.~\ref{sec:#1}}
\newcommand{\appr}[1]{Appendix~\ref{app:#1}}
 
\newcommand{\eqr}[1]{Eq.~\ref{eq:#1}}

\newcommand{\tabr}[1]{\mbox{Table~\ref{tab:#1}}}

\newcommand{\ecosw}{$\sqrt{e}\cos\omega$}
\newcommand{\esinw}{$\sqrt{e}\sin\omega$}

%% file: targets_table.tex
\begin{deluxetable*}{rrrrlccc}
\tablecaption{\label{tab:targets} Properties of the heartbeat systems and the primary stars for the \nsystot\ systems studied here.} 
\tablewidth{0pt}
\tablehead{\multicolumn{1}{c}{KID} & \multicolumn{1}{c}{$P$\tablenotemark{a}} & \multicolumn{1}{c}{$A_{\rm HB}$\tablenotemark{b}} & \multicolumn{1}{c}{$K_{\rm p}$}  & \multicolumn{1}{c}{$T_{\rm eff}$\tablenotemark{c}} &  \multicolumn{1}{c}{$\log g$\tablenotemark{c}} &  \multicolumn{1}{c}{$R_1$\tablenotemark{c}}         &  \multicolumn{1}{c}{$M_1$\tablenotemark{c}} \\ 
                                     & \multicolumn{1}{c}{[d]} & \multicolumn{1}{c}{[ppm]} & \multicolumn{1}{c}{[mag]}         & \multicolumn{1}{c}{[K]}            &                                 &  \multicolumn{1}{c}{[R$_{\sun}$]} &  \multicolumn{1}{c}{[M$_{\sun}$]} }
\startdata
\rule{0pt}{3ex}   4659476$\ \, $ &    58.99637$\pm$3.7e-04 &   520 &  13.22 & 6384$^{+ 155}_{-174}$  & 3.97$^{+0.24}_{-0.13}$  & 1.96$^{+0.42}_{-0.63}$ & 1.31$^{+0.20}_{-0.22}$ \\  
\rule{0pt}{3ex}   5017127$\ \, $ &   20.006404$\pm$7.8e-05 &   410 &  12.51 & 6440$^{+ 155}_{-175}$  & 4.13$^{+0.19}_{-0.13}$  & 1.58$^{+0.35}_{-0.38}$ & 1.25$^{+0.15}_{-0.19}$ \\  
\rule{0pt}{3ex}   5090937$\ \, $ &    8.800693$\pm$2.4e-05 &  1520 &  11.04 & 8092$^{+ 224}_{-336}$  & 3.73$^{+0.41}_{-0.11}$  & 3.27$^{+0.82}_{-1.52}$ & 2.07$^{+0.34}_{-0.50}$ \\  
\rule{0pt}{3ex}   5790807$\ \, $ &    79.99625$\pm$5.4e-04 &  2470 &   9.95 & 6796$^{+  67}_{ -88}$  & 3.88$^{+0.22}_{-0.10}$  & 2.49$^{+0.39}_{-0.73}$ & 1.72$^{+0.15}_{-0.24}$ \\  
\rule{0pt}{3ex}   5818706$\ \, $ &   14.959941$\pm$5.1e-05 &  1020 &  11.49 & 6375$^{+ 162}_{-178}$  & 4.06$^{+0.25}_{-0.13}$  & 1.64$^{+0.35}_{-0.46}$ & 1.12$^{+0.19}_{-0.14}$ \\  
\rule{0pt}{3ex}   5877364$\ \, $ &    89.64854$\pm$6.4e-04 &  1450 &   8.88 & 7502$^{+ 234}_{-313}$  & 4.09$^{+0.21}_{-0.15}$  & 1.76$^{+0.49}_{-0.44}$ & 1.38$^{+0.20}_{-0.23}$ \\  
\rule{0pt}{3ex}   5960989$\ \, $ &    50.72153$\pm$3.0e-04 &  3270 &  12.51 & 6471$^{+  77}_{ -89}$  & 4.05$^{+0.17}_{-0.12}$  & 1.87$^{+0.35}_{-0.42}$ & 1.43$^{+0.12}_{-0.15}$ \\  
\rule{0pt}{3ex}   6370558$\ \, $ &    60.31658$\pm$3.7e-04 &   390 &  12.28 & 6526$^{+ 182}_{-251}$  & 4.02$^{+0.26}_{-0.18}$  & 1.98$^{+0.63}_{-0.63}$ & 1.48$^{+0.21}_{-0.28}$ \\  
\rule{0pt}{3ex}   6775034$\ \, $ &   10.028547$\pm$2.9e-05 &  1460 &  13.99 & 7187$^{+ 228}_{-304}$  & 4.04$^{+0.27}_{-0.16}$  & 1.81$^{+0.53}_{-0.53}$ & 1.30$^{+0.23}_{-0.19}$ \\  
\rule{0pt}{3ex}   8027591$\ \, $ &    24.27443$\pm$1.0e-04 &   620 &  11.42 & 6279$^{+ 199}_{-221}$  & 3.87$^{+0.41}_{-0.14}$  & 2.28$^{+0.55}_{-0.94}$ & 1.40$^{+0.20}_{-0.28}$ \\  
\rule{0pt}{3ex}   8164262$\ \, $ &    87.45717$\pm$6.4e-04 &  3030 &  13.36 & 7700$^{+ 237}_{-316}$  & 4.02$^{+0.19}_{-0.14}$  & 2.11$^{+0.53}_{-0.58}$ & 1.69$^{+0.20}_{-0.30}$ \\  
\rule{0pt}{3ex}   9016693$\ \, $ &    26.36803$\pm$1.2e-04 &  1280 &  11.63 & 7262$^{+ 201}_{-327}$  & 4.01$^{+0.21}_{-0.17}$  & 2.07$^{+0.52}_{-0.58}$ & 1.60$^{+0.20}_{-0.33}$ \\  
\rule{0pt}{3ex}   9965691$\ \, $ &   15.683195$\pm$5.5e-05 &   890 &  13.10 & 6407$^{+ 174}_{-174}$  & 3.89$^{+0.27}_{-0.12}$  & 2.19$^{+0.47}_{-0.70}$ & 1.35$^{+0.22}_{-0.22}$ \\  
\rule{0pt}{3ex}   9972385\tablenotemark{d}  &    58.42211$\pm$3.5e-04 &  1400 &  11.50 & 6313$^{+ 170}_{-170}$  & 3.92$^{+0.32}_{-0.11}$  & 1.82$^{+0.39}_{-0.59}$ & 1.01$^{+0.16}_{-0.14}$ \\  
\rule{0pt}{3ex}  10334122$\ \, $ &    37.95286$\pm$2.0e-04 &   890 &  12.85 & 6363$^{+ 144}_{-192}$  & 4.334$^{+0.088}_{-0.143}$  & 1.20$^{+0.27}_{-0.17}$ & 1.14$^{+0.14}_{-0.13}$ \\  
\rule{0pt}{3ex}  11071278$\ \, $ &    55.88522$\pm$3.3e-04 &  1800 &  11.37 & 6215$^{+ 202}_{-247}$  & 3.85$^{+0.49}_{-0.12}$  & 2.19$^{+0.48}_{-1.03}$ & 1.23$^{+0.19}_{-0.28}$ \\  
\rule{0pt}{3ex}  11122789\tablenotemark{d}  &    3.238154$\pm$5.7e-06 &   260 &   9.64 & 7161$^{+ 172}_{-237}$  & 3.82$^{+0.42}_{-0.10}$  & 2.51$^{+0.49}_{-1.14}$ & 1.52$^{+0.20}_{-0.37}$ \\  
\rule{0pt}{3ex}  11403032$\ \, $ &    7.631634$\pm$2.0e-05 &  1850 &  11.50 & 6657$^{+ 149}_{-199}$  & 3.74$^{+0.28}_{-0.10}$  & 2.87$^{+0.45}_{-1.05}$ & 1.65$^{+0.20}_{-0.36}$ \\  
\rule{0pt}{3ex}  11409673\tablenotemark{d}  &   12.317869$\pm$3.9e-05 &  3820 &  12.88 & 7516$^{+  75}_{ -82}$  & 4.097$^{+0.095}_{-0.116}$  & 1.88$^{+0.35}_{-0.24}$ & 1.62$^{+0.12}_{-0.12}$ \\  
\rule{0pt}{3ex}  11649962$\ \, $ &   10.562737$\pm$3.1e-05 &  1220 &  11.41 & 6756$^{+ 151}_{-219}$  & 4.274$^{+0.092}_{-0.138}$  & 1.37$^{+0.31}_{-0.18}$ & 1.29$^{+0.15}_{-0.20}$ \\  
\rule{0pt}{3ex}  11923629$\ \, $ &   17.973284$\pm$6.7e-05 &   410 &  12.26 & 6250$^{+ 169}_{-169}$  & 3.93$^{+0.33}_{-0.11}$  & 1.77$^{+0.34}_{-0.59}$ & 0.97$^{+0.14}_{-0.12}$ \\  
\rule{0pt}{3ex}  12255108$\ \, $ &    9.131526$\pm$2.5e-05 &  2230 &  11.59 & 7577$^{+ 237}_{-316}$  & 4.04$^{+0.15}_{-0.15}$  & 2.11$^{+0.54}_{-0.49}$ & 1.79$^{+0.18}_{-0.29}$ 
\enddata
\tablenotetext{a}{Photometric period, taken from the \ik\ EB catalog \citep{kirk16}.}\tablenotetext{b}{Photometric amplitude of the heartbeat signal at periastron, defined as the full flux variation, in ppm, of the phase folded and binned light curve (see Appendix B). The typical uncertainty is a few percents.}\tablenotetext{c}{Parameters taken from the revised KIC \citep{huber14}.}\tablenotetext{d}{RV non-variable star.}
\end{deluxetable*}

%% file: orbits_table.tex
\begin{deluxetable*}{rrcrrcrc}
\tablecaption{\label{tab:orbits} Orbital properties of the 19 heartbeat systems measured here.} 
\tablewidth{0pt}
\tablehead{\multicolumn{1}{c}{KID} & \multicolumn{1}{c}{$P$}   & \multicolumn{1}{c}{$T_0$}    & \multicolumn{1}{c}{$K$}     &  \multicolumn{1}{c}{$\gamma$} &  \multicolumn{1}{c}{$e$} & \multicolumn{1}{c}{$\omega$} & \multicolumn{1}{c}{$f(m)$}         \\   
                                     & \multicolumn{1}{c}{[d]} & \multicolumn{1}{c}{[BJD-2457000]} & \multicolumn{1}{c}{[\kms]} &  \multicolumn{1}{c}{[\kms]}   &                           &  \multicolumn{1}{c}{[rad]}    & \multicolumn{1}{c}{[$M_{\sun}$]}   }                   
\startdata
\rule{0pt}{3ex}   4659476 &    58.99637$\pm$3.7e-04 &     229.552$^{+0.096}_{-0.097}$ &   51.21$^{+ 0.99}_{- 0.99}$ &     17.1$^{+ 1.0}_{- 1.0}$ &   0.745$^{+ 0.011}_{- 0.011}$ &  -2.884$^{+ 0.048}_{- 0.048}$ &   0.243$^{+ 0.018}_{- 0.017}$   \\  
\rule{0pt}{3ex}   5017127 &   20.006404$\pm$7.8e-05 &     252.075$^{+0.033}_{-0.034}$ &   41.59$^{+ 0.33}_{- 0.33}$ &   -10.25$^{+0.29}_{-0.29}$ &  0.5504$^{+0.0050}_{-0.0050}$ &  -0.779$^{+ 0.022}_{- 0.022}$ &  0.0868$^{+0.0023}_{-0.0023}$   \\  
\rule{0pt}{3ex}   5090937 &    8.800693$\pm$2.4e-05 &     245.879$^{+0.092}_{-0.081}$ &   36.34$^{+ 0.42}_{- 0.40}$ &   -20.12$^{+0.28}_{-0.28}$ &   0.241$^{+ 0.013}_{- 0.013}$ &  -0.470$^{+ 0.070}_{- 0.067}$ &  0.0400$^{+0.0017}_{-0.0016}$   \\  
\rule{0pt}{3ex}   5790807 &    79.99625$\pm$5.4e-04 &     200.708$^{+0.053}_{-0.058}$ &   24.39$^{+ 0.30}_{- 0.28}$ &   -27.39$^{+0.27}_{-0.26}$ &  0.8573$^{+0.0030}_{-0.0031}$ &   2.728$^{+ 0.026}_{- 0.026}$ &  0.0164$^{+0.00067}_{-0.00065}$   \\  
\rule{0pt}{3ex}   5818706 &   14.959941$\pm$5.1e-05 &     232.392$^{+0.016}_{-0.016}$ &   48.16$^{+ 0.26}_{- 0.26}$ &    25.96$^{+0.16}_{-0.16}$ &  0.4525$^{+0.0038}_{-0.0039}$ &  -1.615$^{+ 0.010}_{- 0.010}$ &  0.1228$^{+0.0023}_{-0.0022}$   \\  
\rule{0pt}{3ex}   5877364 &    89.64854$\pm$6.4e-04 &     237.839$^{+0.088}_{-0.120}$ &    33.2$^{+  2.0}_{-  1.3}$ &     3.87$^{+0.19}_{-0.20}$ &  0.8875$^{+0.0031}_{-0.0031}$ &  -1.452$^{+ 0.018}_{- 0.018}$ &  0.0334$^{+0.00080}_{-0.00070}$   \\  
\rule{0pt}{3ex}   5960989 &    50.72153$\pm$3.0e-04 &     214.915$^{+0.068}_{-0.063}$ &    39.6$^{+  1.4}_{-  1.2}$ &   -24.38$^{+0.90}_{-0.85}$ &   0.813$^{+ 0.017}_{- 0.015}$ &   0.661$^{+ 0.058}_{- 0.059}$ &  0.0645$^{+0.0043}_{-0.0040}$   \\  
\rule{0pt}{3ex}   6370558 &    60.31658$\pm$3.7e-04 &      247.28$^{+ 0.13}_{- 0.13}$ &   12.97$^{+ 0.53}_{- 0.22}$ &  -31.319$^{+0.098}_{-0.107}$ &   0.821$^{+ 0.015}_{- 0.012}$ &  -2.475$^{+ 0.025}_{- 0.024}$ &  0.0025$^{+0.00020}_{-0.00013}$   \\  
\rule{0pt}{3ex}   6775034 &   10.028547$\pm$2.9e-05 &     221.759$^{+0.064}_{-0.051}$ &    39.1$^{+  1.4}_{-  1.1}$ &     11.5$^{+ 1.6}_{- 1.7}$ &   0.556$^{+ 0.047}_{- 0.037}$ &   0.213$^{+ 0.061}_{- 0.054}$ &  0.0356$^{+0.0039}_{-0.0034}$   \\  
\rule{0pt}{3ex}   8027591 &    24.27443$\pm$1.0e-04 &     228.334$^{+0.058}_{-0.060}$ &   38.83$^{+ 0.74}_{- 0.72}$ &     4.51$^{+0.59}_{-0.58}$ &  0.5854$^{+0.0082}_{-0.0083}$ &   0.502$^{+ 0.034}_{- 0.033}$ &  0.0785$^{+0.0049}_{-0.0046}$   \\  
\rule{0pt}{3ex}   8164262 &    87.45717$\pm$6.4e-04 &      243.08$^{+ 0.14}_{- 0.27}$ &    22.9$^{+  9.5}_{-  4.7}$ &     14.4$^{+ 1.4}_{- 1.4}$ &   0.857$^{+ 0.026}_{- 0.065}$ &    1.61$^{+  0.29}_{-  0.28}$ &  0.0148$^{+0.0044}_{-0.0031}$   \\  
\rule{0pt}{3ex}   9016693 &    26.36803$\pm$1.2e-04 &      268.97$^{+ 0.10}_{- 0.11}$ &    56.3$^{+  2.4}_{-  2.2}$ &      8.3$^{+ 1.1}_{- 1.0}$ &   0.596$^{+ 0.018}_{- 0.018}$ &   1.892$^{+ 0.085}_{- 0.093}$ &   0.253$^{+ 0.028}_{- 0.024}$   \\  
\rule{0pt}{3ex}   9965691 &   15.683195$\pm$5.5e-05 &     224.082$^{+0.013}_{-0.013}$ &   36.83$^{+ 0.13}_{- 0.13}$ &   -33.53$^{+0.12}_{-0.12}$ &  0.4733$^{+0.0032}_{-0.0032}$ &  0.7870$^{+0.0094}_{-0.0093}$ &  0.0555$^{+0.00057}_{-0.00057}$   \\  
\rule{0pt}{3ex}  10334122 &    37.95286$\pm$2.0e-04 &      223.79$^{+ 0.60}_{- 0.48}$ &    40.3$^{+  3.7}_{-  3.1}$ &    -14.3$^{+ 1.6}_{- 1.6}$ &   0.534$^{+ 0.060}_{- 0.058}$ &    1.96$^{+  0.16}_{-  0.16}$ &   0.155$^{+ 0.031}_{- 0.028}$   \\  
\rule{0pt}{3ex}  11071278 &    55.88522$\pm$3.3e-04 &      227.73$^{+ 0.14}_{- 0.15}$ &    38.7$^{+  7.1}_{-  4.2}$ &     3.34$^{+0.61}_{-0.75}$ &   0.755$^{+ 0.015}_{- 0.013}$ &  -2.725$^{+ 0.034}_{- 0.035}$ &   0.094$^{+ 0.023}_{- 0.014}$   \\  
\rule{0pt}{3ex}  11403032 &    7.631634$\pm$2.0e-05 &     229.849$^{+0.078}_{-0.084}$ &   30.18$^{+ 0.60}_{- 0.60}$ &    14.74$^{+0.40}_{-0.40}$ &   0.288$^{+ 0.013}_{- 0.013}$ &  -0.556$^{+ 0.082}_{- 0.083}$ &  0.0191$^{+0.0011}_{-0.0011}$   \\  
\rule{0pt}{3ex}  11649962 &   10.562737$\pm$3.1e-05 &    223.4759$^{+0.0096}_{-0.0094}$ &   65.39$^{+ 0.34}_{- 0.36}$ &   -14.02$^{+0.32}_{-0.34}$ &  0.5206$^{+0.0035}_{-0.0035}$ &  2.8229$^{+0.0086}_{-0.0083}$ &  0.1905$^{+0.0041}_{-0.0039}$   \\  
\rule{0pt}{3ex}  11923629 &   17.973284$\pm$6.7e-05 &     223.607$^{+0.038}_{-0.039}$ &   35.84$^{+ 0.22}_{- 0.21}$ &   -16.93$^{+0.15}_{-0.15}$ &  0.3629$^{+0.0058}_{-0.0059}$ &   2.280$^{+ 0.019}_{- 0.019}$ &  0.0694$^{+0.0012}_{-0.0012}$   \\  
\rule{0pt}{3ex}  12255108 &    9.131526$\pm$2.5e-05 &     221.060$^{+0.086}_{-0.110}$ &    48.8$^{+  1.2}_{-  1.1}$ &    -7.15$^{+0.98}_{-0.94}$ &   0.296$^{+ 0.016}_{- 0.015}$ &   2.647$^{+ 0.102}_{- 0.096}$ &  0.0957$^{+0.0080}_{-0.0073}$   
\enddata
\end{deluxetable*}

%% file: stats_table.tex
\begin{deluxetable*}{rrrcll}
\tablecaption{\label{tab:stats} Statistical quantities describing the RV Keplerian model fits. Columns include (from left to right): KIC ID, best fit $\chi^2$, number of degrees of freedom, number of fitting iterations, the additive systematic RV uncertainty, and the RV residuals scatter.} 
\tablewidth{0pt}
\tablehead{\multicolumn{1}{c}{KID} & \multicolumn{1}{c}{$\chi^2$} &  \multicolumn{1}{c}{$\nu$} &  \multicolumn{1}{c}{\#Iter} &  \multicolumn{1}{c}{RV Jitter} &  \multicolumn{1}{c}{Res.~StD}  \\  
                                     &                                &                              &                               &  \multicolumn{1}{c}{[\kms]}  &  \multicolumn{1}{c}{[\kms]}  }      
\startdata
  4659476 &  4.7 &      2  &  2 &    1.9 &    2.2 \\ 
  5017127 &  6.3 &      5  &  2 &   0.63 &   0.40 \\ 
  5090937 &  4.8 &      6  &  1 &  -- &    0.61 \\ 
  5790807 &  4.1 &      7  &  1 &  -- &    0.21 \\ 
  5818706 &  5.1 &      6  &  1 &  -- &    0.31 \\ 
  5877364 & 10.8 &      5  &  1 &  -- &    0.38 \\ 
  5960989 &  9.9 &      5  &  1 &  -- &     2.6 \\ 
  6370558 &  3.2 &      2  &  1 &  -- &   0.042 \\ 
  6775034 &  3.1 &      4  &  2 &   0.99 &   0.57 \\ 
  8027591 &  6.5 &      5  &  2 &    1.0 &   0.71 \\ 
  8164262 &  7.1 &      5  &  2 &    2.4 &    1.4 \\ 
  9016693 &  4.8 &      3  &  2 &    2.3 &    1.7 \\ 
  9965691 &  3.6 &      3  &  1 &  -- &    0.20 \\ 
 10334122 &  2.4 &      2  &  2 &    3.6 &    3.0 \\ 
 11071278 &  5.3 &      3  &  1 &  -- &    0.23 \\ 
 11403032 &  7.1 &      7  &  2 &   0.89 &   0.78 \\ 
 11649962 &  4.1 &      5  &  1 &  -- &    0.35 \\ 
 11923629 &  4.9 &      3  &  2 &   0.33 &   0.16 \\ 
 12255108 & 10.3 &      6  &  1 &  -- &     2.4
\enddata
\end{deluxetable*}

%% file: figures_orbits1.tex

\def\figwidth{3.4in}

\stepcounter{figure} 

\begin{table}[h]
\begin{center}
\begin{tabular}{ll}
 \includegraphics[width=\figwidth]{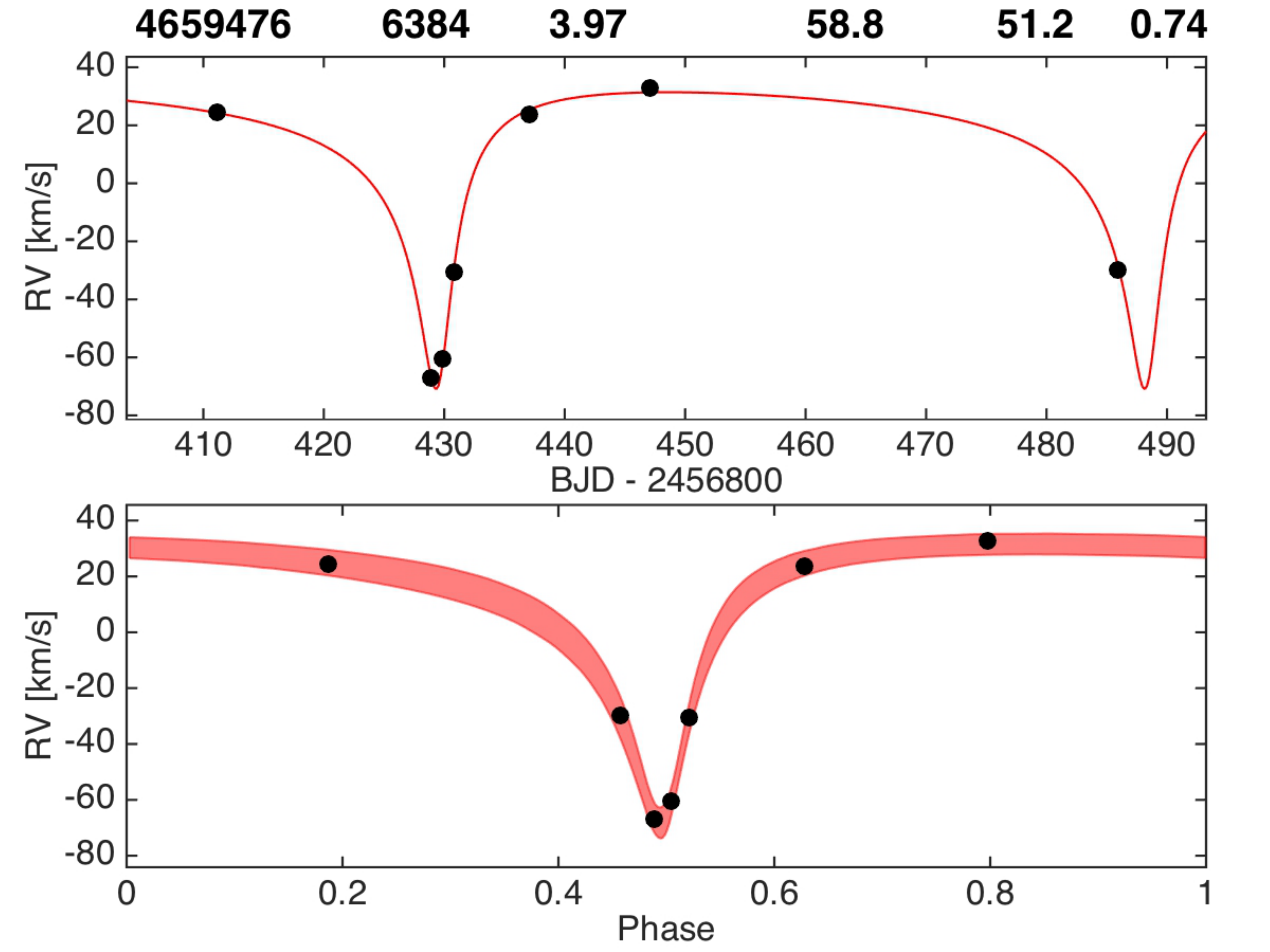} \vspace{5mm}&
 \includegraphics[width=\figwidth]{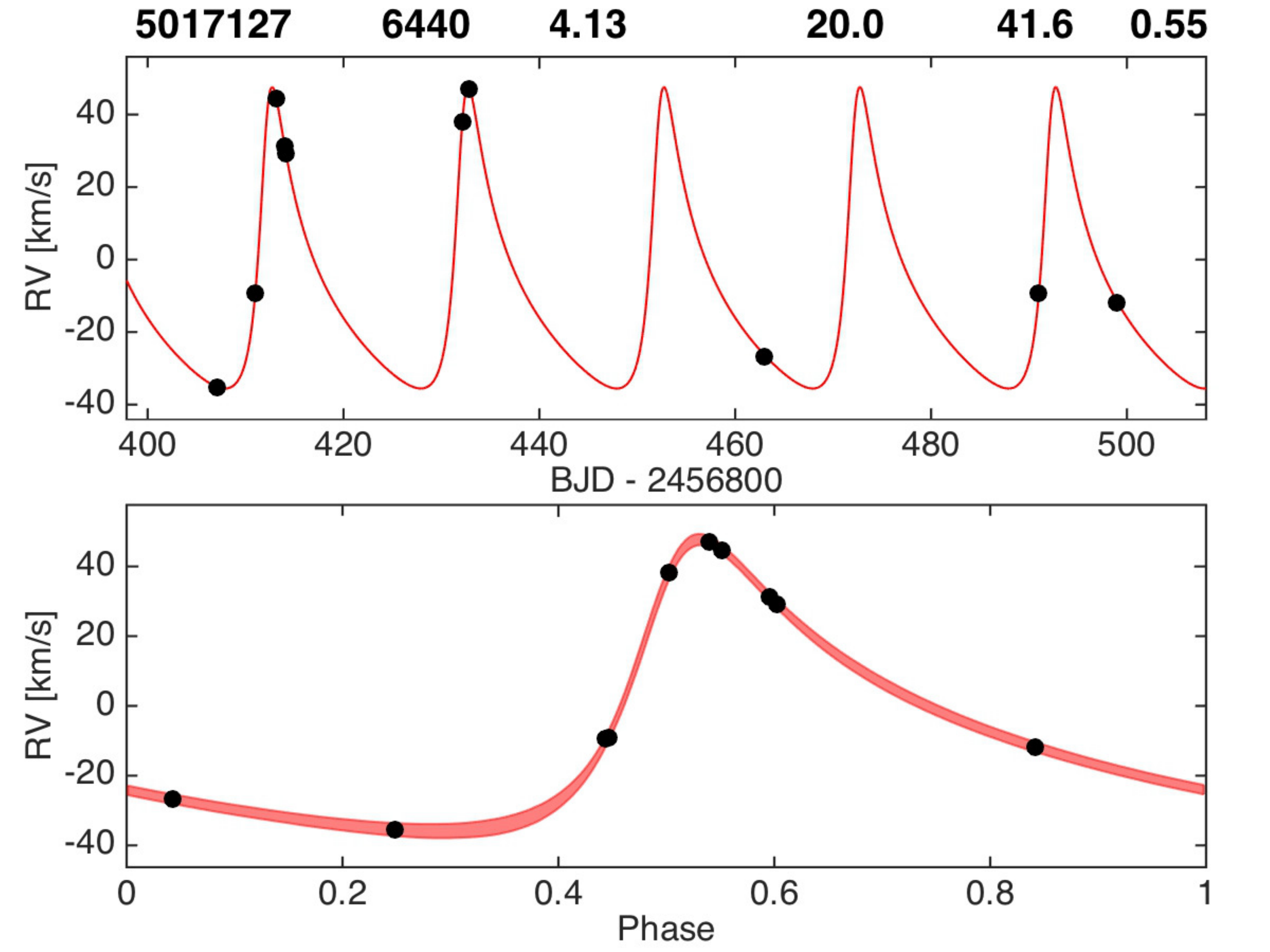} \\
\includegraphics[width=\figwidth]{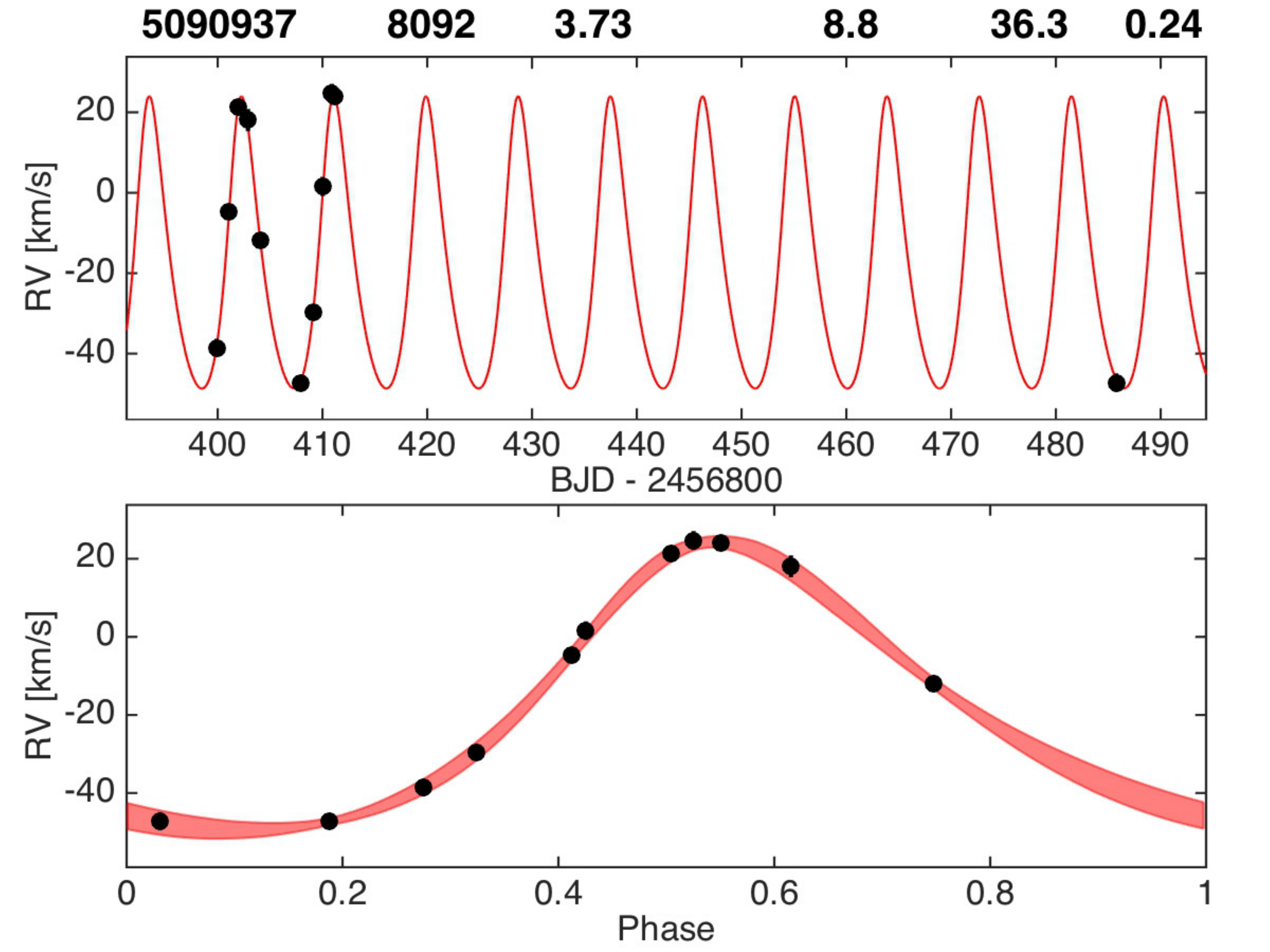} \vspace{5mm}&
 \includegraphics[width=\figwidth]{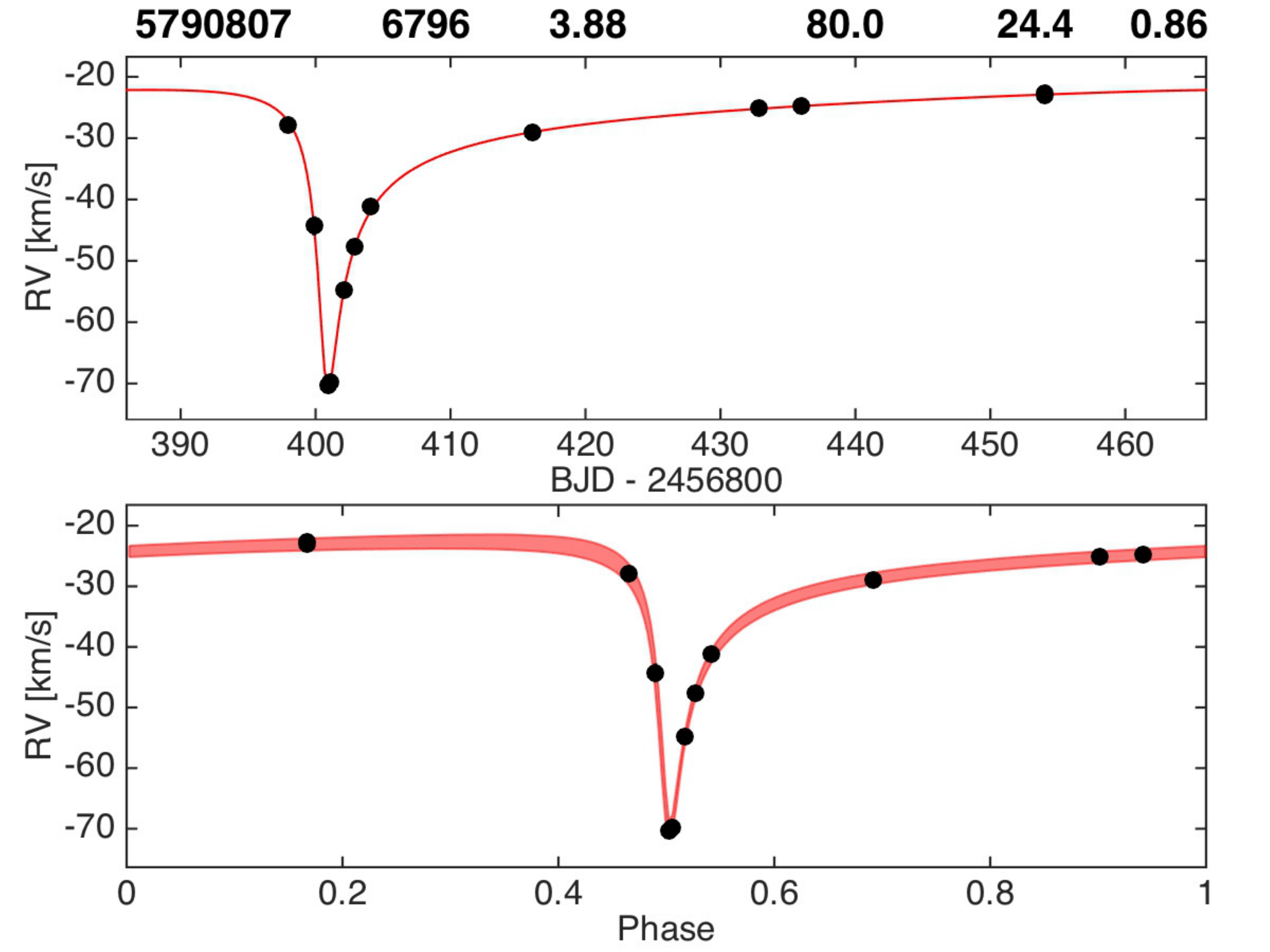} 
\end{tabular}
\begin{minipage}{60em}
{\bf Figure \thefigure.} RV curves of KID 4659476, KID 5017127, KID 5090937, and KID 5790807. RV measurements are shown in black, including error bars which are typically smaller than the marker size. Each system is shown in two panels, top panel shows the RVs as a function of time and the bottom panel the phase-folded RV curve with periastron at phase 0.5. The fitted Keplerian model is shown as a red solid line in the top panels, and by a \sig{3} contour plot in the bottom panels. The title for each plot lists (from left to right) KIC ID, \teff\ (K), \logg, $P$ (d), $K$ (\kms), and $e$. 
\end{minipage}
\label{fig:orbits1} 
\end{center}
\end{table}

%% file: figures_orbits2.tex

\def\figwidth{3.4in}

\stepcounter{figure} 

\begin{table}[h]
\begin{center}
\begin{tabular}{ll} 
 \includegraphics[width=\figwidth]{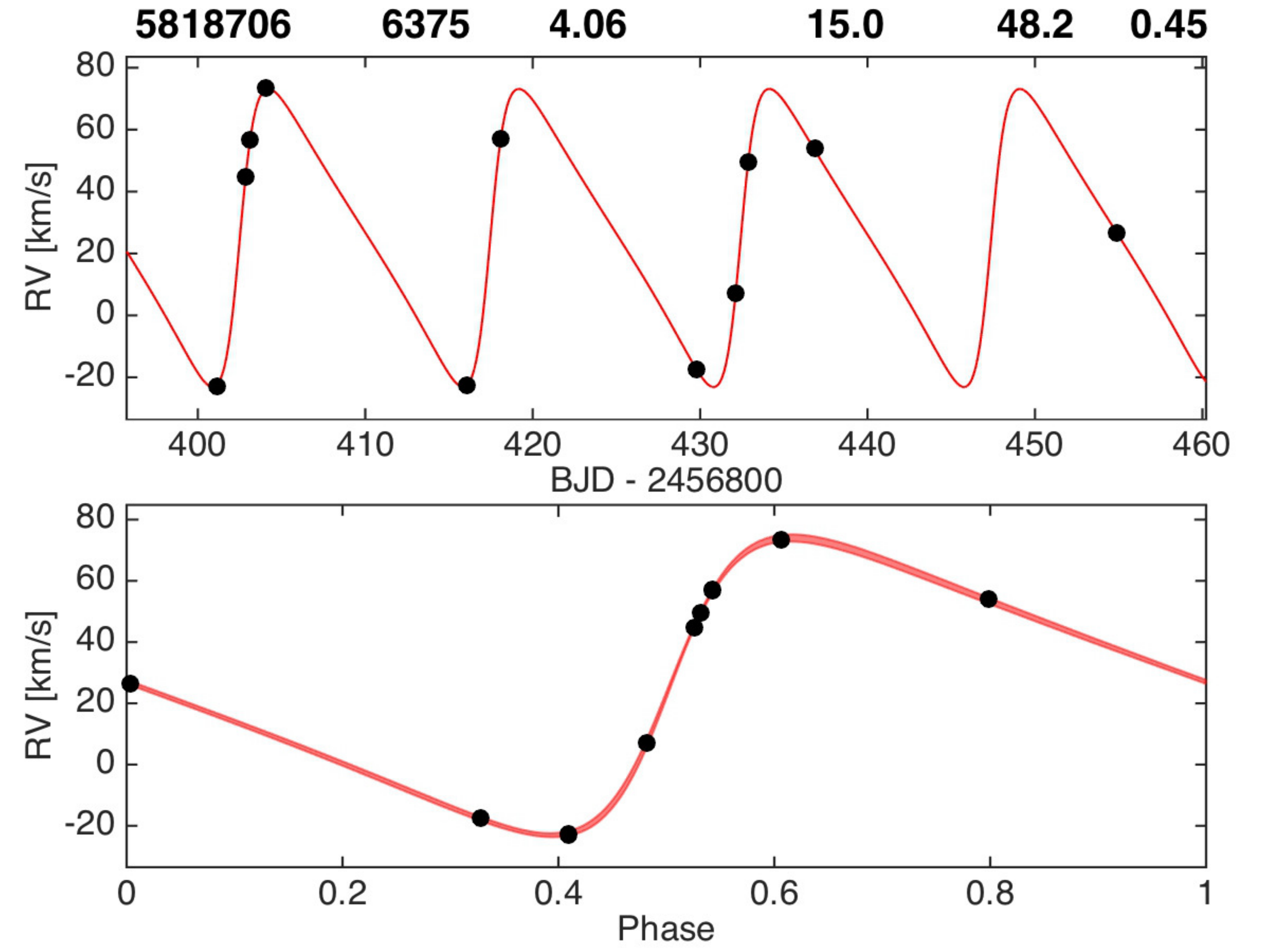}  \vspace{5mm}&
 \includegraphics[width=\figwidth]{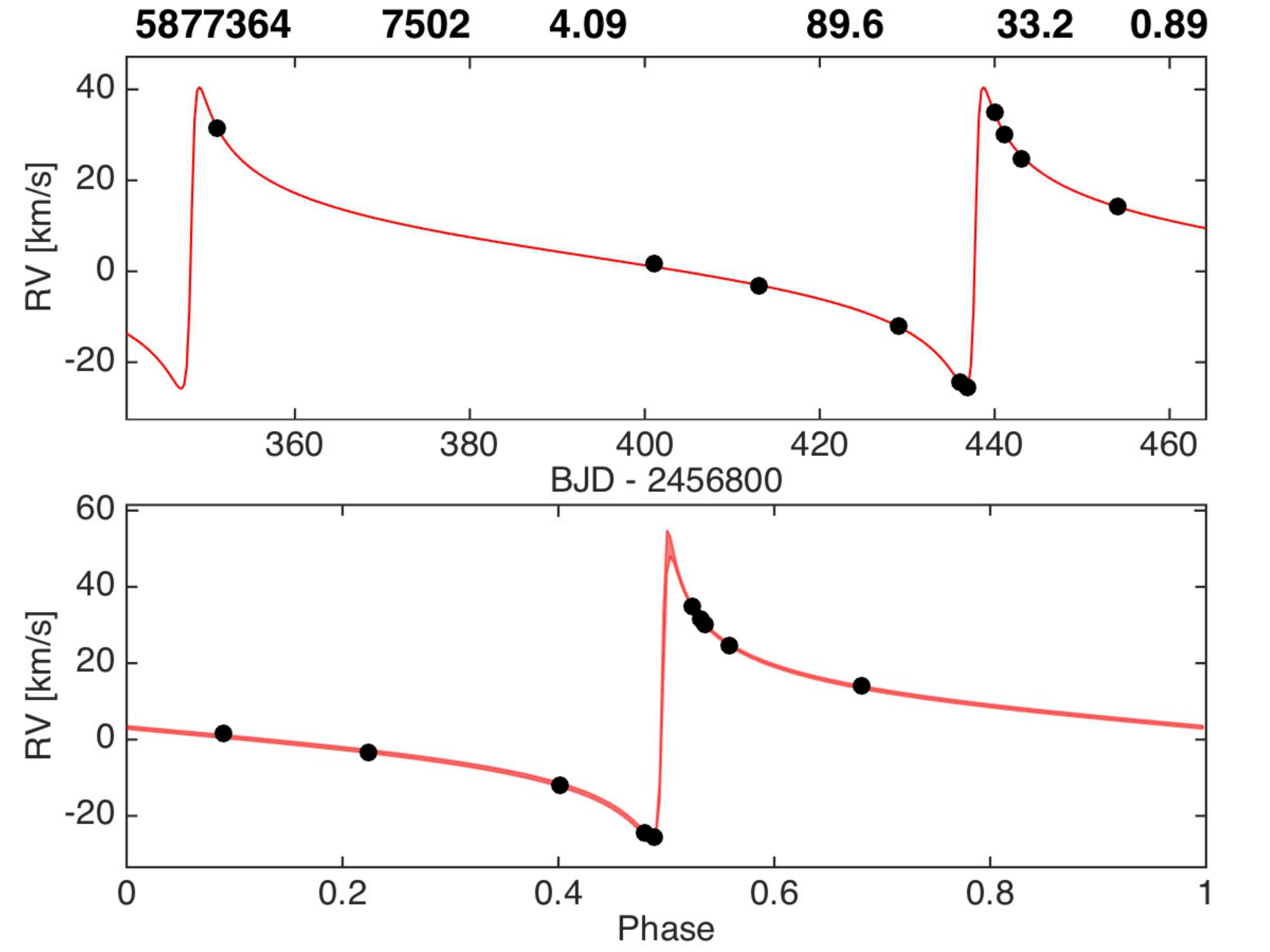}   \\
 \includegraphics[width= \figwidth]{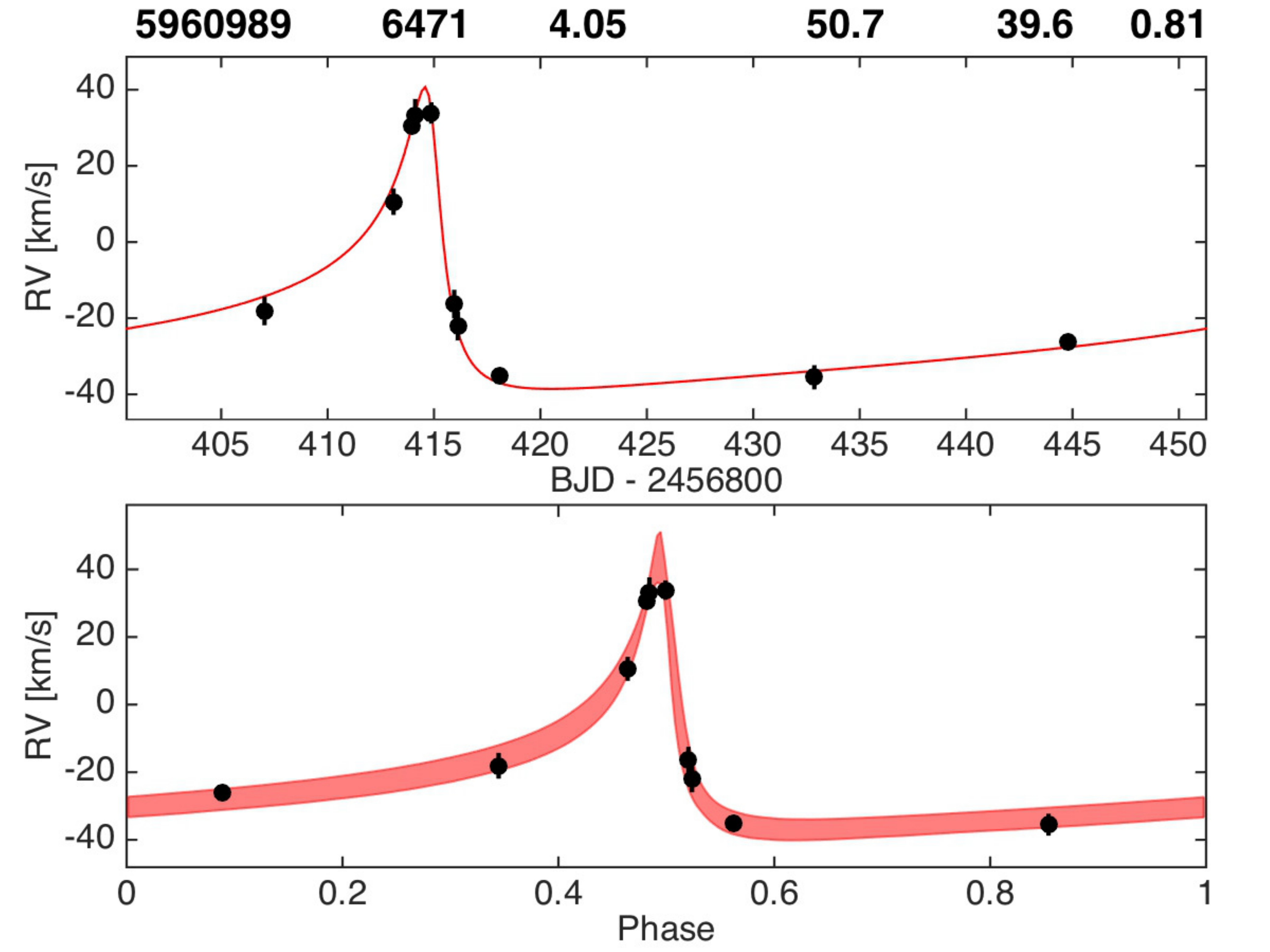} \vspace{5mm}&
 \includegraphics[width= \figwidth]{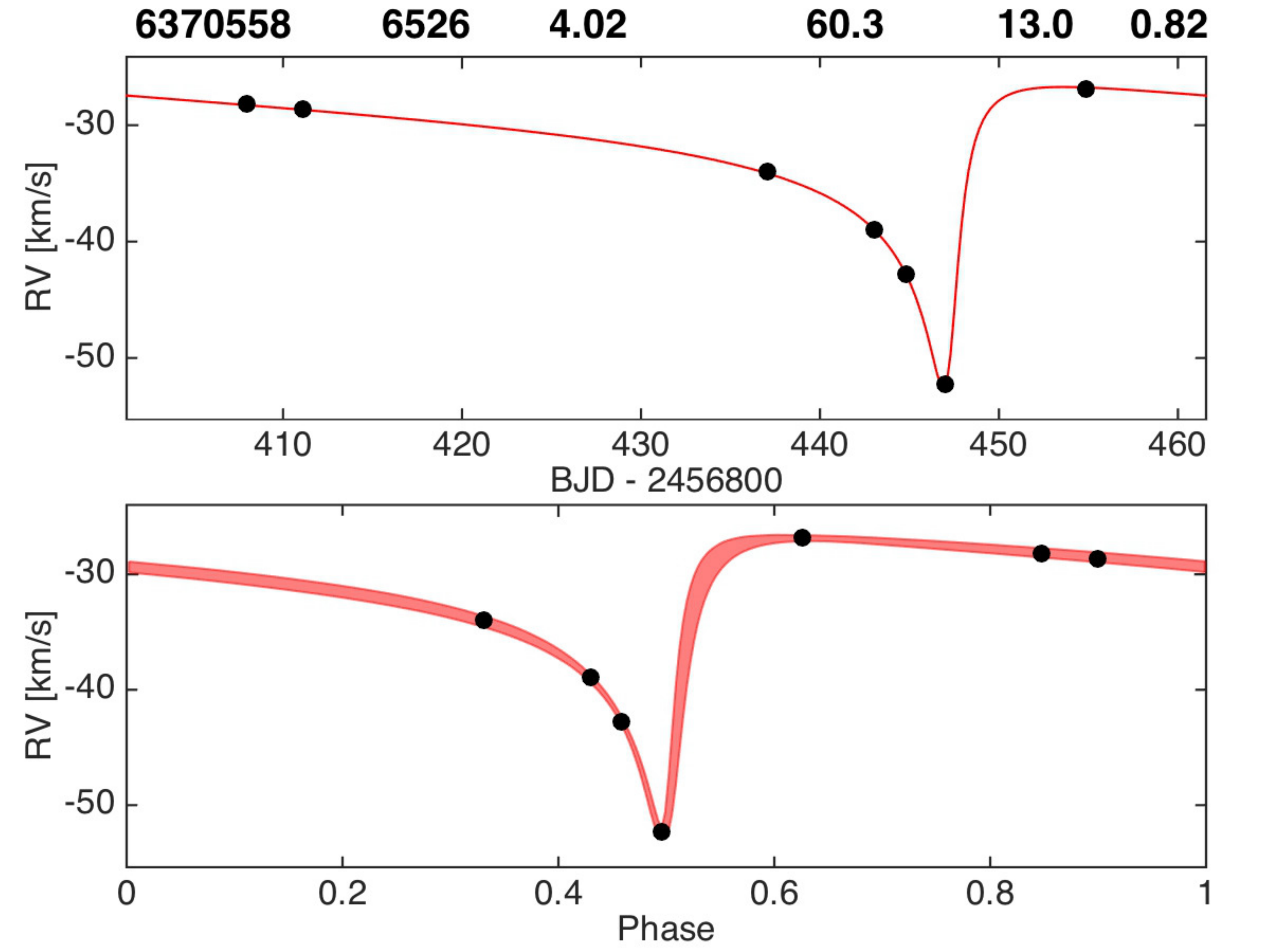} 
\end{tabular}
\begin{minipage}{60em}
{\bf Figure \thefigure.} Similar to Fig.~1 for KID 5818706, KID 5877364, KID 5960989, and KID 6370558. The title for each plot lists (from left to right) KIC ID, \teff\ (K), \logg, $P$ (d), $K$ (\kms), and $e$.
\end{minipage}
\label{fig:orbits2} 
\end{center}
\end{table}

%% file: figures_orbits3.tex

\def\figwidth{3.4in}

\stepcounter{figure} 

\begin{table}[h]
\begin{center}
\begin{tabular}{ll} 
\includegraphics[width= \figwidth]{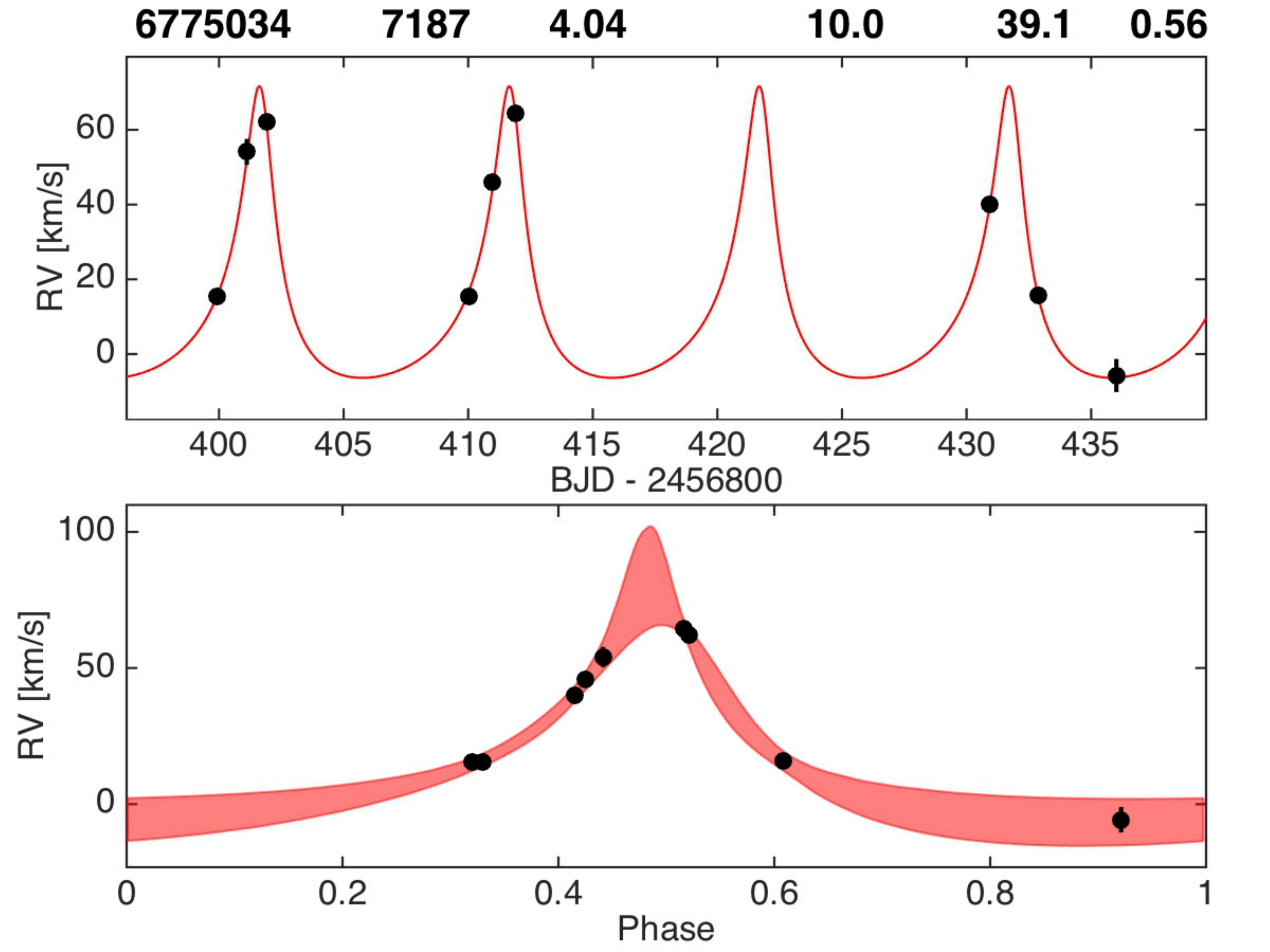} \vspace{5mm}&
\includegraphics[width=\figwidth]{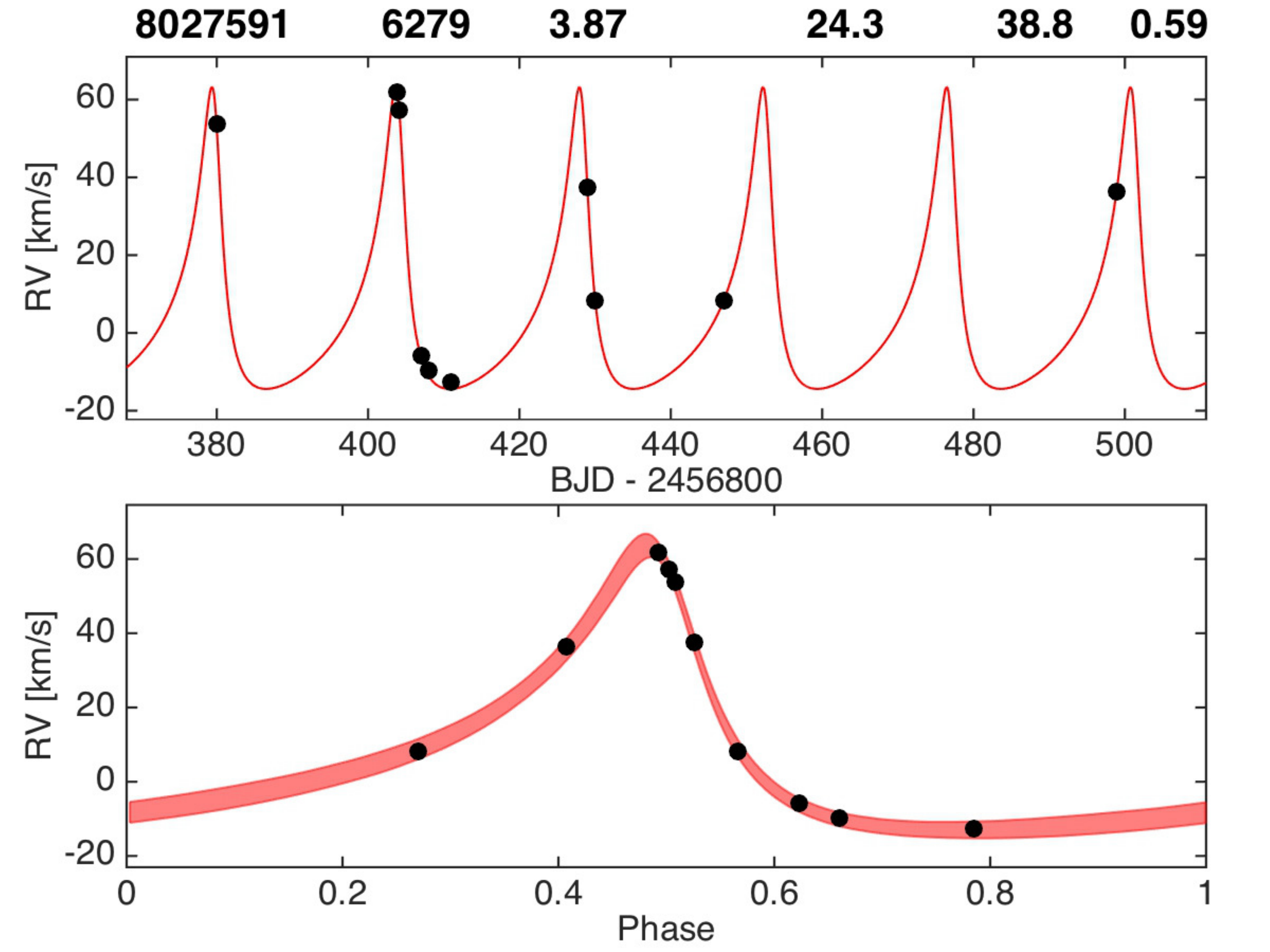}  \\
 \includegraphics[width=\figwidth]{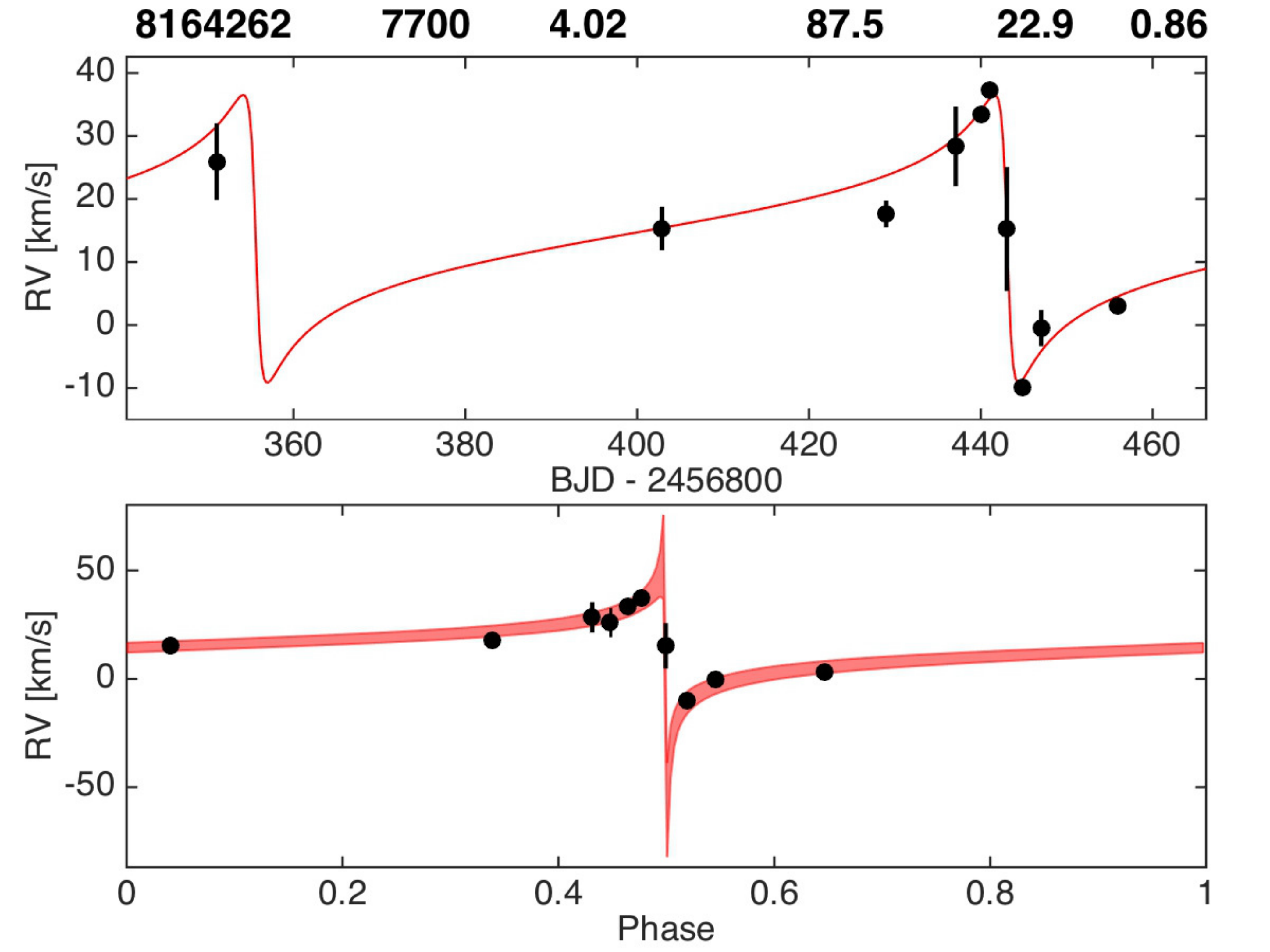} \vspace{5mm}&
\includegraphics[width=\figwidth]{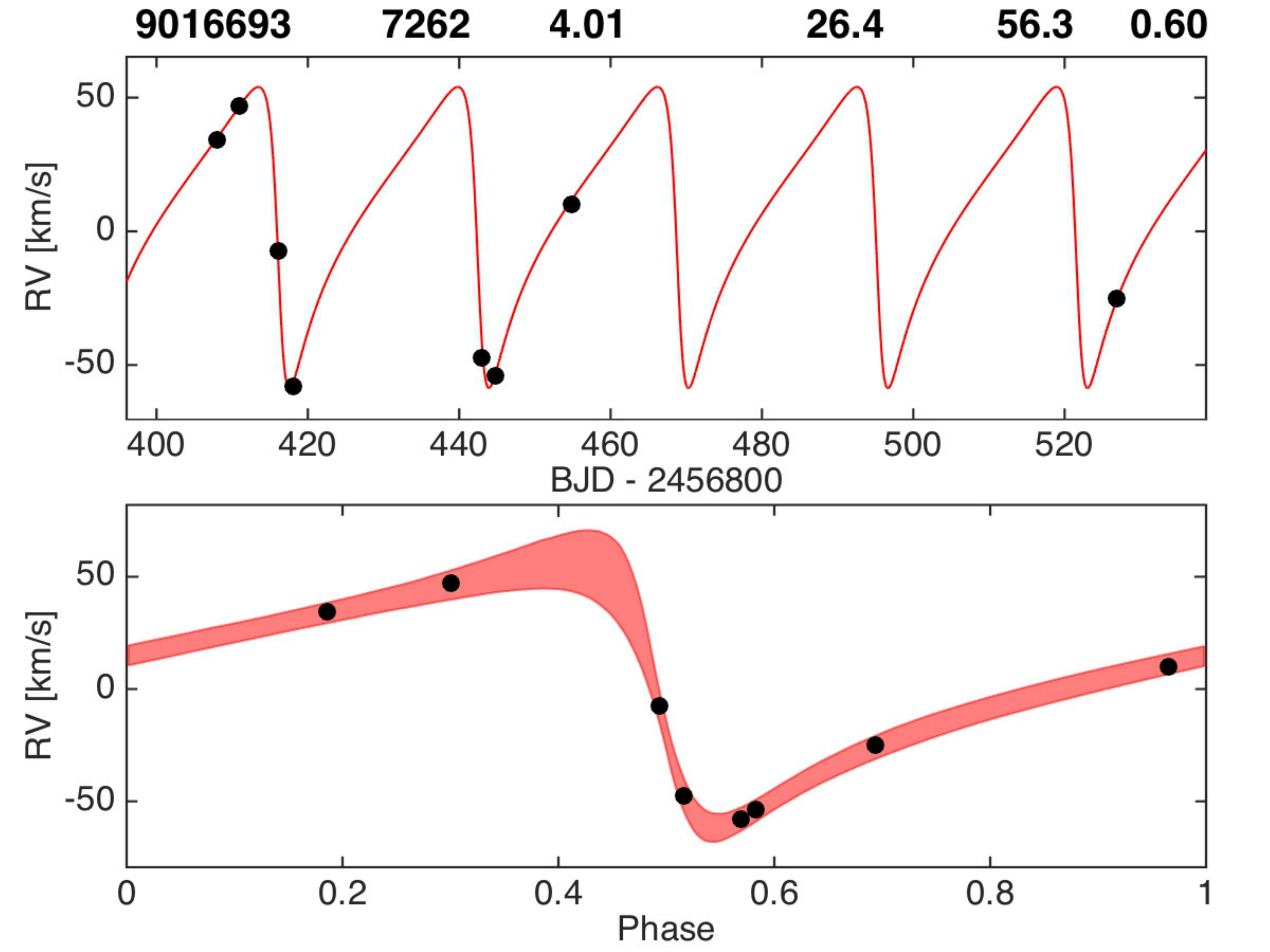}  
\end{tabular}
\begin{minipage}{60em}
{\bf Figure \thefigure.} Similar to Fig.~1 for KID 6775034, KID 8027591, KID 8164262, and KID 9016693. The title for each plot lists (from left to right) KIC ID, \teff\ (K), \logg, $P$ (d), $K$ (\kms), and $e$.
\end{minipage}
\label{fig:orbits3}
\end{center}
\end{table}

%% file: figures_orbits4.tex

\def\figwidth{3.4in}

\stepcounter{figure} 

\begin{table}[h]
\begin{center}
\begin{tabular}{ll} 
\includegraphics[width=\figwidth]{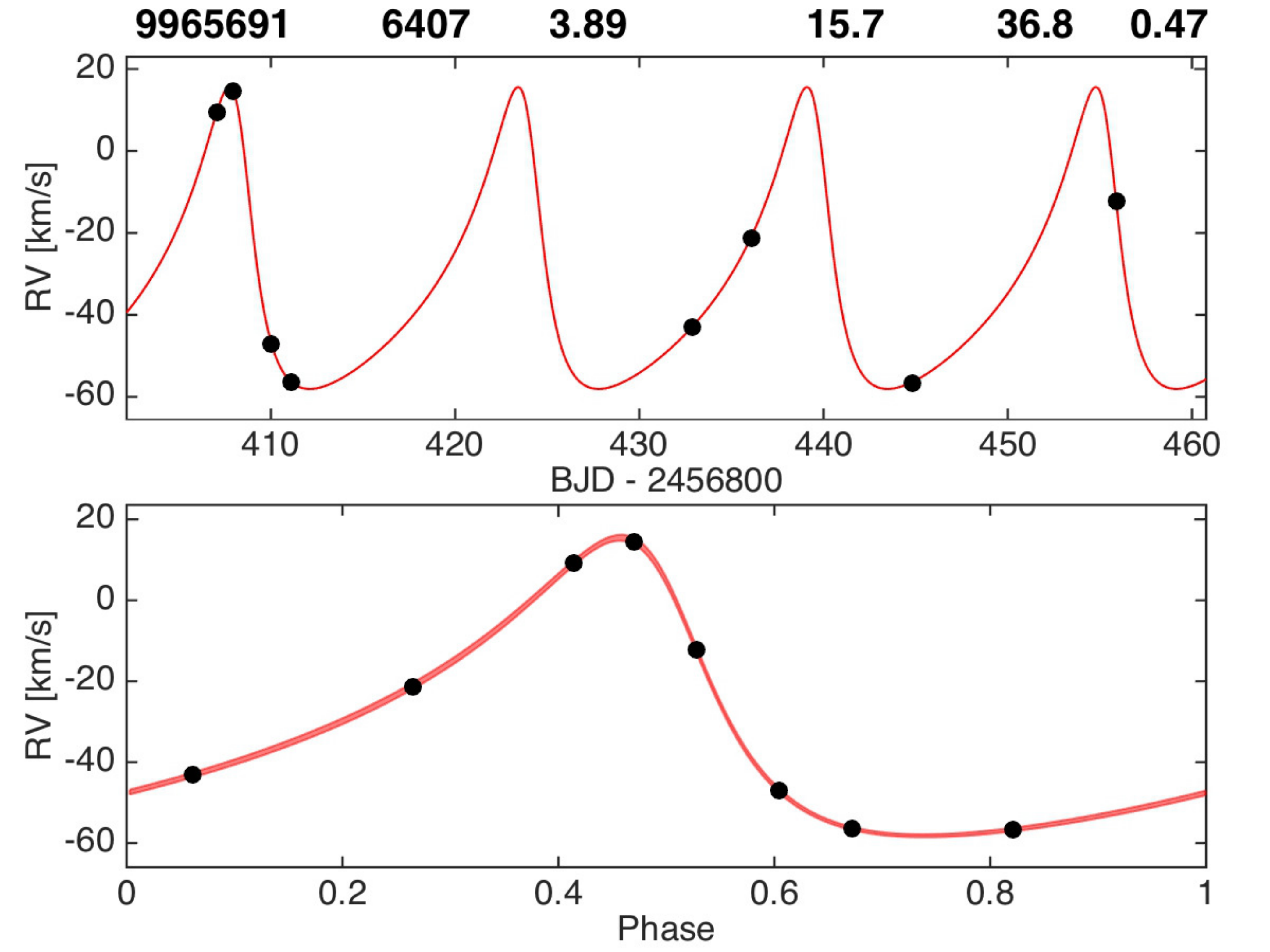}  \vspace{5mm}&
\includegraphics[width=\figwidth]{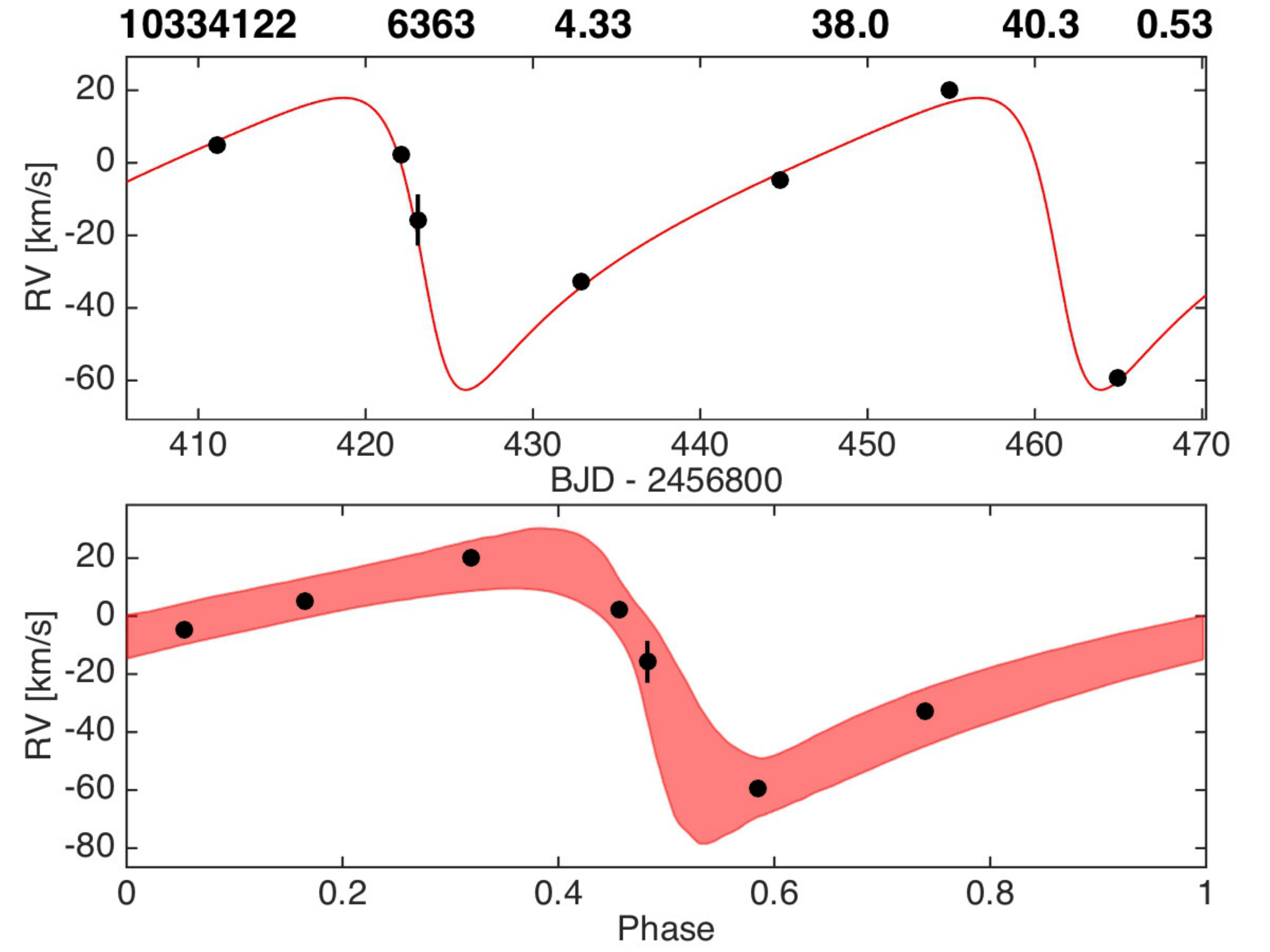}  \\
\includegraphics[width=\figwidth]{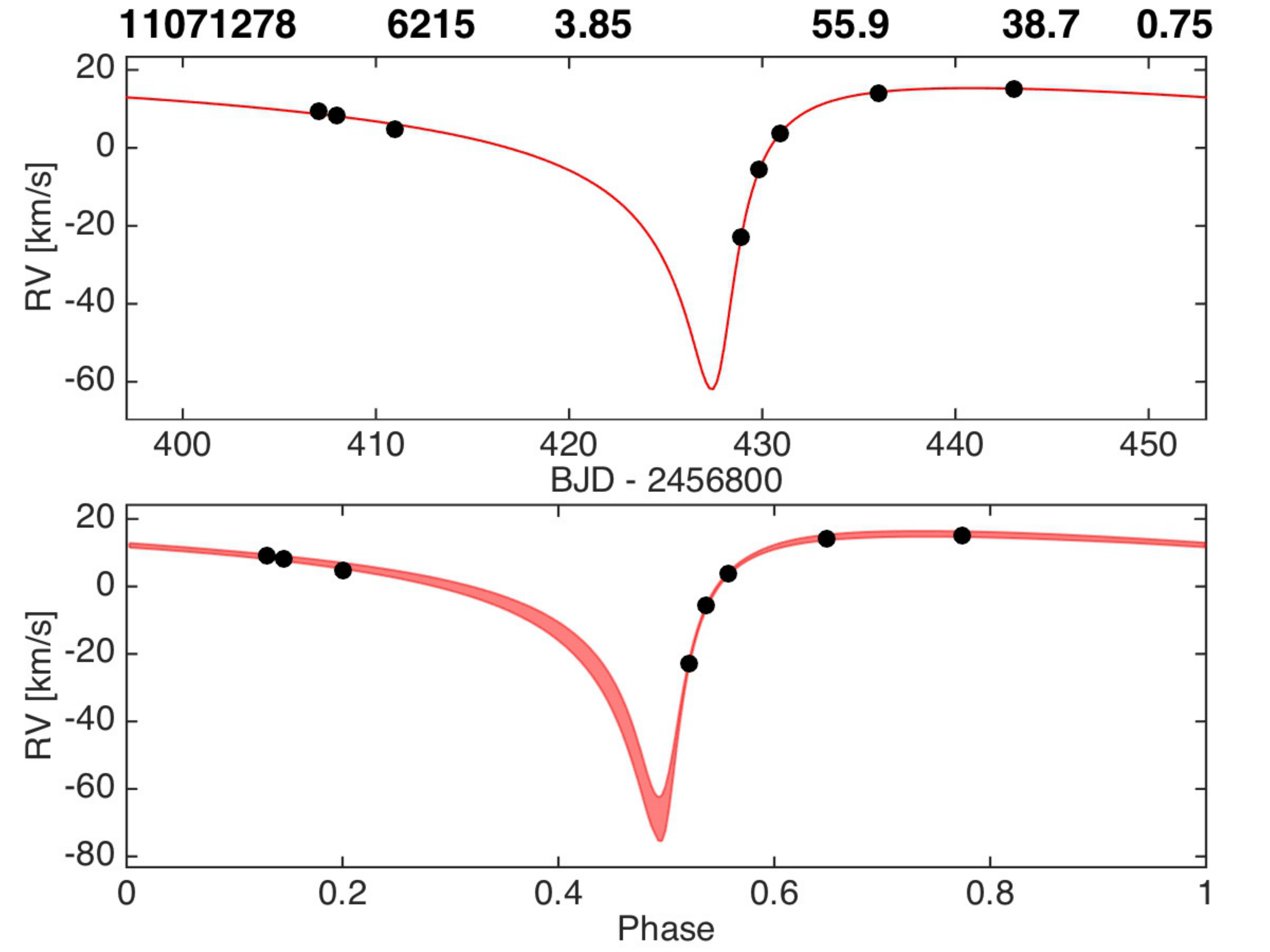}  \vspace{5mm}&
\includegraphics[width=\figwidth]{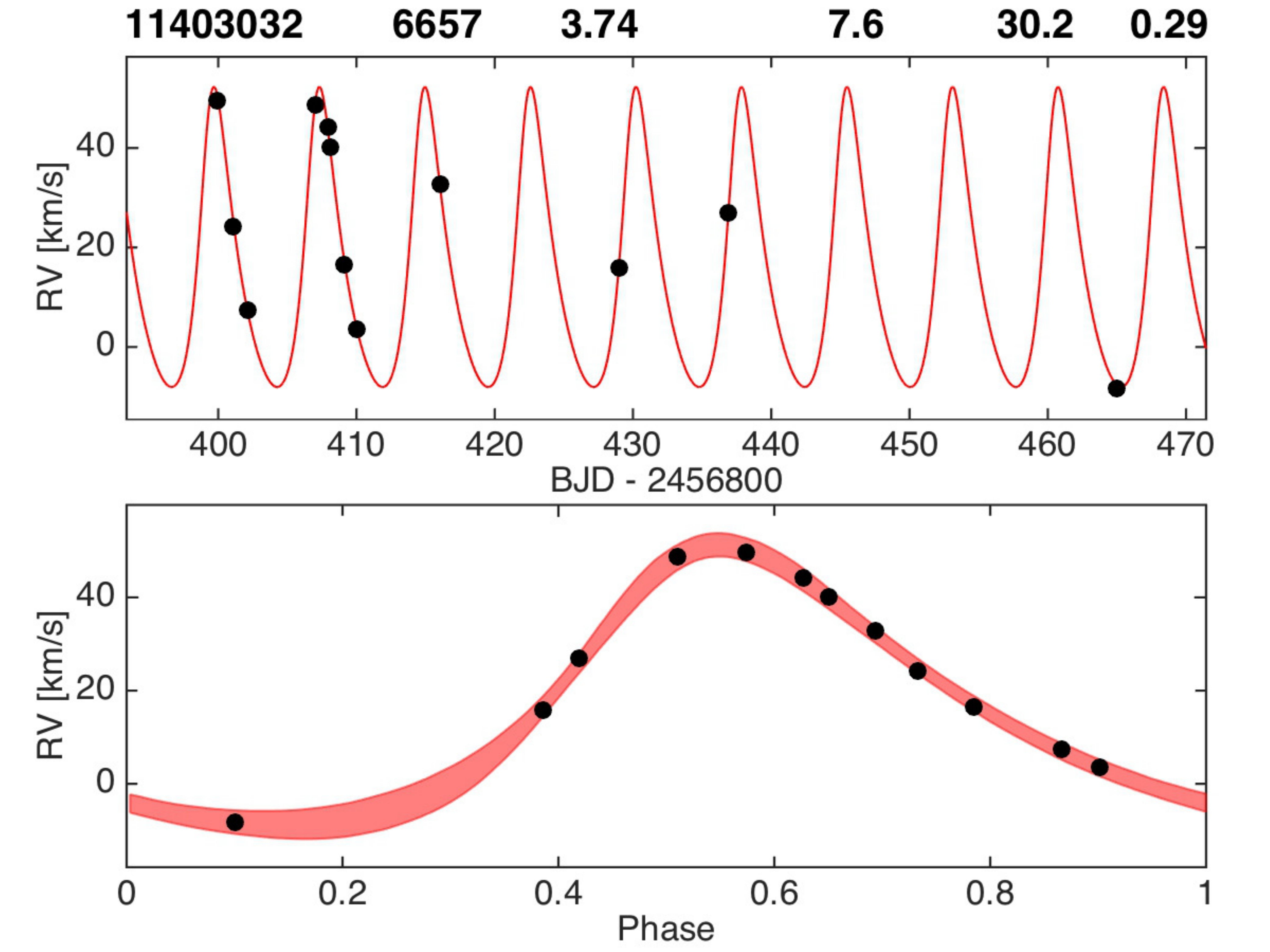}
\end{tabular}
\begin{minipage}{60em}
{\bf Figure \thefigure.} Similar to Fig.~1 for KID 9965691, KID 10334122, KID 11071278, KID 11403032. The title for each plot lists (from left to right) KIC ID, \teff\ (K), \logg, $P$ (d), $K$ (\kms), and $e$.
\end{minipage}
\label{fig:orbits3}
\end{center}
\end{table}

%% file: figures_orbits5.tex

\def\figwidth{3.4in}

\stepcounter{figure} 

\begin{table}[h]
\begin{center}
\begin{tabular}{ll} 
\includegraphics[width=\figwidth]{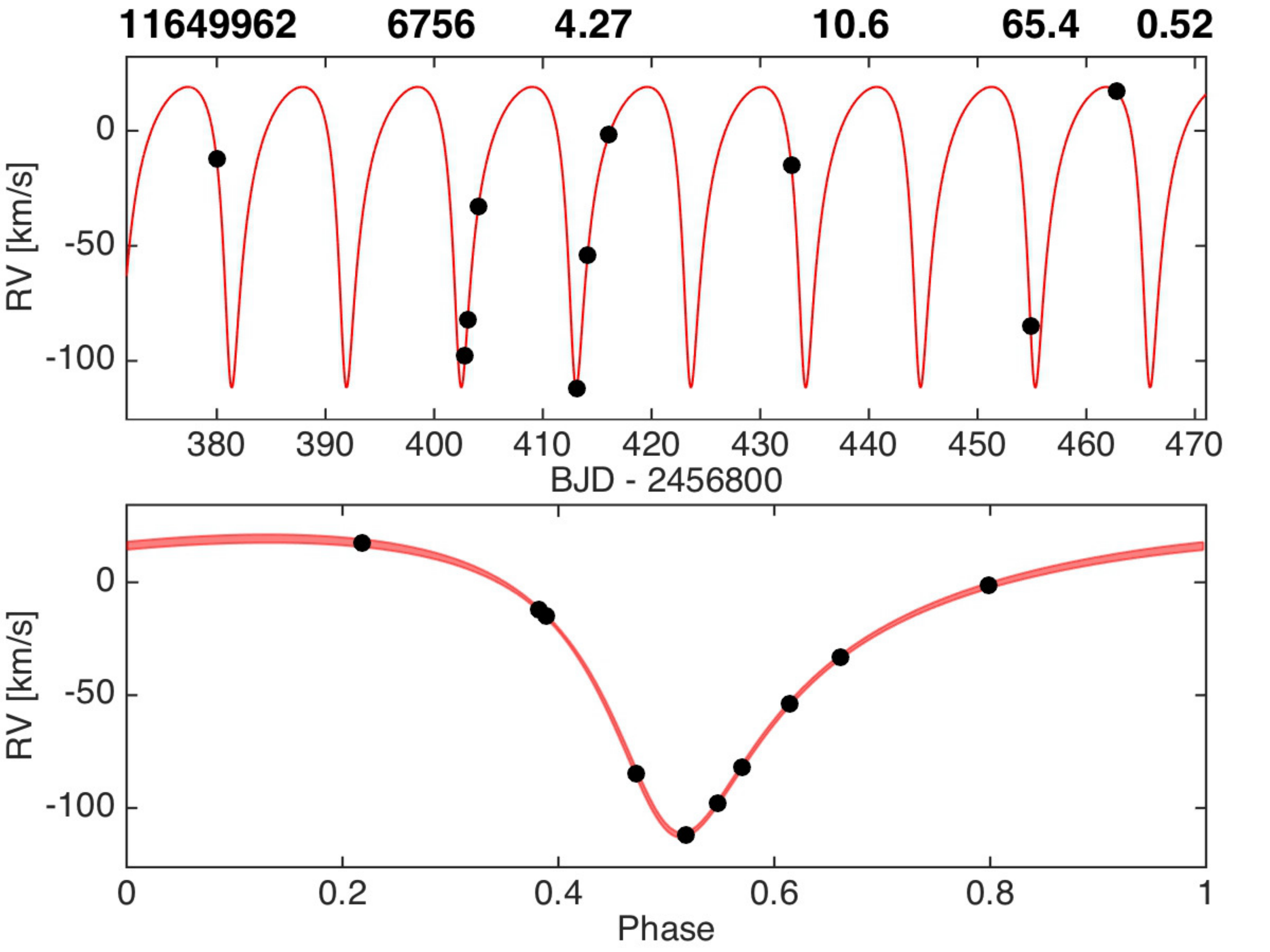}  \vspace{5mm}&
\includegraphics[width=\figwidth]{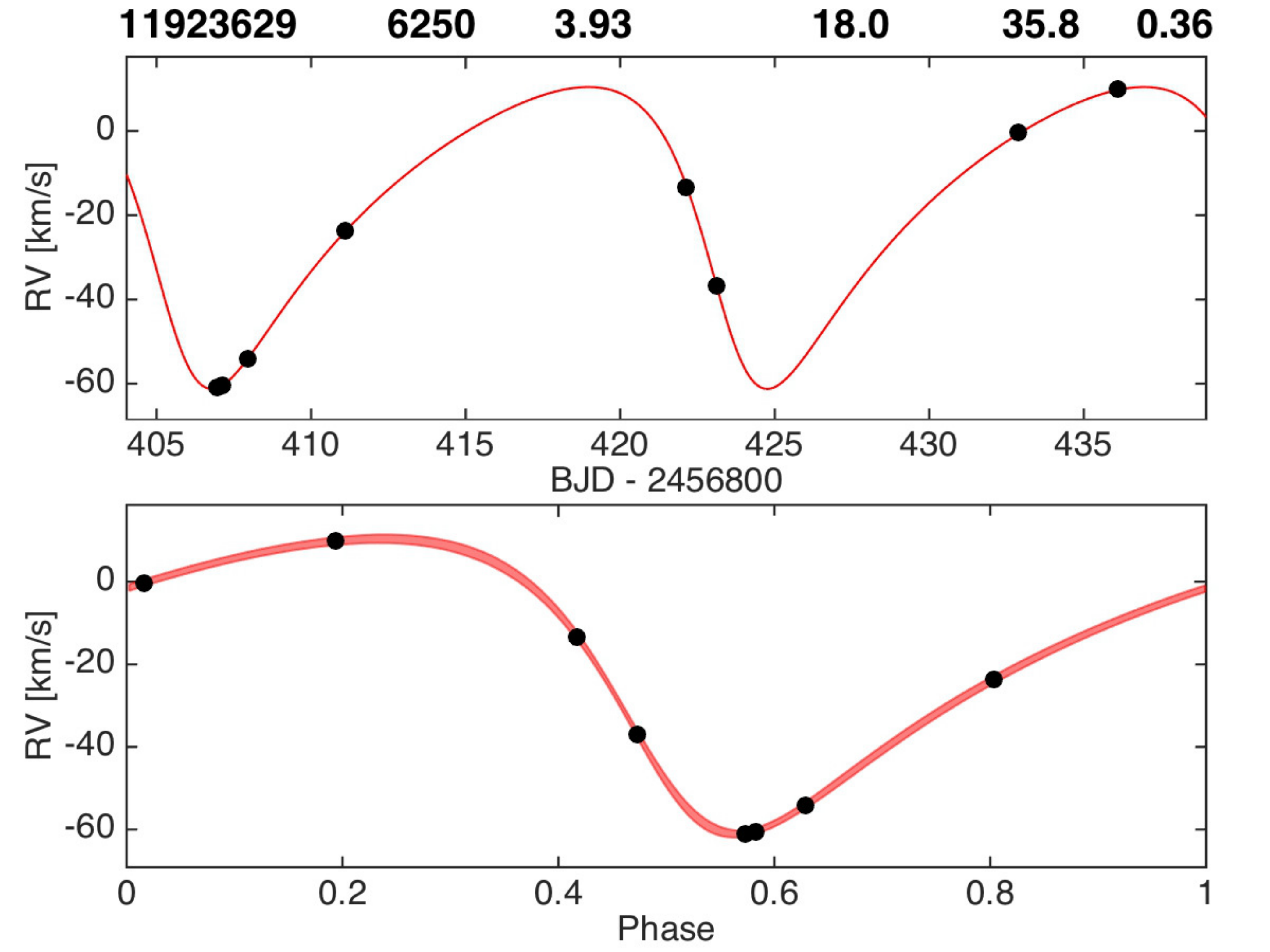}  \\
\includegraphics[width=\figwidth]{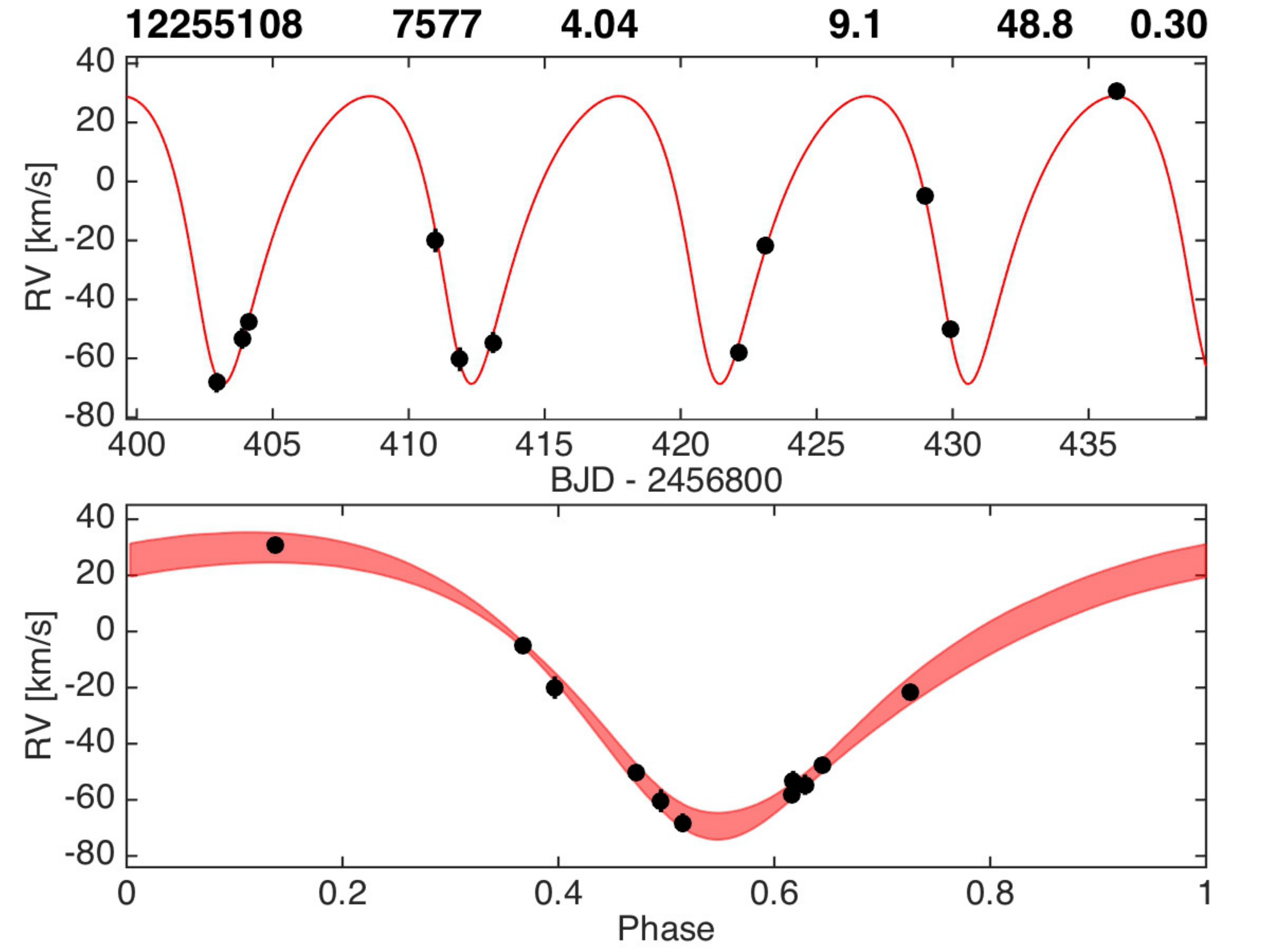}   
\end{tabular}
\begin{minipage}{60em}
{\bf Figure \thefigure.} Similar to Fig.~1 for KID 11649962, KID 11923629, and KID 12255108. The title for each plot lists (from left to right) KIC ID, \teff\ (K), \logg, $P$ (d), $K$ (\kms), and $e$.
\end{minipage}
\label{fig:orbits3}
\end{center}
\end{table}

%% file: const_table.tex
\begin{deluxetable}{crcrc}
\tablecaption{\label{tab:const} Heartbeat systems showing no radial velocity variability.} 
\tablewidth{0pt}
\tablehead{\multicolumn{1}{c}{KID} & \multicolumn{1}{c}{$P$} & \multicolumn{1}{c}{\#RVs} & \multicolumn{1}{c}{RV mean}  &  \multicolumn{1}{c}{RV scatter}   \\   
                                                              & \multicolumn{1}{c}{[d]} &                                              & \multicolumn{1}{c}{[\kms]}       & \multicolumn{1}{c}{[\kms]}  }  
\startdata
9972385     & 58.42  & \, 9 &  -0.55  \ \ & 0.47 \\
11122789 \, & 3.24   &  20  & -19.73  \ \ & 1.42 \\
11409673 \, & 12.32  &  10  &  -5.44  \ \ & 0.88 
\enddata
\end{deluxetable}

%% file: m2_table.tex
\begin{deluxetable*}{rccccc}
\tablecaption{\label{tab:m2} Masses of the two stars of the heartbeat systems measured here. The primary mass is taken from the KIC and the secondary is estimated from the orbital properties. } 
\tablewidth{0pt}
\tablehead{\multicolumn{1}{c}{KID} & \multicolumn{1}{c}{$M_1$}         & \multicolumn{1}{c}{$f(m)$}         & \multicolumn{1}{c}{$M_2$\tablenotemark{a}} & \multicolumn{1}{c}{$M_2$\tablenotemark{b}}    & \multicolumn{1}{c}{$M_2$\tablenotemark{c}}  \\   
                                     & \multicolumn{1}{c}{[$M_{\sun}$]} & \multicolumn{1}{c}{[$M_{\sun}$]}  & \multicolumn{1}{c}{[$M_{\sun}$]}           & \multicolumn{1}{c}{[$M_{\sun}$]}              & \multicolumn{1}{c}{[$M_{\sun}$]}  }  
\startdata
\rule{0pt}{3ex}   4659476 &  1.31$^{+ 0.20}_{-0.22}$ & 0.243$^{+0.018}_{-0.017}$ & 1.162$^{+0.094}_{-0.106}$ &  1.44$^{+ 0.11}_{- 0.13}$ &  1.45$^{+ 1.76}_{- 0.27}$ \\  
\rule{0pt}{3ex}   5017127 &  1.25$^{+ 0.15}_{-0.19}$ & 0.0868$^{+0.0023}_{-0.0023}$ & 0.692$^{+0.047}_{-0.060}$ & 0.839$^{+0.055}_{-0.071}$ &  0.84$^{+ 0.85}_{- 0.14}$ \\  
\rule{0pt}{3ex}   5090937 &  2.07$^{+ 0.34}_{-0.50}$ & 0.0400$^{+0.0017}_{-0.0016}$ & 0.676$^{+0.066}_{-0.103}$ & 0.805$^{+0.077}_{-0.121}$ &  0.81$^{+ 0.66}_{- 0.15}$ \\  
\rule{0pt}{3ex}   5790807 &  1.72$^{+ 0.15}_{-0.24}$ & 0.01640$^{+0.00067}_{-0.00065}$ & 0.426$^{+0.024}_{-0.038}$ & 0.503$^{+0.028}_{-0.044}$ & 0.502$^{+0.392}_{-0.076}$ \\  
\rule{0pt}{3ex}   5818706 &  1.12$^{+ 0.19}_{-0.14}$ & 0.1228$^{+0.0023}_{-0.0022}$ & 0.760$^{+0.069}_{-0.056}$ & 0.930$^{+0.081}_{-0.066}$ &  0.95$^{+ 1.02}_{- 0.17}$ \\  
\rule{0pt}{3ex}   5877364 &  1.38$^{+ 0.20}_{-0.23}$ & 0.03337$^{+0.00080}_{-0.00070}$ & 0.492$^{+0.043}_{-0.049}$ & 0.588$^{+0.050}_{-0.058}$ &  0.59$^{+ 0.50}_{- 0.10}$ \\  
\rule{0pt}{3ex}   5960989 &  1.43$^{+ 0.12}_{-0.15}$ & 0.0645$^{+0.0043}_{-0.0040}$ & 0.667$^{+0.035}_{-0.042}$ & 0.804$^{+0.042}_{-0.049}$ &  0.81$^{+ 0.76}_{- 0.13}$ \\  
\rule{0pt}{3ex}   6370558 &  1.48$^{+ 0.21}_{-0.28}$ & 0.00254$^{+0.00020}_{-0.00013}$ & 0.195$^{+0.018}_{-0.025}$ & 0.229$^{+0.021}_{-0.029}$ & 0.231$^{+0.158}_{-0.040}$ \\  
\rule{0pt}{3ex}   6775034 &  1.30$^{+ 0.23}_{-0.19}$ & 0.0356$^{+0.0039}_{-0.0034}$ & 0.499$^{+0.052}_{-0.047}$ & 0.597$^{+0.061}_{-0.055}$ &  0.61$^{+ 0.52}_{- 0.11}$ \\  
\rule{0pt}{3ex}   8027591 &  1.40$^{+ 0.20}_{-0.28}$ & 0.0785$^{+0.0049}_{-0.0046}$ & 0.715$^{+0.059}_{-0.083}$ & 0.864$^{+0.070}_{-0.098}$ &  0.87$^{+ 0.83}_{- 0.16}$ \\  
\rule{0pt}{3ex}   8164262 &  1.69$^{+ 0.20}_{-0.30}$ & 0.0148$^{+0.0044}_{-0.0031}$ & 0.426$^{+0.041}_{-0.048}$ & 0.504$^{+0.048}_{-0.056}$ & 0.512$^{+0.392}_{-0.090}$ \\  
\rule{0pt}{3ex}   9016693 &  1.60$^{+ 0.20}_{-0.33}$ & 0.253$^{+0.028}_{-0.024}$ & 1.323$^{+0.099}_{-0.143}$ &  1.63$^{+ 0.12}_{- 0.17}$ &  1.63$^{+ 1.95}_{- 0.31}$ \\  
\rule{0pt}{3ex}   9965691 &  1.35$^{+ 0.22}_{-0.22}$ & 0.05550$^{+0.00057}_{-0.00057}$ & 0.598$^{+0.056}_{-0.059}$ & 0.718$^{+0.065}_{-0.069}$ &  0.73$^{+ 0.66}_{- 0.13}$ \\  
\rule{0pt}{3ex}  10334122 &  1.14$^{+ 0.14}_{-0.13}$ & 0.155$^{+0.031}_{-0.028}$ & 0.903$^{+0.074}_{-0.067}$ & 1.113$^{+0.091}_{-0.081}$ &  1.13$^{+ 1.30}_{- 0.21}$ \\  
\rule{0pt}{3ex}  11071278 &  1.23$^{+ 0.19}_{-0.28}$ & 0.094$^{+0.023}_{-0.014}$ & 0.747$^{+0.077}_{-0.098}$ & 0.909$^{+0.093}_{-0.115}$ &  0.92$^{+ 0.94}_{- 0.18}$ \\  
\rule{0pt}{3ex}  11403032 &  1.65$^{+ 0.20}_{-0.36}$ & 0.0191$^{+0.0011}_{-0.0011}$ & 0.443$^{+0.034}_{-0.062}$ & 0.525$^{+0.039}_{-0.072}$ & 0.523$^{+0.405}_{-0.091}$ \\  
\rule{0pt}{3ex}  11649962 &  1.29$^{+ 0.15}_{-0.20}$ & 0.1905$^{+0.0041}_{-0.0039}$ & 1.007$^{+0.060}_{-0.083}$ & 1.240$^{+0.071}_{-0.098}$ &  1.24$^{+ 1.45}_{- 0.22}$ \\  
\rule{0pt}{3ex}  11923629 &  0.97$^{+ 0.14}_{-0.12}$ & 0.0694$^{+0.0012}_{-0.0012}$ & 0.543$^{+0.043}_{-0.040}$ & 0.659$^{+0.050}_{-0.048}$ &  0.67$^{+ 0.67}_{- 0.12}$ \\  
\rule{0pt}{3ex}  12255108 &  1.79$^{+ 0.18}_{-0.29}$ & 0.0957$^{+0.0080}_{-0.0073}$ & 0.899$^{+0.058}_{-0.086}$ & 1.086$^{+0.069}_{-0.100}$ &  1.08$^{+ 1.05}_{- 0.18}$ 
\enddata
\tablenotetext{a}{Minimum mass estimate, assuming $\sin^{3}(i)$ = 1 meaning a completely edge on configuration with $i = 90$ deg.}\tablenotetext{b}{Assuming the median value of $\sin^{3}(i)$, of 0.6495, where $i$ is distributed as $\sin(i)$.}\tablenotetext{c}{Using the distribution of $\sin^3(i)$, where $i$ is distributed as $\sin(i)$.}\end{deluxetable*}

%% file: hblit_table.tex
\begin{deluxetable*}{llrl}
\tablecaption{\label{tab:hblit} Known heartbeat systems with RV measured orbit with $P < 200$~d.} 
\tablewidth{0pt}
\tablehead{
\multicolumn{1}{c}{Reference} & \multicolumn{1}{c}{Name} & \multicolumn{1}{c}{$P$}   & \multicolumn{1}{c}{$e$\tablenotemark{a} \, \, \, \, } \\   
                              &                          & \multicolumn{1}{c}{[d]} &                         
}  
\startdata
Beck et al.~2014    &   KID 8912308    & 20.17   &$0.23\pm0.01$\\
$-$  &   KID 2697935\tablenotemark{b}    & 21.50   &$0.41\pm0.02$\\
$-$  &   KID 8095275    & 23.00   &$0.32\pm0.01$\\
$-$  &   KID 2720096    & 26.70   &$0.49\pm0.01$\\
$-$  &   KID 9408183    & 49.70   &$0.42\pm0.01$\\
$-$  &   KID 5006817    & 94.81   &$0.7069\pm0.0002$\\
$-$  &   KID 2444348    & 103.50  &$0.48\pm0.01$\\
$-$  &   KID 9163796    & 121.30  &$0.69\pm0.01$\\
$-$  &   KID 10614012   & 132.13  &$0.71\pm0.01$\\
$-$  &   KID 8210370    & 153.50  &$0.70\pm0.01$\\
$-$  &   KID 9540226    & 175.43  &$0.39\pm0.01$\\
Hambleton et al.~2013  & KID 4544587  &  2.19  & $0.288\pm0.026$ \\  
Hambleton et al.~2016  & KID 3749404  & 20.31  & $0.658\pm0.005$ \\  
Hareter et al.~2014    & HD 51844     & 33.50  & $0.484\pm0.020$ \\ 
Maceroni et al.~2009   & HD 174884    &  3.66 &$0.2939\pm0.0005$\\ 
Schmid et al.~2015     & KID 10080943 & 15.34  & $0.44\pm0.05$ \\ 
Smullen et al.~2015\tablenotemark{c}    & KID 3230227  & 7.05  & $0.60\pm0.04$ \\
$-$                    & KID 4248941  & 8.65  & $0.34\pm0.04$\\ 
$-$                    & KID 8719324  & 10.24 & $0.64\pm0.05$ \\
$-$                    & KID 11494130 & 18.97 & $0.49\pm0.05$ \\
Welsh et al.~2011      & KOI 54       & 41.81  & $0.8315\pm0.0032$
\enddata 
\tablenotetext{a}{Values and errors are as given by the relevant paper.}
\tablenotetext{b}{See also \cite{lillo15}.}
\tablenotetext{c}{Two other systems included in that work are not listed in the table: For KID 9899216 the orbital parameters are not well constrained, and for KID 3749404 we adopt the parameters given by Hambleton et al.~2016.} 
\end{deluxetable*}

%% file: psynch_table.tex
\begin{deluxetable}{rrrr}
\tablecaption{\label{tab:psynch} Predicted pseudo-synchronous stellar rotation period ($P_{\rm ps}$; rightmost column). The table also lists the orbital period ($P$) and eccentricity ($e$).} 
\tablewidth{0pt}
\tablehead{\multicolumn{1}{c}{KID} & \multicolumn{1}{c}{$P$}  & \multicolumn{1}{c}{$e$} & \multicolumn{1}{c}{$P_{\rm ps}$}   \\  
                                     & \multicolumn{1}{c}{[d]}  &                          & \multicolumn{1}{c}{[d]}      }     
\startdata
   4659476  &    58.83045$\pm$3.7e-04 &   0.745$^{+ 0.011}_{- 0.011}$ & $   6.99 \pm  0.48 $ \\
   5017127  &   20.006404$\pm$7.8e-05 &  0.5504$^{+0.0050}_{-0.0050}$ & $   5.96 \pm  0.13 $ \\
   5090937  &    8.800693$\pm$2.4e-05 &   0.241$^{+ 0.013}_{- 0.013}$ & $   6.51 \pm  0.26 $ \\
   5790807  &    79.99625$\pm$5.5e-04 &  0.8573$^{+0.0030}_{-0.0031}$ & $   3.84 \pm  0.12 $ \\
   5818706  &   14.959941$\pm$5.2e-05 &  0.4525$^{+0.0038}_{-0.0039}$ & $  6.233 \pm 0.097 $ \\
   5877364  &    89.64854$\pm$6.4e-04 &  0.8875$^{+0.0031}_{-0.0031}$ & $   2.99 \pm  0.12 $ \\
   5960989  &    50.72153$\pm$3.0e-04 &   0.813$^{+ 0.017}_{- 0.015}$ & $   3.70 \pm  0.46 $ \\
   6370558  &    60.31658$\pm$3.7e-04 &   0.821$^{+ 0.015}_{- 0.012}$ & $   4.11 \pm  0.46 $ \\
   6775034  &   10.028547$\pm$2.9e-05 &   0.556$^{+ 0.047}_{- 0.037}$ & $   2.93 \pm  0.54 $ \\
   8027591  &    24.27443$\pm$1.0e-04 &  0.5854$^{+0.0082}_{-0.0083}$ & $   6.32 \pm  0.24 $ \\
   8164262  &    87.45717$\pm$6.4e-04 &   0.857$^{+ 0.026}_{- 0.065}$ & $    4.2 \pm   1.9 $ \\
   9016693  &    26.36803$\pm$1.2e-04 &   0.596$^{+ 0.018}_{- 0.018}$ & $   6.58 \pm  0.54 $ \\
   9965691  &   15.683195$\pm$5.6e-05 &  0.4733$^{+0.0032}_{-0.0032}$ & $  6.111 \pm 0.081 $ \\
  10334122  &    37.95286$\pm$1.9e-04 &   0.534$^{+ 0.060}_{- 0.058}$ & $   12.0 \pm   3.1 $ \\
  11071278  &    55.88522$\pm$3.3e-04 &   0.755$^{+ 0.015}_{- 0.013}$ & $   6.24 \pm  0.56 $ \\
  11403032  &    7.631634$\pm$2.0e-05 &   0.288$^{+ 0.013}_{- 0.013}$ & $   5.05 \pm  0.22 $ \\
  11649962  &   10.562737$\pm$3.2e-05 &  0.5206$^{+0.0035}_{-0.0035}$ & $  3.506 \pm 0.052 $ \\
  11923629  &   17.973284$\pm$6.7e-05 &  0.3629$^{+0.0058}_{-0.0059}$ & $   9.77 \pm  0.21 $ \\
  12255108  &    9.131526$\pm$2.5e-05 &   0.296$^{+ 0.016}_{- 0.015}$ & $   5.92 \pm  0.32 $
\enddata
\end{deluxetable}

%% file: RVs_table.tex
\begin{center} 
\begin{longtable}{rcrr} 
\caption{} 
\label{tab:rvs} \\ 
\hline 
\multicolumn{1}{c}{KID} & \multicolumn{1}{c}{Time}           & \multicolumn{1}{c}{RV}        & \multicolumn{1}{c}{RV Error} \\ 
                         & \multicolumn{1}{c}{[BJD-2457000]}  & \multicolumn{1}{c}{[\kms]}   & \multicolumn{1}{c}{[\kms]}  \\
\hline\\
\endfirsthead
\multicolumn{4}{c}{{\tablename \ \thetable{} -- continued}} \\
\hline
\multicolumn{1}{c}{KID} & \multicolumn{1}{c}{Time}            & \multicolumn{1}{c}{RV}       & \multicolumn{1}{c}{RV Error} \\ 
                         & \multicolumn{1}{c}{[BJD-2,457,000]} & \multicolumn{1}{c}{[\kms]}  & \multicolumn{1}{c}{[\kms]}  \\
\hline
\endhead
\hline\\
\endfoot
\\ \hline \hline
\endlastfoot
  4659476 &      211.105843 &  $      24.36 \, \, \, $ & $     0.33 \, \, \, $ \\ 
  4659476 &      228.908432 &  $     -67.05 \, \, \, $ & $     0.36 \, \, \, $ \\ 
  4659476 &      229.814142 &  $     -60.52 \, \, \, $ & $     0.71 \, \, \, $ \\ 
  4659476 &      230.797930 &  $     -30.64 \, \, \, $ & $     0.83 \, \, \, $ \\ 
  4659476 &      237.071357 &  $      23.80 \, \, \, $ & $     0.80 \, \, \, $ \\ 
  4659476 &      247.033747 &  $      32.74 \, \, \, $ & $     0.62 \, \, \, $ \\ 
  4659476 &      285.868678 &  $      -29.7 \, \, \, \,  \,  \,  $ & $      2.0 \, \, \, \, \,  \,  $ \\ 
\\ 
  5017127 &      207.040503 &  $     -35.38 \, \, \, $ & $     0.30 \, \, \, $ \\ 
  5017127 &      210.987355 &  $     -9.197 $ & $    0.092 $ \\ 
  5017127 &      213.098794 &  $      44.56 \, \, \, $ & $     0.45 \, \, \, $ \\ 
  5017127 &      213.981495 &  $      31.32 \, \, \, $ & $     0.36 \, \, \, $ \\ 
  5017127 &      214.120710 &  $      29.26 \, \, \, $ & $     0.33 \, \, \, $ \\ 
  5017127 &      232.117274 &  $      38.14 \, \, \, $ & $     0.28 \, \, \, $ \\ 
  5017127 &      232.866650 &  $      47.18 \, \, \, $ & $     0.27 \, \, \, $ \\ 
  5017127 &      262.931572 &  $     -26.66 \, \, \, $ & $     0.34 \, \, \, $ \\ 
  5017127 &      290.957067 &  $      -9.37 \, \, \, $ & $     0.74 \, \, \, $ \\ 
  5017127 &      298.918703 &  $     -11.84 \, \, \, $ & $     0.19 \, \, \, $ \\ 
\\ 
  5090937 &      199.894859 &  $      -38.6 \, \, \, \,  \,  \,  $ & $      1.7 \, \, \, \, \,  \,  $ \\ 
  5090937 &      201.101216 &  $      -4.68 \, \, \, $ & $     0.70 \, \, \, $ \\ 
  5090937 &      201.914717 &  $      21.21 \, \, \, $ & $     0.85 \, \, \, $ \\ 
  5090937 &      202.890882 &  $       18.1 \, \, \, \,  \,  \,  $ & $      2.2 \, \, \, \, \,  \,  $ \\ 
  5090937 &      204.052374 &  $     -11.90 \, \, \, $ & $     0.50 \, \, \, $ \\ 
  5090937 &      207.930832 &  $     -47.28 \, \, \, $ & $     0.36 \, \, \, $ \\ 
  5090937 &      209.128617 &  $      -29.7 \, \, \, \,  \,  \,  $ & $      1.4 \, \, \, \, \,  \,  $ \\ 
  5090937 &      210.020181 &  $        1.6 \, \, \, \,  \,  \,  $ & $      1.8 \, \, \, \, \,  \,  $ \\ 
  5090937 &      210.897994 &  $       24.7 \, \, \, \,  \,  \,  $ & $      1.9 \, \, \, \, \,  \,  $ \\ 
  5090937 &      211.116562 &  $      24.03 \, \, \, $ & $     0.56 \, \, \, $ \\ 
  5090937 &      285.755862 &  $      -47.2 \, \, \, \,  \,  \,  $ & $      1.8 \, \, \, \, \,  \,  $ \\ 
\\ 
  5790807 &      197.931757 &  $     -27.85 \, \, \, $ & $     0.79 \, \, \, $ \\ 
  5790807 &      199.889515 &  $      -44.3 \, \, \, \,  \,  \,  $ & $      1.2 \, \, \, \, \,  \,  $ \\ 
  5790807 &      200.940686 &  $     -70.26 \, \, \, $ & $     0.46 \, \, \, $ \\ 
  5790807 &      201.125563 &  $     -69.78 \, \, \, $ & $     0.70 \, \, \, $ \\ 
  5790807 &      202.109996 &  $     -54.73 \, \, \, $ & $     0.37 \, \, \, $ \\ 
  5790807 &      202.889852 &  $     -47.65 \, \, \, $ & $     0.80 \, \, \, $ \\ 
  5790807 &      204.050364 &  $     -41.22 \, \, \, $ & $     0.75 \, \, \, $ \\ 
  5790807 &      216.066338 &  $     -29.00 \, \, \, $ & $     0.77 \, \, \, $ \\ 
  5790807 &      232.860373 &  $     -25.08 \, \, \, $ & $     0.43 \, \, \, $ \\ 
  5790807 &      236.009942 &  $      -24.8 \, \, \, \,  \,  \,  $ & $      1.0 \, \, \, \, \,  \,  $ \\ 
  5790807 &      254.051394 &  $     -23.07 \, \, \, $ & $     0.82 \, \, \, $ \\ 
  5790807 &      254.052170 &  $      -22.7 \, \, \, \,  \,  \,  $ & $      1.1 \, \, \, \, \,  \,  $ \\ 
\\ 
  5818706 &      201.120880 &  $     -23.01 \, \, \, $ & $     0.87 \, \, \, $ \\ 
  5818706 &      202.857127 &  $      44.87 \, \, \, $ & $     0.26 \, \, \, $ \\ 
  5818706 &      203.105341 &  $      56.71 \, \, \, $ & $     0.56 \, \, \, $ \\ 
  5818706 &      204.064768 &  $      73.34 \, \, \, $ & $     0.56 \, \, \, $ \\ 
  5818706 &      216.069677 &  $     -22.60 \, \, \, $ & $     0.34 \, \, \, $ \\ 
  5818706 &      218.074777 &  $      57.05 \, \, \, $ & $     0.72 \, \, \, $ \\ 
  5818706 &      229.812243 &  $     -17.60 \, \, \, $ & $     0.28 \, \, \, $ \\ 
  5818706 &      232.115750 &  $       7.01 \, \, \, $ & $     0.61 \, \, \, $ \\ 
  5818706 &      232.865028 &  $      49.40 \, \, \, $ & $     0.36 \, \, \, $ \\ 
  5818706 &      236.865600 &  $      54.07 \, \, \, $ & $     0.88 \, \, \, $ \\ 
  5818706 &      254.882565 &  $      26.51 \, \, \, $ & $     0.23 \, \, \, $ \\ 
\\ 
  5877364 &      151.057619 &  $      31.50 \, \, \, $ & $     0.25 \, \, \, $ \\ 
  5877364 &      201.098937 &  $       1.58 \, \, \, $ & $     0.53 \, \, \, $ \\ 
  5877364 &      213.101516 &  $      -3.29 \, \, \, $ & $     0.87 \, \, \, $ \\ 
  5877364 &      228.975561 &  $     -12.11 \, \, \, $ & $     0.25 \, \, \, $ \\ 
  5877364 &      236.004642 &  $     -24.33 \, \, \, $ & $     0.35 \, \, \, $ \\ 
  5877364 &      236.862834 &  $     -25.46 \, \, \, $ & $     0.29 \, \, \, $ \\ 
  5877364 &      239.984825 &  $      34.87 \, \, \, $ & $     0.21 \, \, \, $ \\ 
  5877364 &      241.069924 &  $      30.06 \, \, \, $ & $     0.35 \, \, \, $ \\ 
  5877364 &      243.039448 &  $      24.74 \, \, \, $ & $     0.31 \, \, \, $ \\ 
  5877364 &      254.050374 &  $      14.17 \, \, \, $ & $     0.48 \, \, \, $ \\ 
\\ 
  5960989 &      207.034701 &  $      -18.1 \, \, \, \,  \,  \,  $ & $      3.2 \, \, \, \, \,  \,  $ \\ 
  5960989 &      213.096418 &  $       10.6 \, \, \, \,  \,  \,  $ & $      2.9 \, \, \, \, \,  \,  $ \\ 
  5960989 &      213.979452 &  $      30.58 \, \, \, $ & $     0.95 \, \, \, $ \\ 
  5960989 &      214.118817 &  $       33.2 \, \, \, \,  \,  \,  $ & $      3.8 \, \, \, \, \,  \,  $ \\ 
  5960989 &      214.876039 &  $       33.9 \, \, \, \,  \,  \,  $ & $      2.2 \, \, \, \, \,  \,  $ \\ 
  5960989 &      215.950510 &  $      -16.3 \, \, \, \,  \,  \,  $ & $      3.2 \, \, \, \, \,  \,  $ \\ 
  5960989 &      216.120802 &  $      -22.0 \, \, \, \,  \,  \,  $ & $      3.2 \, \, \, \, \,  \,  $ \\ 
  5960989 &      218.071857 &  $      -35.0 \, \, \, \,  \,  \,  $ & $      1.2 \, \, \, \, \,  \,  $ \\ 
  5960989 &      232.869936 &  $      -35.5 \, \, \, \,  \,  \,  $ & $      2.6 \, \, \, \, \,  \,  $ \\ 
  5960989 &      244.798626 &  $      -26.1 \, \, \, \,  \,  \,  $ & $      1.4 \, \, \, \, \,  \,  $ \\ 
\\ 
  6370558 &      207.951542 &  $     -28.20 \, \, \, $ & $     0.31 \, \, \, $ \\ 
  6370558 &      211.092057 &  $     -28.63 \, \, \, $ & $     0.22 \, \, \, $ \\ 
  6370558 &      237.075829 &  $     -33.95 \, \, \, $ & $     0.25 \, \, \, $ \\ 
  6370558 &      243.037098 &  $     -38.94 \, \, \, $ & $     0.27 \, \, \, $ \\ 
  6370558 &      244.793176 &  $     -42.76 \, \, \, $ & $     0.13 \, \, \, $ \\ 
  6370558 &      247.025000 &  $    -52.285 $ & $    0.090 $ \\ 
  6370558 &      254.883490 &  $    -26.856 $ & $    0.093 $ \\ 
\\ 
  6775034 &      199.901981 &  $      15.48 \, \, \, $ & $     0.42 \, \, \, $ \\ 
  6775034 &      201.112854 &  $       54.1 \, \, \, \,  \,  \,  $ & $      2.9 \, \, \, \, \,  \,  $ \\ 
  6775034 &      201.912073 &  $       62.3 \, \, \, \,  \,  \,  $ & $      1.2 \, \, \, \, \,  \,  $ \\ 
  6775034 &      210.024418 &  $      15.33 \, \, \, $ & $     0.66 \, \, \, $ \\ 
  6775034 &      210.979572 &  $       45.9 \, \, \, \,  \,  \,  $ & $      1.5 \, \, \, \, \,  \,  $ \\ 
  6775034 &      211.887561 &  $      64.40 \, \, \, $ & $     0.64 \, \, \, $ \\ 
  6775034 &      230.935338 &  $      40.08 \, \, \, $ & $     0.93 \, \, \, $ \\ 
  6775034 &      232.874112 &  $      15.80 \, \, \, $ & $     0.64 \, \, \, $ \\ 
  6775034 &      236.012002 &  $       -5.7 \, \, \, \,  \,  \,  $ & $      3.8 \, \, \, \, \,  \,  $ \\ 
\\ 
  8027591 &      179.976502 &  $      53.72 \, \, \, $ & $     0.47 \, \, \, $ \\ 
  8027591 &      203.869961 &  $     61.865 $ & $    0.060 $ \\ 
  8027591 &      204.118427 &  $      57.29 \, \, \, $ & $     0.41 \, \, \, $ \\ 
  8027591 &      207.035903 &  $      -5.81 \, \, \, $ & $     0.44 \, \, \, $ \\ 
  8027591 &      207.946487 &  $      -9.78 \, \, \, $ & $     0.50 \, \, \, $ \\ 
  8027591 &      210.984387 &  $     -12.55 \, \, \, $ & $     0.57 \, \, \, $ \\ 
  8027591 &      228.974524 &  $      37.41 \, \, \, $ & $     0.36 \, \, \, $ \\ 
  8027591 &      229.935800 &  $       8.19 \, \, \, $ & $     0.27 \, \, \, $ \\ 
  8027591 &      247.029096 &  $       8.21 \, \, \, $ & $     0.65 \, \, \, $ \\ 
  8027591 &      298.920298 &  $      36.40 \, \, \, $ & $     0.42 \, \, \, $ \\ 
\\ 
  8164262 &      151.059455 &  $       25.9 \, \, \, \,  \,  \,  $ & $      5.7 \, \, \, \, \,  \,  $ \\ 
  8164262 &      202.892222 &  $       15.3 \, \, \, \,  \,  \,  $ & $      3.1 \, \, \, \, \,  \,  $ \\ 
  8164262 &      228.978804 &  $       17.6 \, \, \, \,  \,  \,  $ & $      1.8 \, \, \, \, \,  \,  $ \\ 
  8164262 &      237.073545 &  $       28.4 \, \, \, \,  \,  \,  $ & $      6.0 \, \, \, \, \,  \,  $ \\ 
  8164262 &      239.988444 &  $      33.46 \, \, \, $ & $     0.72 \, \, \, $ \\ 
  8164262 &      241.068023 &  $      37.22 \, \, \, $ & $     0.78 \, \, \, $ \\ 
  8164262 &      243.031959 &  $       15.2 \, \, \, \,  \,  \,  $ & $      9.5 \, \, \, \, \,  \,  $ \\ 
  8164262 &      244.791056 &  $      -9.82 \, \, \, $ & $     0.54 \, \, \, $ \\ 
  8164262 &      247.027160 &  $       -0.5 \, \, \, \,  \,  \,  $ & $      2.5 \, \, \, \, \,  \,  $ \\ 
  8164262 &      255.883316 &  $       3.04 \, \, \, $ & $     0.86 \, \, \, $ \\ 
\\ 
  9016693 &      207.936227 &  $       34.4 \, \, \, \,  \,  \,  $ & $      1.0 \, \, \, \, \,  \,  $ \\ 
  9016693 &      210.973901 &  $       46.9 \, \, \, \,  \,  \,  $ & $      1.3 \, \, \, \, \,  \,  $ \\ 
  9016693 &      216.062924 &  $      -7.44 \, \, \, $ & $     0.50 \, \, \, $ \\ 
  9016693 &      218.048802 &  $      -57.9 \, \, \, \,  \,  \,  $ & $      1.0 \, \, \, \, \,  \,  $ \\ 
  9016693 &      243.035362 &  $     -47.37 \, \, \, $ & $     0.53 \, \, \, $ \\ 
  9016693 &      244.794447 &  $     -53.90 \, \, \, $ & $     0.55 \, \, \, $ \\ 
  9016693 &      254.878619 &  $       9.98 \, \, \, $ & $     0.45 \, \, \, $ \\ 
  9016693 &      326.819095 &  $     -24.94 \, \, \, $ & $     0.45 \, \, \, $ \\ 
\\ 
  9965691 &      207.049627 &  $       9.30 \, \, \, $ & $     0.24 \, \, \, $ \\ 
  9965691 &      207.933480 &  $      14.49 \, \, \, $ & $     0.32 \, \, \, $ \\ 
  9965691 &      210.032164 &  $     -46.99 \, \, \, $ & $     0.23 \, \, \, $ \\ 
  9965691 &      211.102710 &  $     -56.37 \, \, \, $ & $     0.17 \, \, \, $ \\ 
  9965691 &      232.880033 &  $     -43.05 \, \, \, $ & $     0.45 \, \, \, $ \\ 
  9965691 &      236.084435 &  $     -21.29 \, \, \, $ & $     0.38 \, \, \, $ \\ 
  9965691 &      244.800398 &  $     -56.62 \, \, \, $ & $     0.18 \, \, \, $ \\ 
  9965691 &      255.887813 &  $     -12.27 \, \, \, $ & $     0.28 \, \, \, $ \\ 
\\ 
  9972385 &      199.886543 &  $      -0.85 \, \, \, $ & $     0.54 \, \, \, $ \\ 
  9972385 &      199.887608 &  $      -1.14 \, \, \, $ & $     0.34 \, \, \, $ \\ 
  9972385 &      213.106998 &  $      -1.04 \, \, \, $ & $     0.86 \, \, \, $ \\ 
  9972385 &      228.976389 &  $      -0.44 \, \, \, $ & $     0.94 \, \, \, $ \\ 
  9972385 &      236.869200 &  $      -0.76 \, \, \, $ & $     0.54 \, \, \, $ \\ 
  9972385 &      239.986204 &  $       0.06 \, \, \, $ & $     0.73 \, \, \, $ \\ 
  9972385 &      241.070574 &  $      -0.23 \, \, \, $ & $     0.18 \, \, \, $ \\ 
  9972385 &      244.802941 &  $       0.39 \, \, \, $ & $     0.41 \, \, \, $ \\ 
  9972385 &      298.947904 &  $      -0.97 \, \, \, $ & $     0.24 \, \, \, $ \\ 
\\ 
 10334122 &      211.100578 &  $       5.02 \, \, \, $ & $     0.83 \, \, \, $ \\ 
 10334122 &      222.121926 &  $       2.40 \, \, \, $ & $     0.41 \, \, \, $ \\ 
 10334122 &      223.121539 &  $      -15.8 \, \, \, \,  \,  \,  $ & $      6.5 \, \, \, \, \,  \,  $ \\ 
 10334122 &      232.877807 &  $      -32.9 \, \, \, \,  \,  \,  $ & $      1.1 \, \, \, \, \,  \,  $ \\ 
 10334122 &      244.795948 &  $      -4.75 \, \, \, $ & $     0.73 \, \, \, $ \\ 
 10334122 &      254.876944 &  $       20.0 \, \, \, \,  \,  \,  $ & $      1.2 \, \, \, \, \,  \,  $ \\ 
 10334122 &      264.957672 &  $     -59.42 \, \, \, $ & $     0.81 \, \, \, $ \\ 
\\ 
 11071278 &      207.025944 &  $       9.26 \, \, \, $ & $     0.46 \, \, \, $ \\ 
 11071278 &      207.944808 &  $       8.27 \, \, \, $ & $     0.49 \, \, \, $ \\ 
 11071278 &      210.982799 &  $       4.75 \, \, \, $ & $     0.78 \, \, \, $ \\ 
 11071278 &      228.905477 &  $     -22.98 \, \, \, $ & $     0.37 \, \, \, $ \\ 
 11071278 &      229.809886 &  $      -5.50 \, \, \, $ & $     0.79 \, \, \, $ \\ 
 11071278 &      230.930693 &  $       3.78 \, \, \, $ & $     0.17 \, \, \, $ \\ 
 11071278 &      236.004933 &  $      14.14 \, \, \, $ & $     0.41 \, \, \, $ \\ 
 11071278 &      243.043690 &  $      15.19 \, \, \, $ & $     0.45 \, \, \, $ \\ 
\\ 
 11122789 &      201.099488 &  $      -19.8 \, \, \, \,  \,  \,  $ & $      1.7 \, \, \, \, \,  \,  $ \\ 
 11122789 &      202.114385 &  $      -20.5 \, \, \, \,  \,  \,  $ & $      2.3 \, \, \, \, \,  \,  $ \\ 
 11122789 &      202.851557 &  $      -17.8 \, \, \, \,  \,  \,  $ & $      3.6 \, \, \, \, \,  \,  $ \\ 
 11122789 &      203.087150 &  $      -20.0 \, \, \, \,  \,  \,  $ & $      1.7 \, \, \, \, \,  \,  $ \\ 
 11122789 &      204.050598 &  $      -18.4 \, \, \, \,  \,  \,  $ & $      2.0 \, \, \, \, \,  \,  $ \\ 
 11122789 &      207.026946 &  $      -20.6 \, \, \, \,  \,  \,  $ & $      5.6 \, \, \, \, \,  \,  $ \\ 
 11122789 &      207.923667 &  $      -21.4 \, \, \, \,  \,  \,  $ & $      3.6 \, \, \, \, \,  \,  $ \\ 
 11122789 &      209.125135 &  $      -19.6 \, \, \, \,  \,  \,  $ & $      1.2 \, \, \, \, \,  \,  $ \\ 
 11122789 &      209.887247 &  $      -18.2 \, \, \, \,  \,  \,  $ & $      1.0 \, \, \, \, \,  \,  $ \\ 
 11122789 &      210.099344 &  $      -22.3 \, \, \, \,  \,  \,  $ & $      3.4 \, \, \, \, \,  \,  $ \\ 
 11122789 &      210.972325 &  $      -18.8 \, \, \, \,  \,  \,  $ & $      1.9 \, \, \, \, \,  \,  $ \\ 
 11122789 &      211.883106 &  $      -18.0 \, \, \, \,  \,  \,  $ & $      2.6 \, \, \, \, \,  \,  $ \\ 
 11122789 &      213.113229 &  $      -22.8 \, \, \, \,  \,  \,  $ & $      2.5 \, \, \, \, \,  \,  $ \\ 
 11122789 &      237.076765 &  $     -19.47 \, \, \, $ & $     0.75 \, \, \, $ \\ 
 11122789 &      244.787644 &  $      -20.0 \, \, \, \,  \,  \,  $ & $      1.3 \, \, \, \, \,  \,  $ \\ 
 11122789 &      247.035372 &  $     -19.91 \, \, \, $ & $     0.67 \, \, \, $ \\ 
 11122789 &      254.871927 &  $      -18.5 \, \, \, \,  \,  \,  $ & $      2.2 \, \, \, \, \,  \,  $ \\ 
 11122789 &      255.880013 &  $     -18.19 \, \, \, $ & $     0.73 \, \, \, $ \\ 
 11122789 &      290.878199 &  $     -20.63 \, \, \, $ & $     0.51 \, \, \, $ \\ 
 11122789 &      294.899399 &  $      -19.5 \, \, \, \,  \,  \,  $ & $      2.6 \, \, \, \, \,  \,  $ \\ 
\\ 
 11403032 &      199.884762 &  $      49.54 \, \, \, $ & $     0.93 \, \, \, $ \\ 
 11403032 &      201.102763 &  $      24.25 \, \, \, $ & $     0.70 \, \, \, $ \\ 
 11403032 &      202.116291 &  $       7.46 \, \, \, $ & $     0.38 \, \, \, $ \\ 
 11403032 &      207.036516 &  $      48.60 \, \, \, $ & $     0.50 \, \, \, $ \\ 
 11403032 &      207.920456 &  $       44.2 \, \, \, \,  \,  \,  $ & $      1.3 \, \, \, \, \,  \,  $ \\ 
 11403032 &      208.103271 &  $      40.17 \, \, \, $ & $     0.85 \, \, \, $ \\ 
 11403032 &      209.128857 &  $      16.56 \, \, \, $ & $     0.43 \, \, \, $ \\ 
 11403032 &      210.020770 &  $       3.54 \, \, \, $ & $     0.47 \, \, \, $ \\ 
 11403032 &      216.061339 &  $      32.79 \, \, \, $ & $     0.70 \, \, \, $ \\ 
 11403032 &      228.979912 &  $      15.85 \, \, \, $ & $     0.12 \, \, \, $ \\ 
 11403032 &      236.863282 &  $      26.94 \, \, \, $ & $     0.51 \, \, \, $ \\ 
 11403032 &      264.959009 &  $      -8.24 \, \, \, $ & $     0.22 \, \, \, $ \\ 
\\ 
 11409673 &      179.973959 &  $      -4.48 \, \, \, $ & $     0.87 \, \, \, $ \\ 
 11409673 &      202.893883 &  $      -5.27 \, \, \, $ & $     0.32 \, \, \, $ \\ 
 11409673 &      203.866661 &  $      -6.14 \, \, \, $ & $     0.27 \, \, \, $ \\ 
 11409673 &      204.119595 &  $      -5.79 \, \, \, $ & $     0.85 \, \, \, $ \\ 
 11409673 &      211.096310 &  $      -5.43 \, \, \, $ & $     0.86 \, \, \, $ \\ 
 11409673 &      213.103745 &  $      -5.62 \, \, \, $ & $     0.53 \, \, \, $ \\ 
 11409673 &      214.114574 &  $      -4.42 \, \, \, $ & $     0.93 \, \, \, $ \\ 
 11409673 &      216.064042 &  $      -6.44 \, \, \, $ & $     0.69 \, \, \, $ \\ 
 11409673 &      228.984199 &  $      -6.09 \, \, \, $ & $     0.42 \, \, \, $ \\ 
 11409673 &      232.870856 &  $      -4.73 \, \, \, $ & $     0.70 \, \, \, $ \\ 
\\ 
 11649962 &      179.977721 &  $     -12.32 \, \, \, $ & $     0.49 \, \, \, $ \\ 
 11649962 &      202.852279 &  $     -97.80 \, \, \, $ & $     0.98 \, \, \, $ \\ 
 11649962 &      203.088023 &  $     -81.86 \, \, \, $ & $     0.59 \, \, \, $ \\ 
 11649962 &      204.054354 &  $     -33.12 \, \, \, $ & $     0.53 \, \, \, $ \\ 
 11649962 &      213.099662 &  $    -111.80 \, \, \, $ & $     0.40 \, \, \, $ \\ 
 11649962 &      214.116290 &  $     -53.90 \, \, \, $ & $     0.74 \, \, \, $ \\ 
 11649962 &      216.067657 &  $      -1.44 \, \, \, $ & $     0.60 \, \, \, $ \\ 
 11649962 &      232.862673 &  $     -14.98 \, \, \, $ & $     0.31 \, \, \, $ \\ 
 11649962 &      254.873138 &  $     -84.86 \, \, \, $ & $     0.78 \, \, \, $ \\ 
 11649962 &      262.750967 &  $       17.3 \, \, \, \,  \,  \,  $ & $      1.2 \, \, \, \, \,  \,  $ \\ 
\\ 
 11923629 &      206.943348 &  $     -60.94 \, \, \, $ & $     0.22 \, \, \, $ \\ 
 11923629 &      207.119024 &  $     -60.45 \, \, \, $ & $     0.15 \, \, \, $ \\ 
 11923629 &      207.948997 &  $     -54.05 \, \, \, $ & $     0.27 \, \, \, $ \\ 
 11923629 &      211.092654 &  $     -23.61 \, \, \, $ & $     0.18 \, \, \, $ \\ 
 11923629 &      222.125231 &  $     -13.35 \, \, \, $ & $     0.15 \, \, \, $ \\ 
 11923629 &      223.125918 &  $     -36.86 \, \, \, $ & $     0.12 \, \, \, $ \\ 
 11923629 &      232.875970 &  $      -0.41 \, \, \, $ & $     0.24 \, \, \, $ \\ 
 11923629 &      236.082087 &  $       9.92 \, \, \, $ & $     0.16 \, \, \, $ \\ 
\\ 
 12255108 &      202.935881 &  $      -68.2 \, \, \, \,  \,  \,  $ & $      2.6 \, \, \, \, \,  \,  $ \\ 
 12255108 &      203.870341 &  $      -53.2 \, \, \, \,  \,  \,  $ & $      2.8 \, \, \, \, \,  \,  $ \\ 
 12255108 &      204.116491 &  $      -47.7 \, \, \, \,  \,  \,  $ & $      1.1 \, \, \, \, \,  \,  $ \\ 
 12255108 &      210.984874 &  $      -20.0 \, \, \, \,  \,  \,  $ & $      3.2 \, \, \, \, \,  \,  $ \\ 
 12255108 &      211.883787 &  $      -60.3 \, \, \, \,  \,  \,  $ & $      3.3 \, \, \, \, \,  \,  $ \\ 
 12255108 &      213.101697 &  $      -54.6 \, \, \, \,  \,  \,  $ & $      2.8 \, \, \, \, \,  \,  $ \\ 
 12255108 &      222.123971 &  $      -58.0 \, \, \, \,  \,  \,  $ & $      2.3 \, \, \, \, \,  \,  $ \\ 
 12255108 &      223.122590 &  $      -21.6 \, \, \, \,  \,  \,  $ & $      2.1 \, \, \, \, \,  \,  $ \\ 
 12255108 &      228.980858 &  $      -4.85 \, \, \, $ & $     0.33 \, \, \, $ \\ 
 12255108 &      229.936210 &  $      -50.1 \, \, \, \,  \,  \,  $ & $      1.4 \, \, \, \, \,  \,  $ \\ 
 12255108 &      236.014967 &  $       30.7 \, \, \, \,  \,  \,  $ & $      1.8 \, \, \, \, \,  \,  $ \\ 
\end{longtable}
\end{center}

%% file: figures_rvlc1.tex

\def\figwidth{2.7in}

\stepcounter{figure} 

\begin{table}[h]
\begin{center}
\begin{tabular}{lr}
 \includegraphics[width=\figwidth]{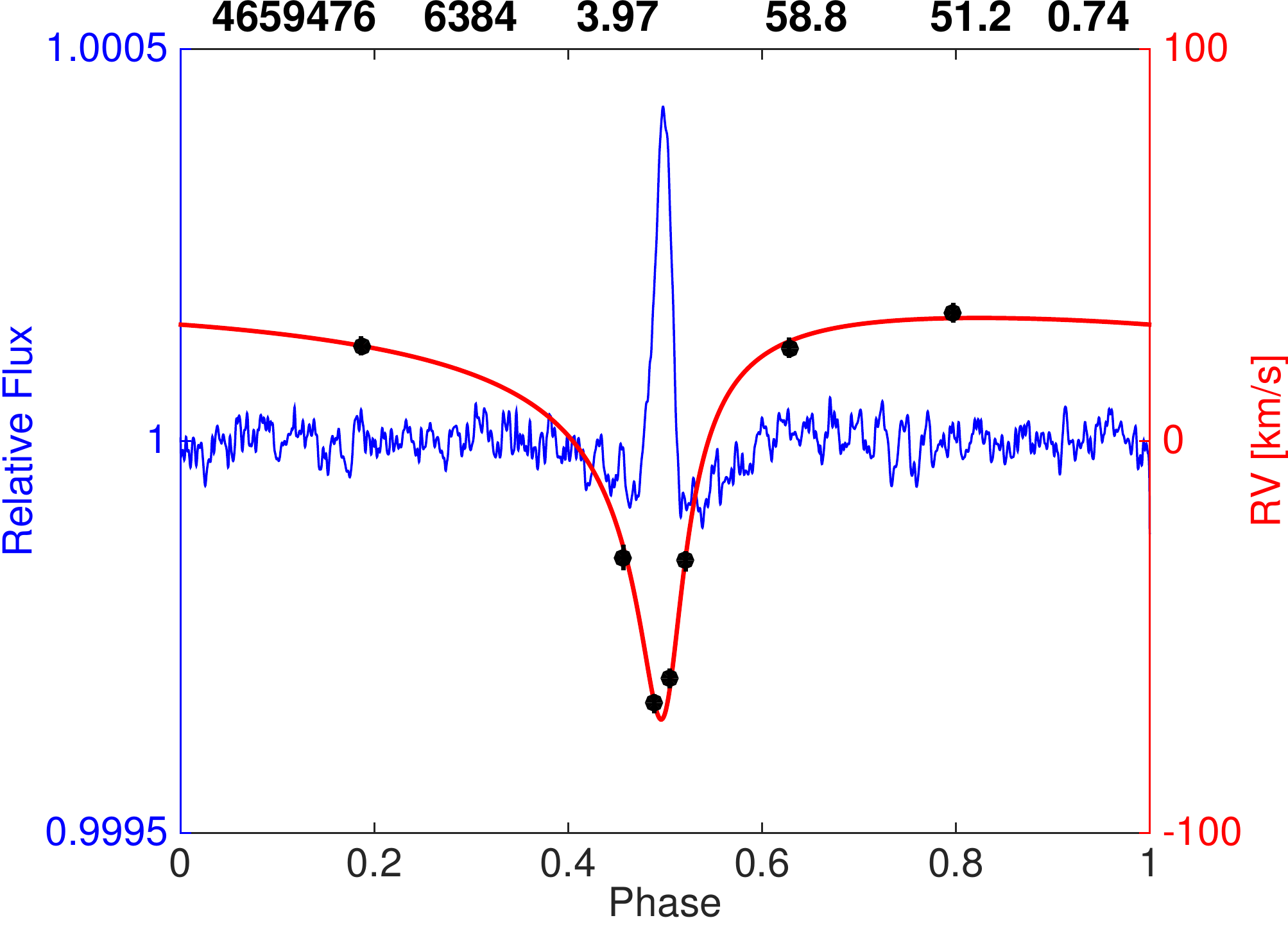} \vspace{3mm}&
 \includegraphics[width=\figwidth]{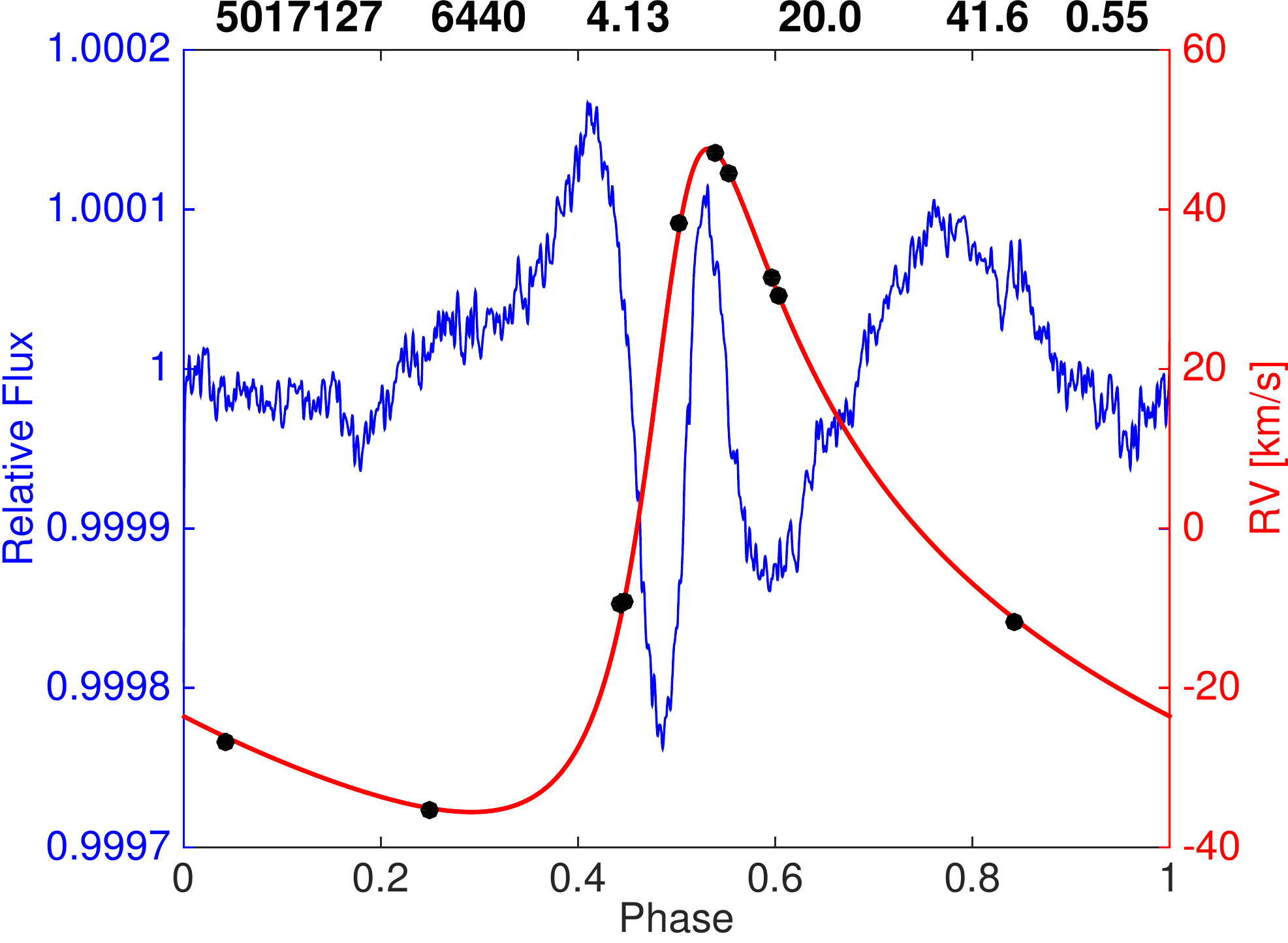}  \\
\includegraphics[width=\figwidth]{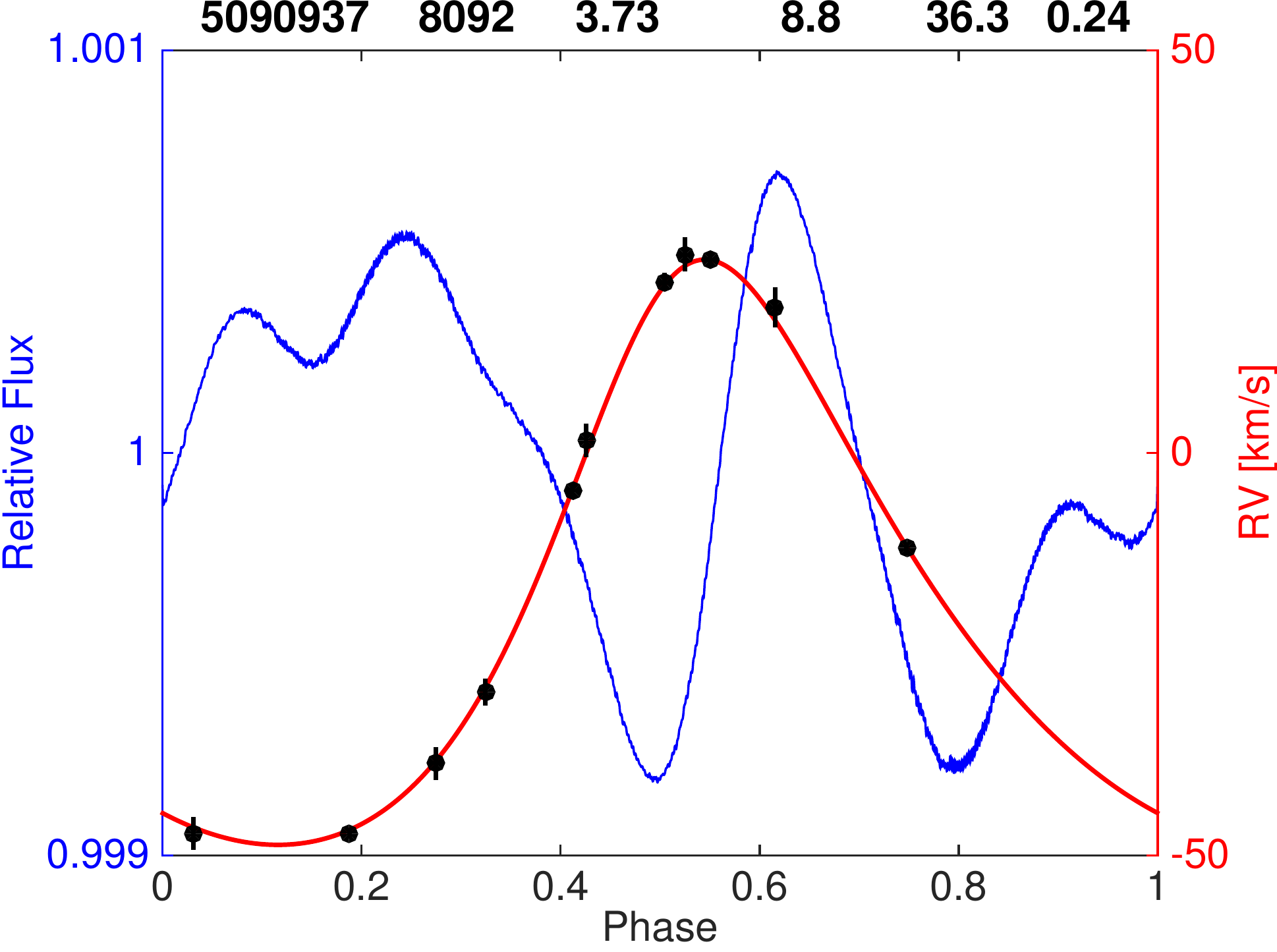} \vspace{3mm}&
 \includegraphics[width=\figwidth]{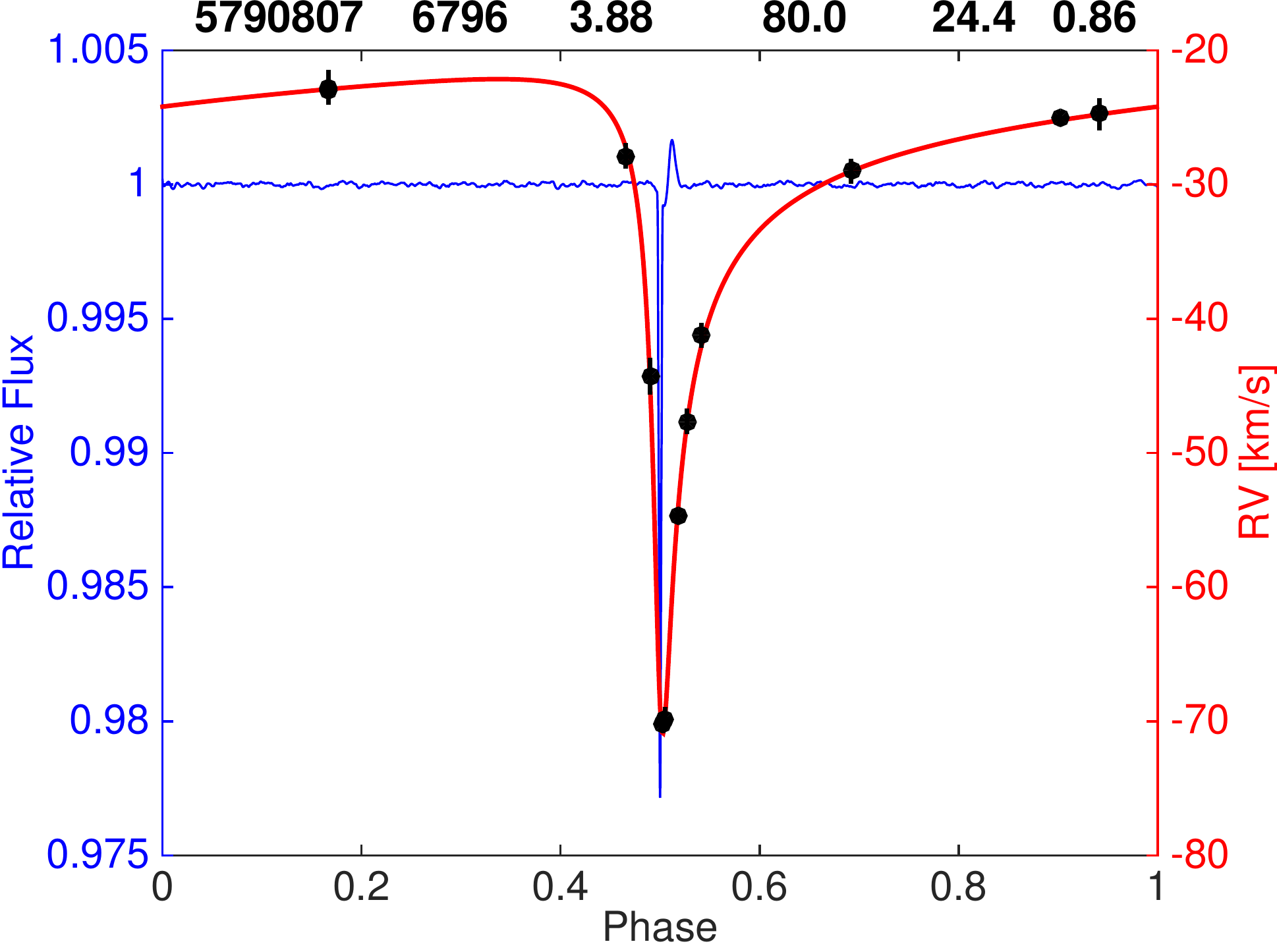} \\
 \includegraphics[width=\figwidth]{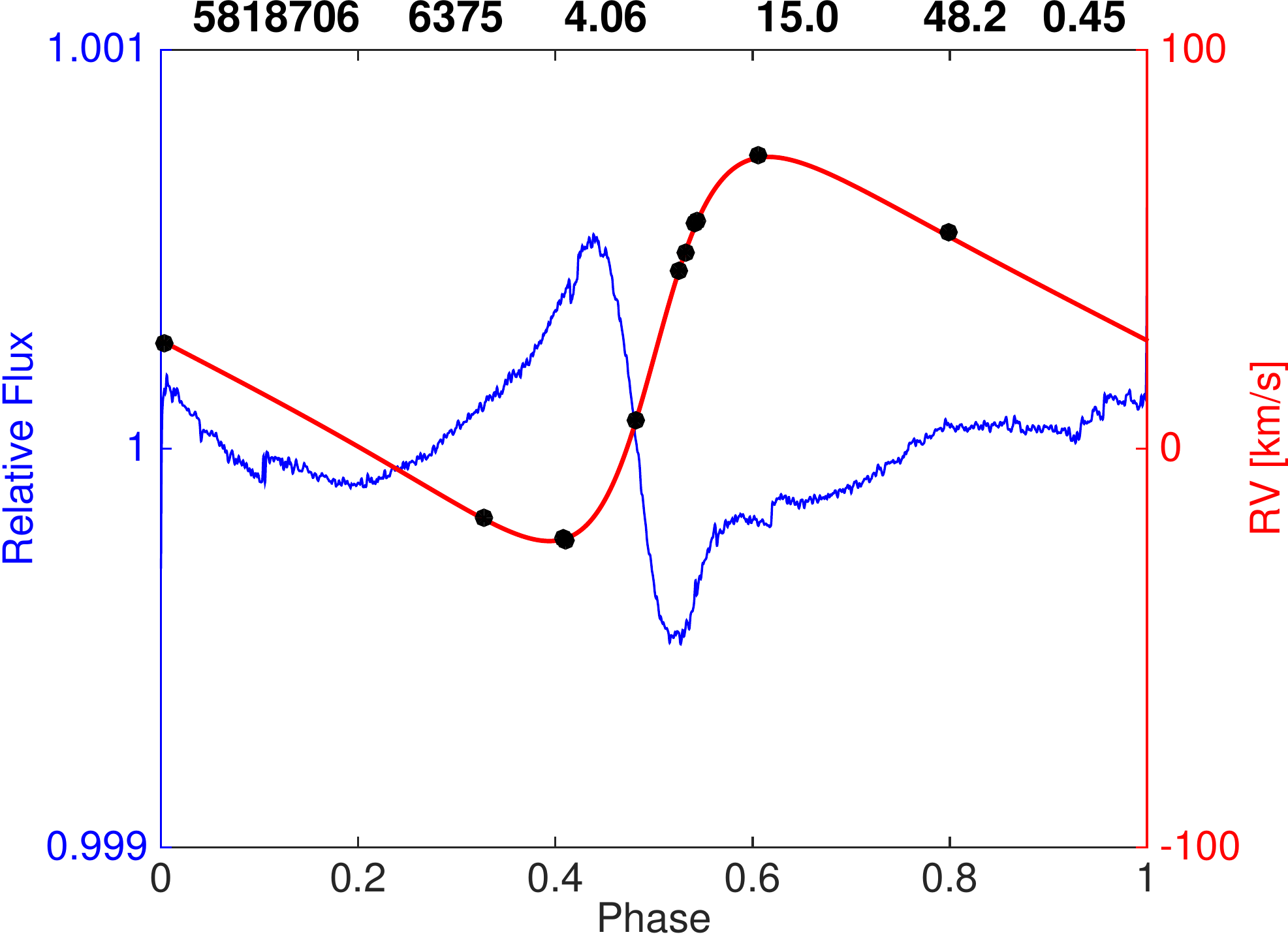} \vspace{3mm}&
 \includegraphics[width= \figwidth]{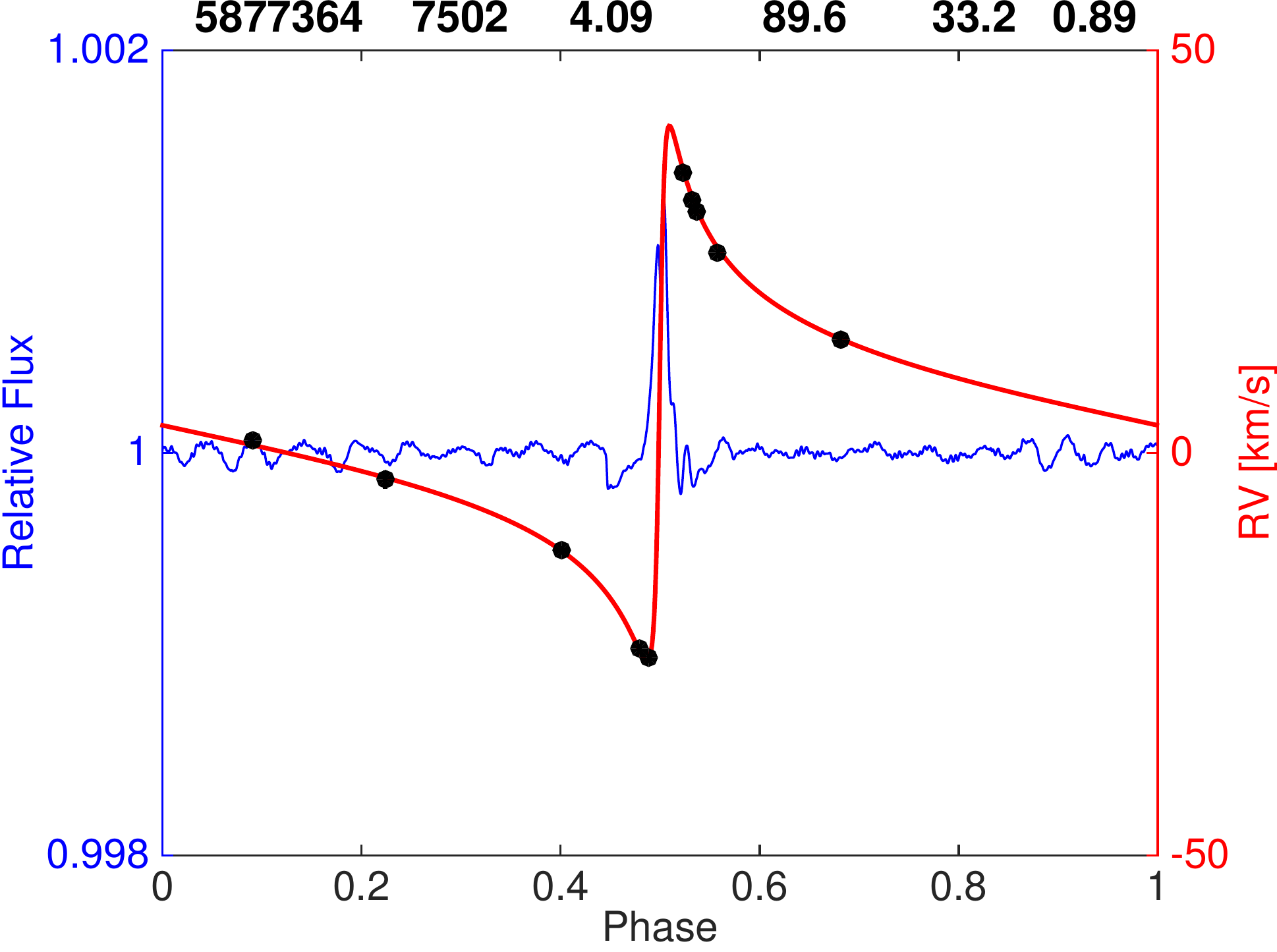} \\
 \includegraphics[width= \figwidth]{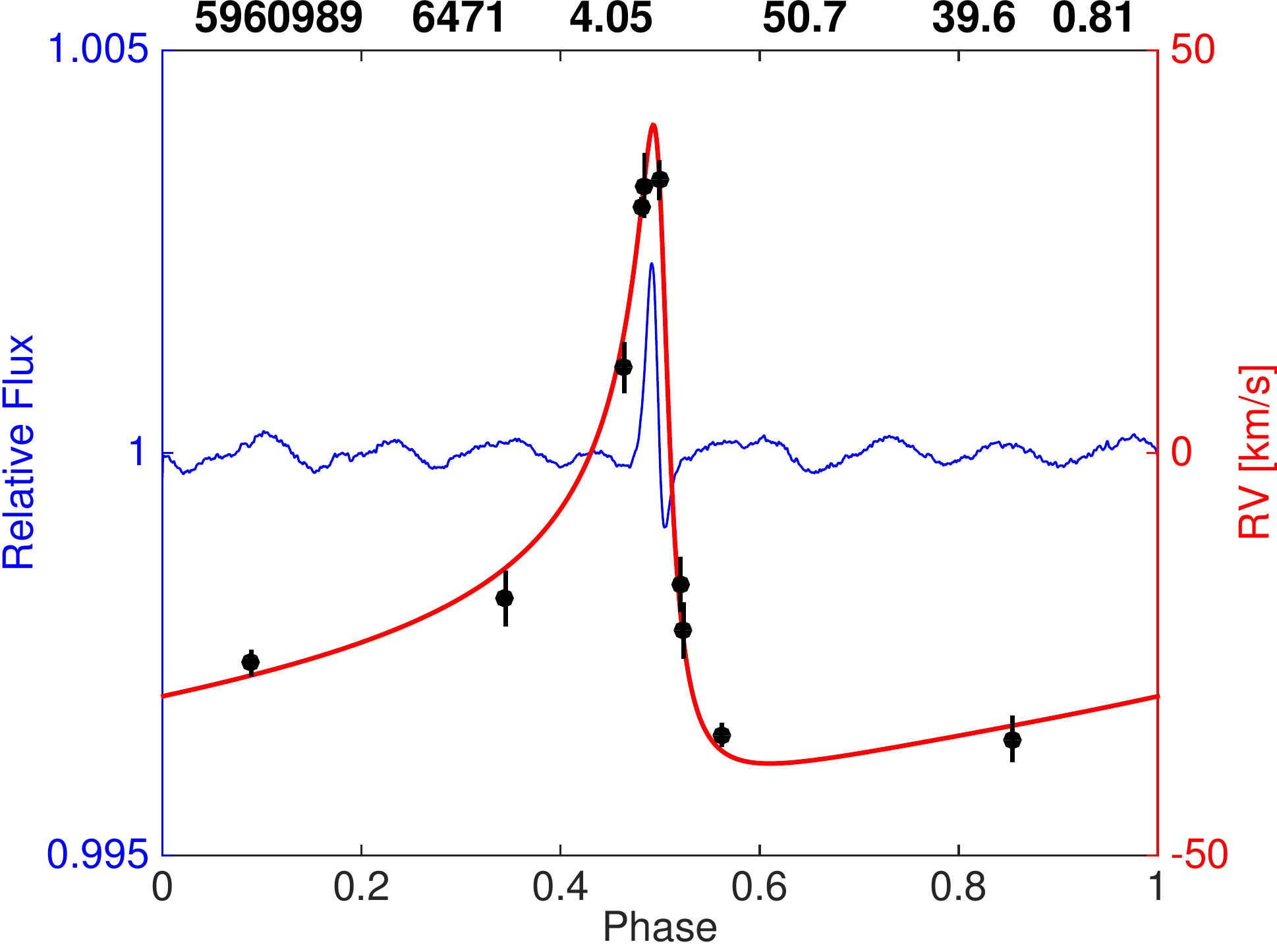} \vspace{3mm}&
 \includegraphics[width= \figwidth]{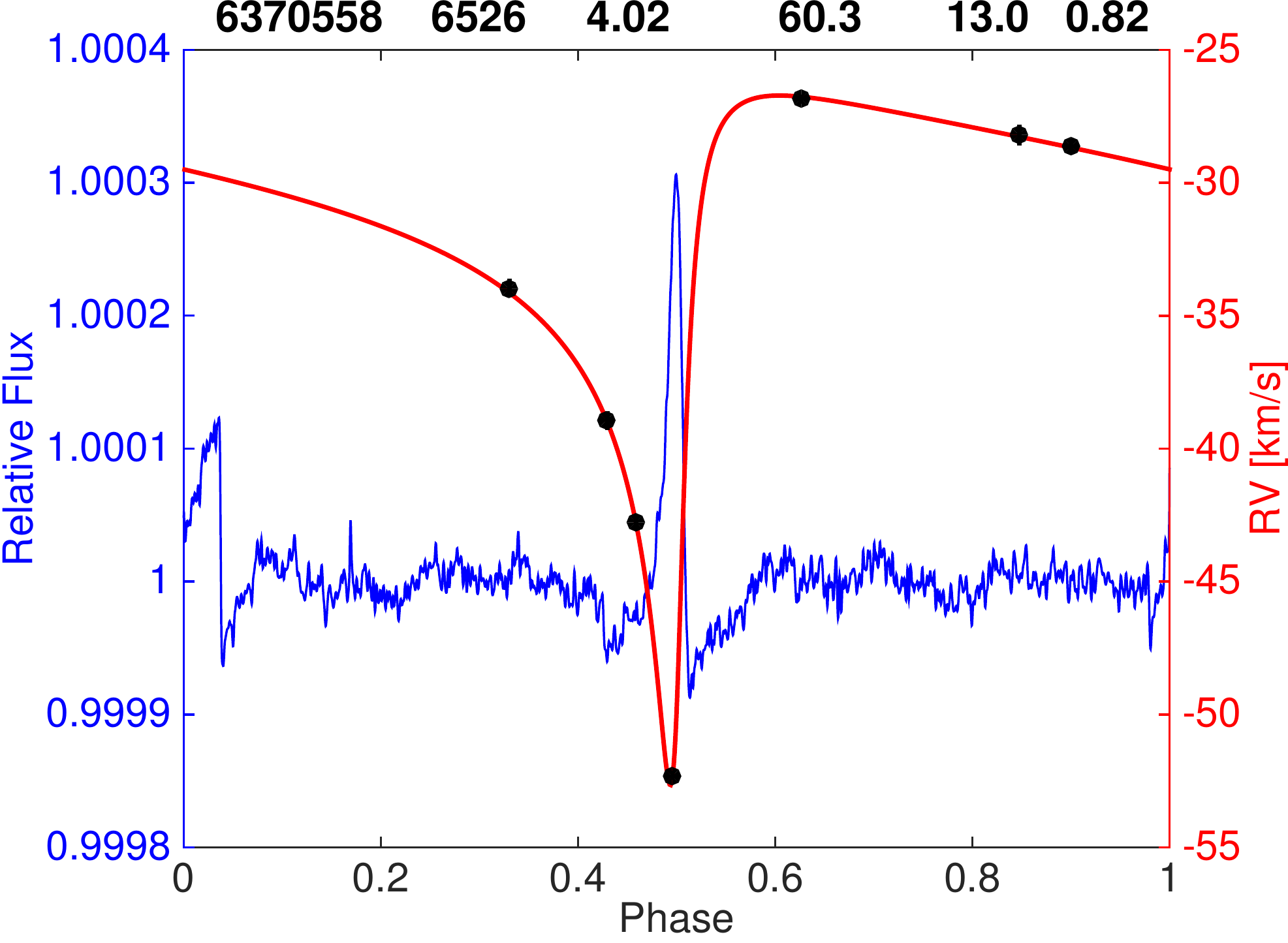} 
\end{tabular}
\begin{minipage}{60em}
Figure \thefigure: The panels show the \ik\ phase-folded relative flux light curve (blue; left Y-axis) and the phase-folded RV curve model (red; right Y-axis) with periastron at phase 0.5, for KID 4659476, KID 5017127, KID 5090937, KID 5790807, KID 58181706, KID 5877364, KID 5960989, and KID 6370558. In each panel the title lists, from left to right, the KIC ID, \teff\ (K), $\logg$, $P$ (d), $K$ (\kms), and $e$. The RV measurements are overplotted in black, including error bars although in some panels the markers are larger than the error bars. The \ik\ light curves shown here were derived by applying a running mean to the \ik\ data, followed by binning with a bin size of 0.0002 in phase.
\end{minipage}
\label{fig:rvlc1} 
\end{center}
\end{table}

%% file: figures_rvlc2.tex

\def\figwidth{2.7in}

\stepcounter{figure} 

\begin{table}[h]
\begin{center}
\begin{tabular}{lcr}
\includegraphics[width=\figwidth]{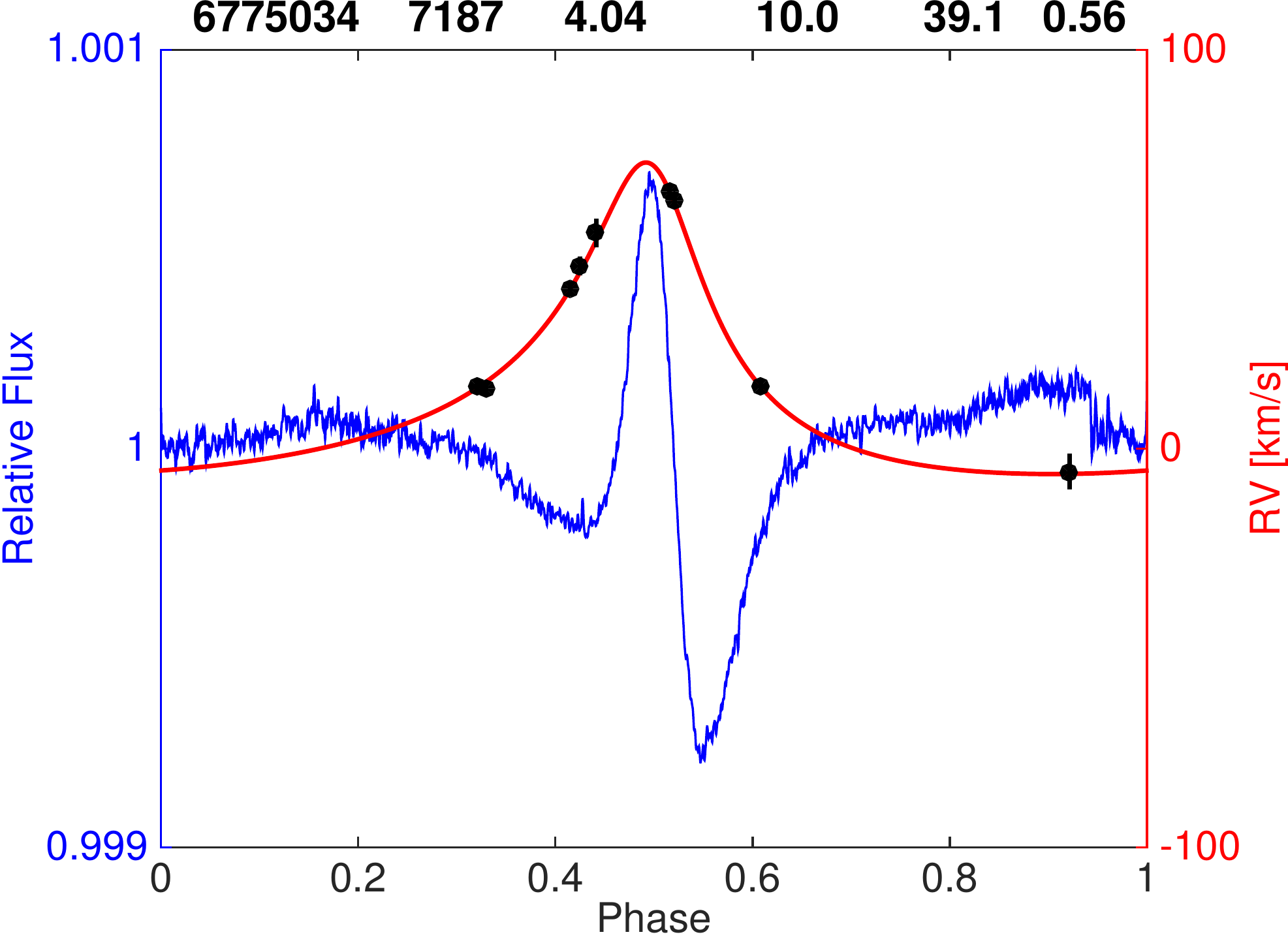}\vspace{3mm}& 
\includegraphics[width=\figwidth]{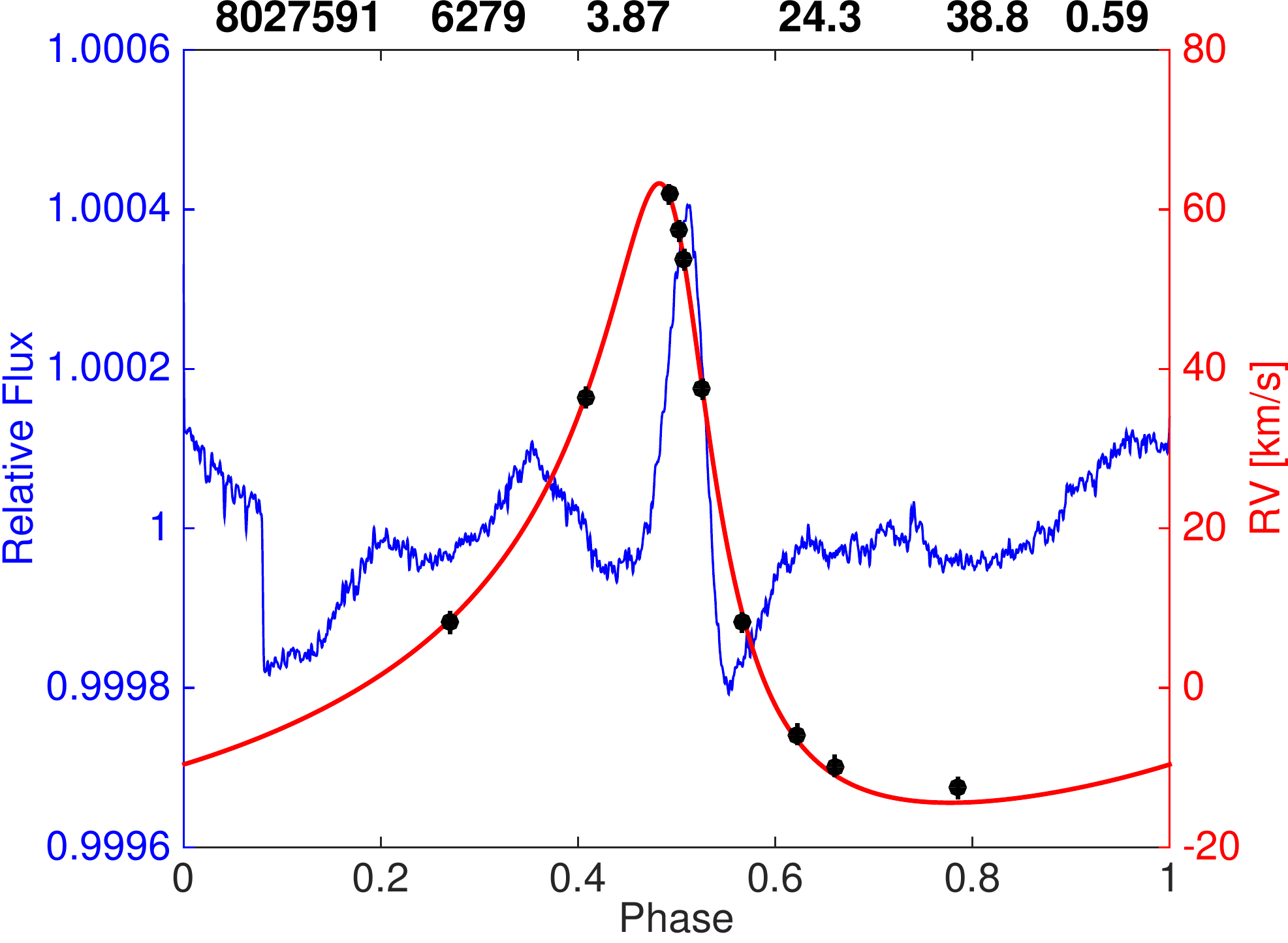}\\
\includegraphics[width=\figwidth]{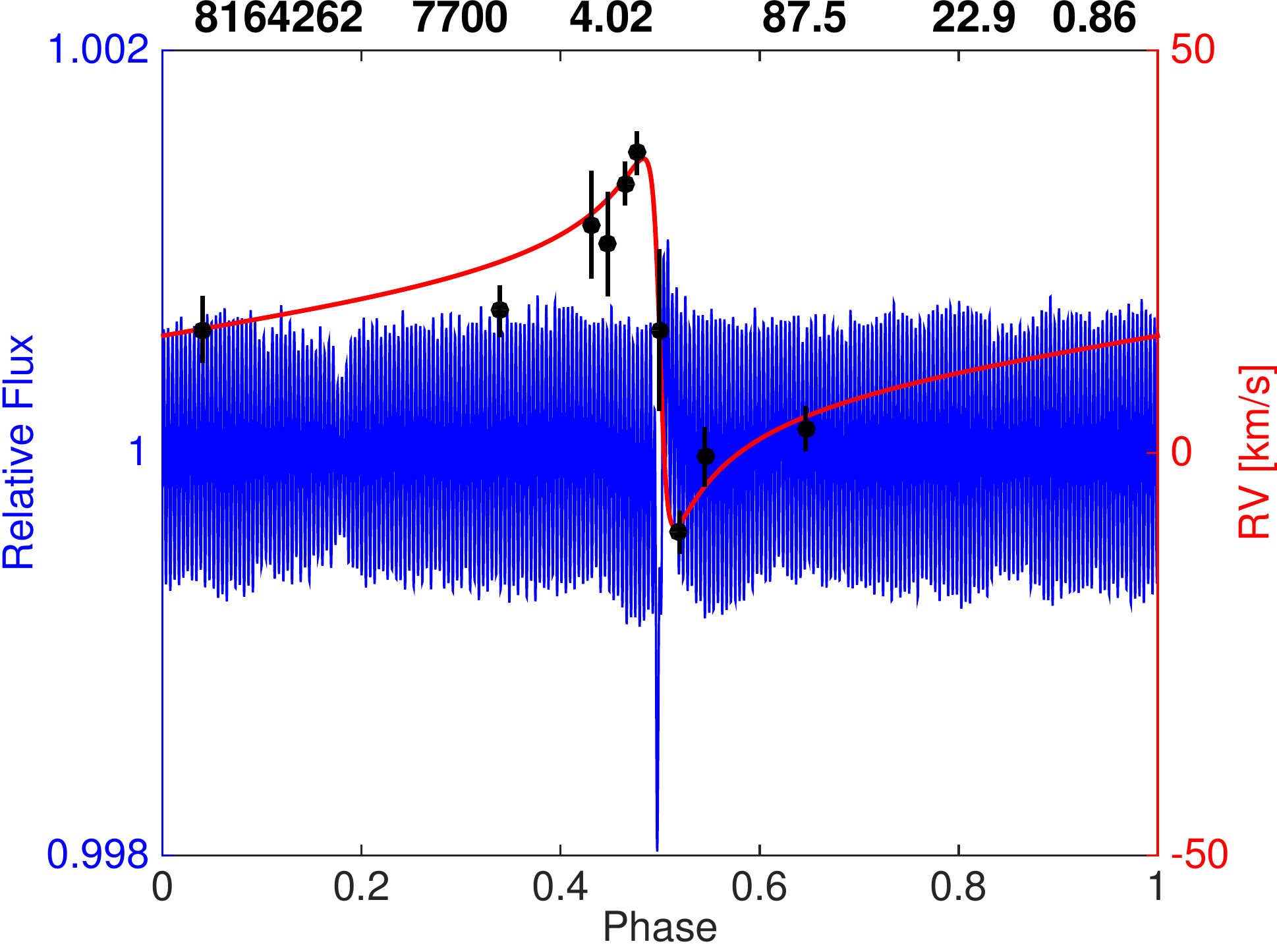}  \vspace{3mm}&
 \includegraphics[width=\figwidth]{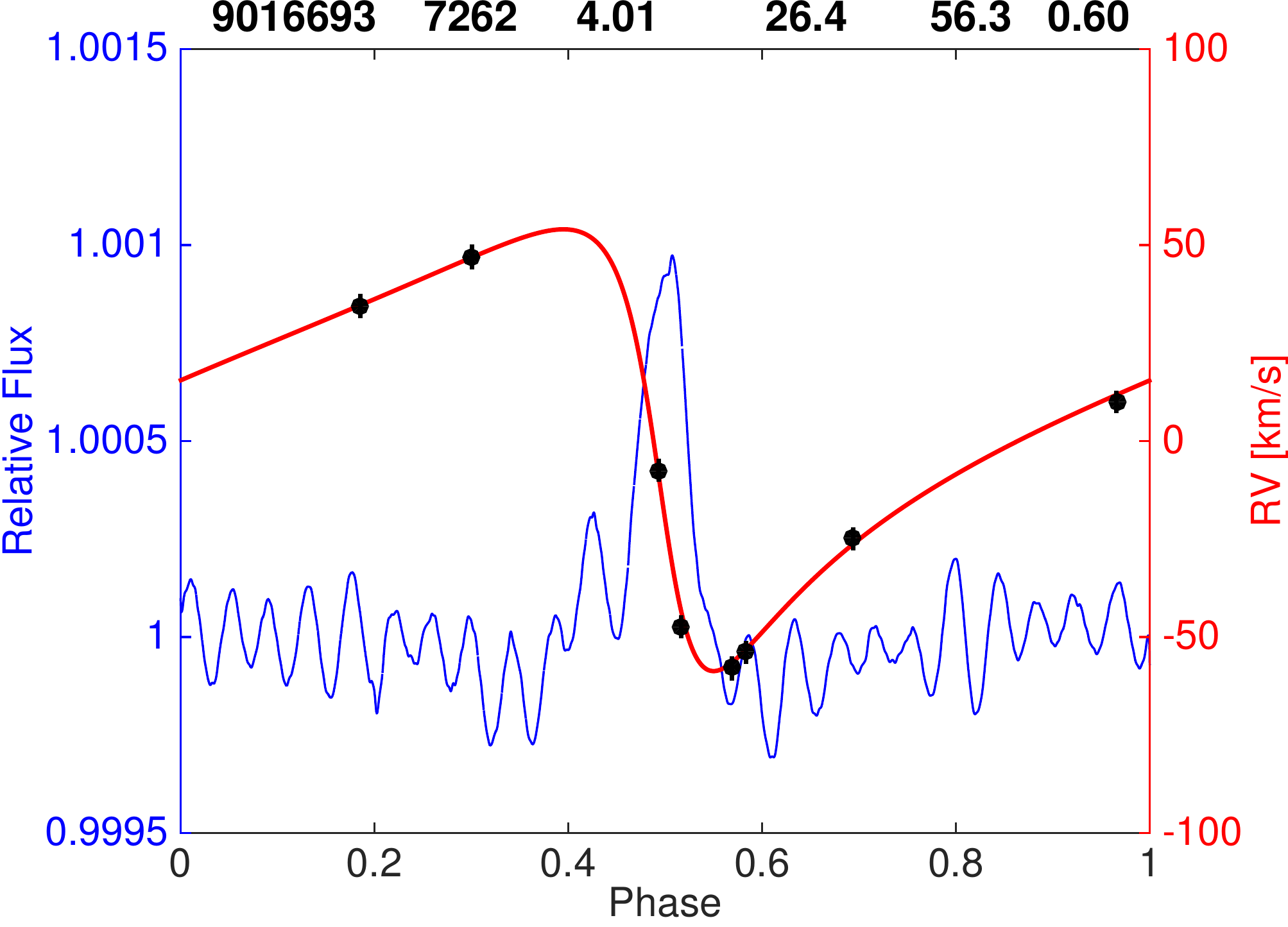} \\
\includegraphics[width=\figwidth]{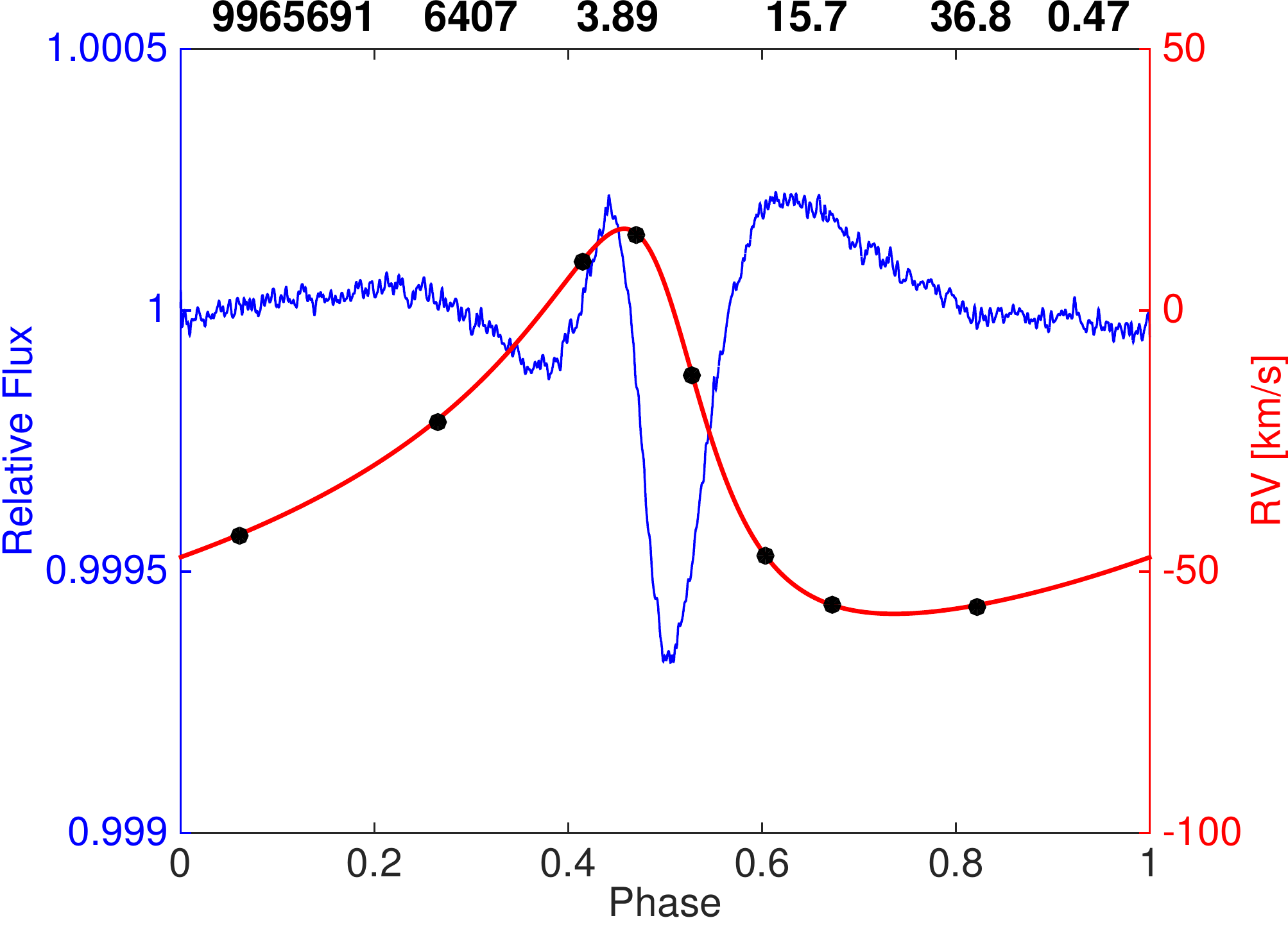}  \vspace{3mm}&
\includegraphics[width=\figwidth]{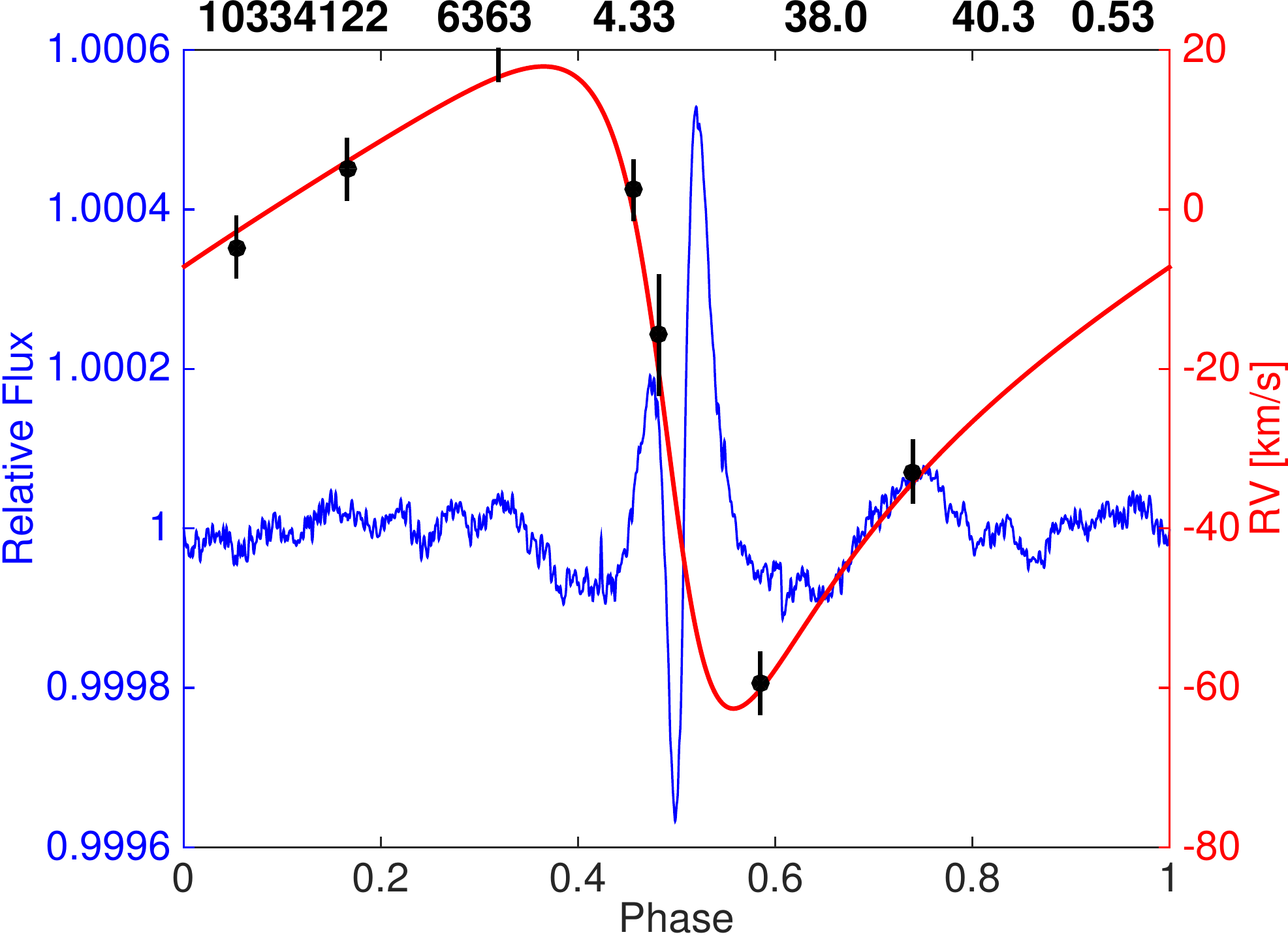} \\
\includegraphics[width=\figwidth]{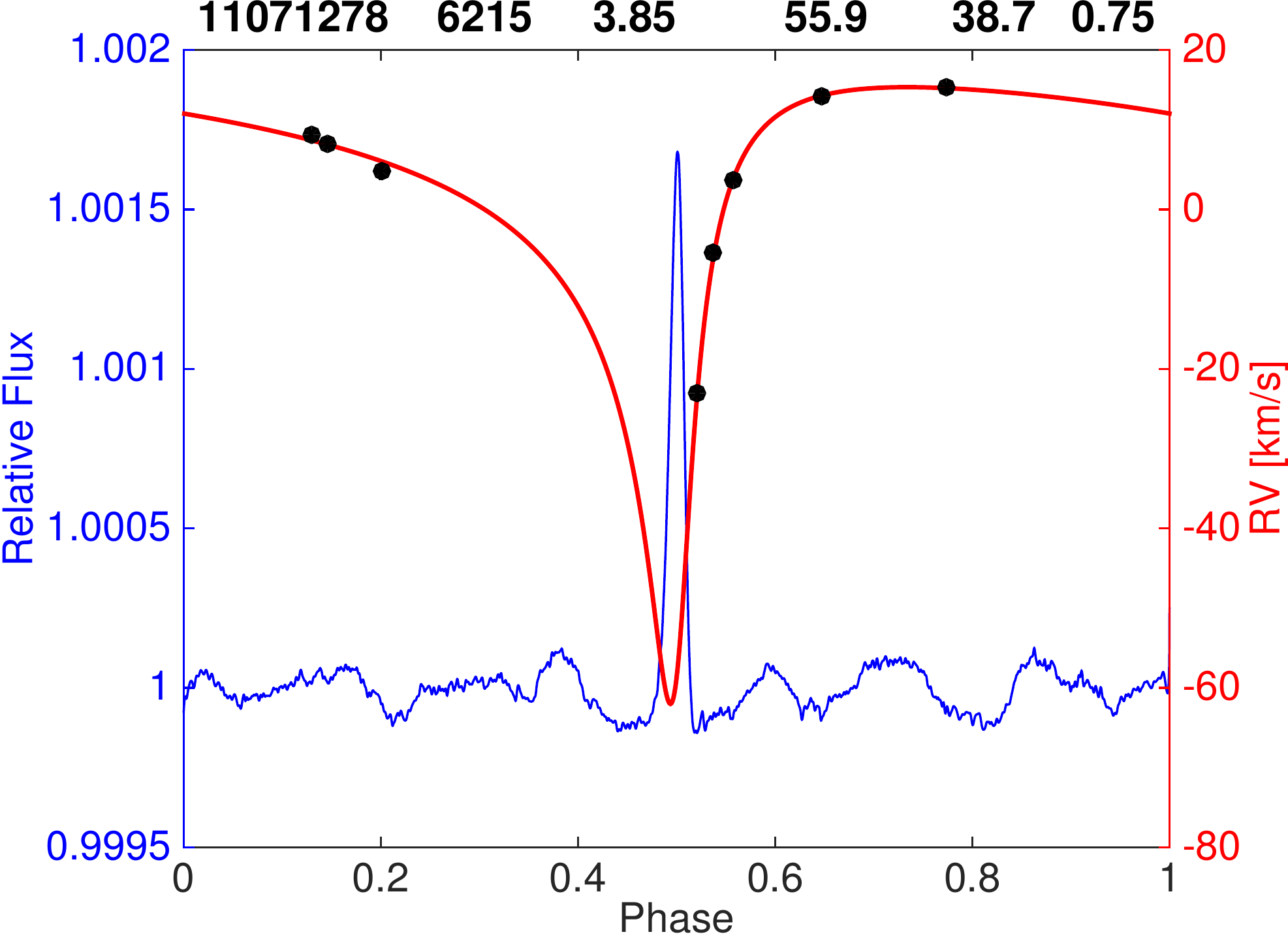} \vspace{3mm}&
\includegraphics[width=\figwidth]{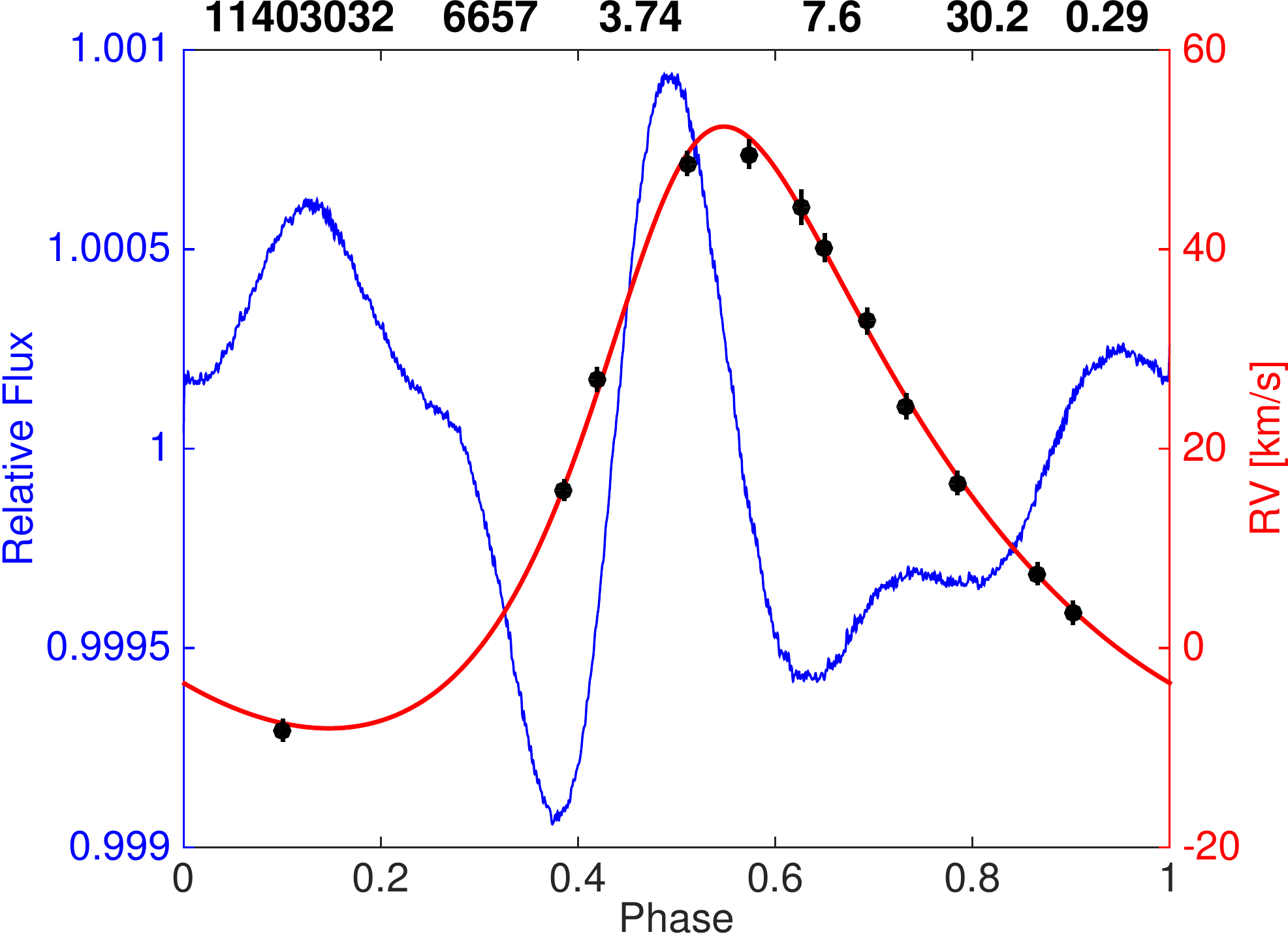} 
\end{tabular}
\begin{minipage}{60em}
Figure \thefigure: Same as Fig.~9 for KID 6775034, KID 8027591, KID 8164262, KID 9016693, KID 9965691, KID 10334122, KID 11071278, and KID 11403032. In each panel the title lists, from left to right, the KIC ID, \teff\ (K), $\logg$, $P$ (d), $K$ (\kms), and $e$.
\end{minipage}
\label{fig:rvlc2}
\end{center}
\end{table}

%% file: figures_rvlc3.tex

\def\figwidth{2.7in}

\stepcounter{figure} 

\begin{table}[h]
\begin{center}
\begin{tabular}{lcr}
\includegraphics[width=\figwidth]{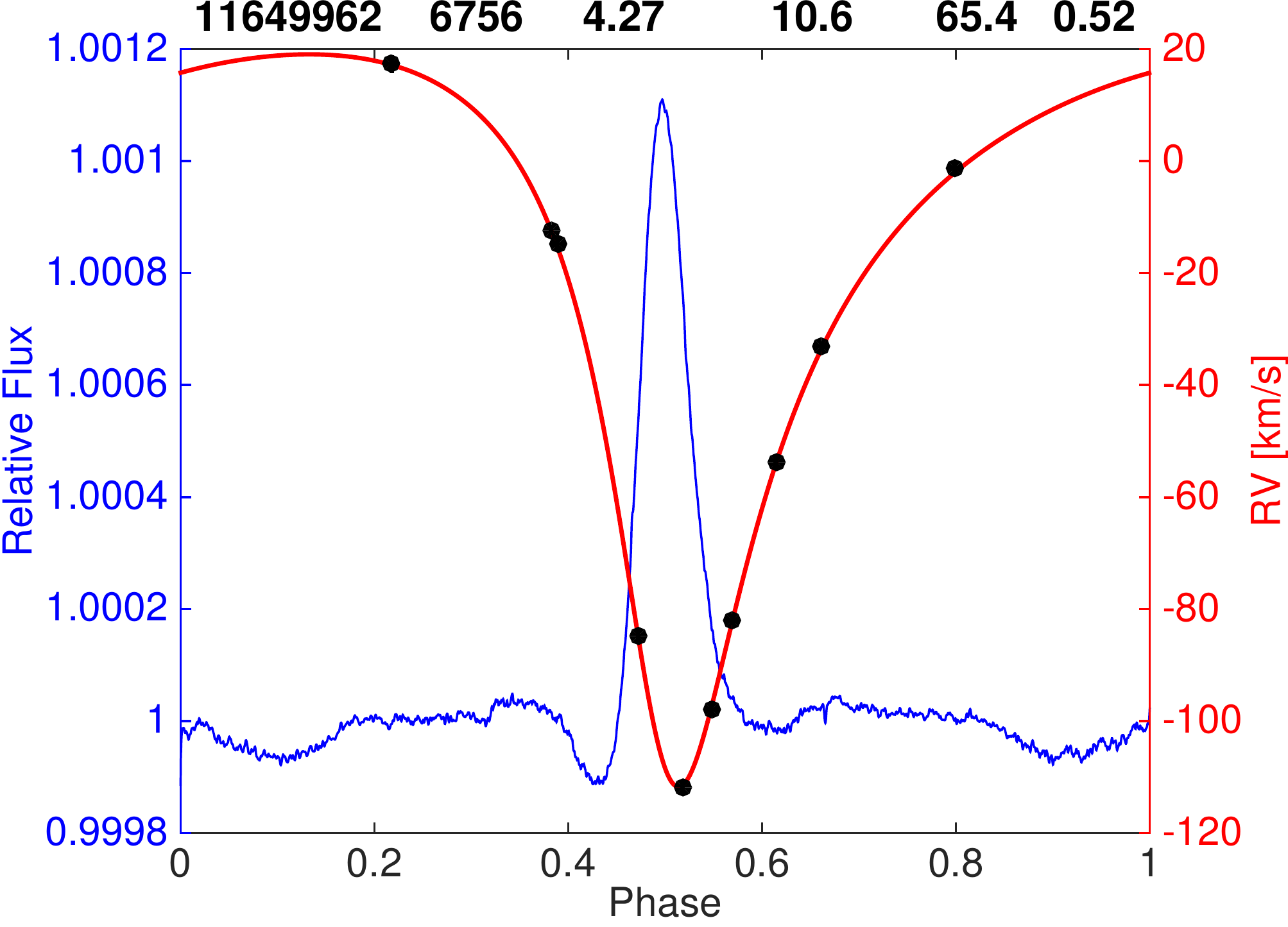} \\
\includegraphics[width=\figwidth]{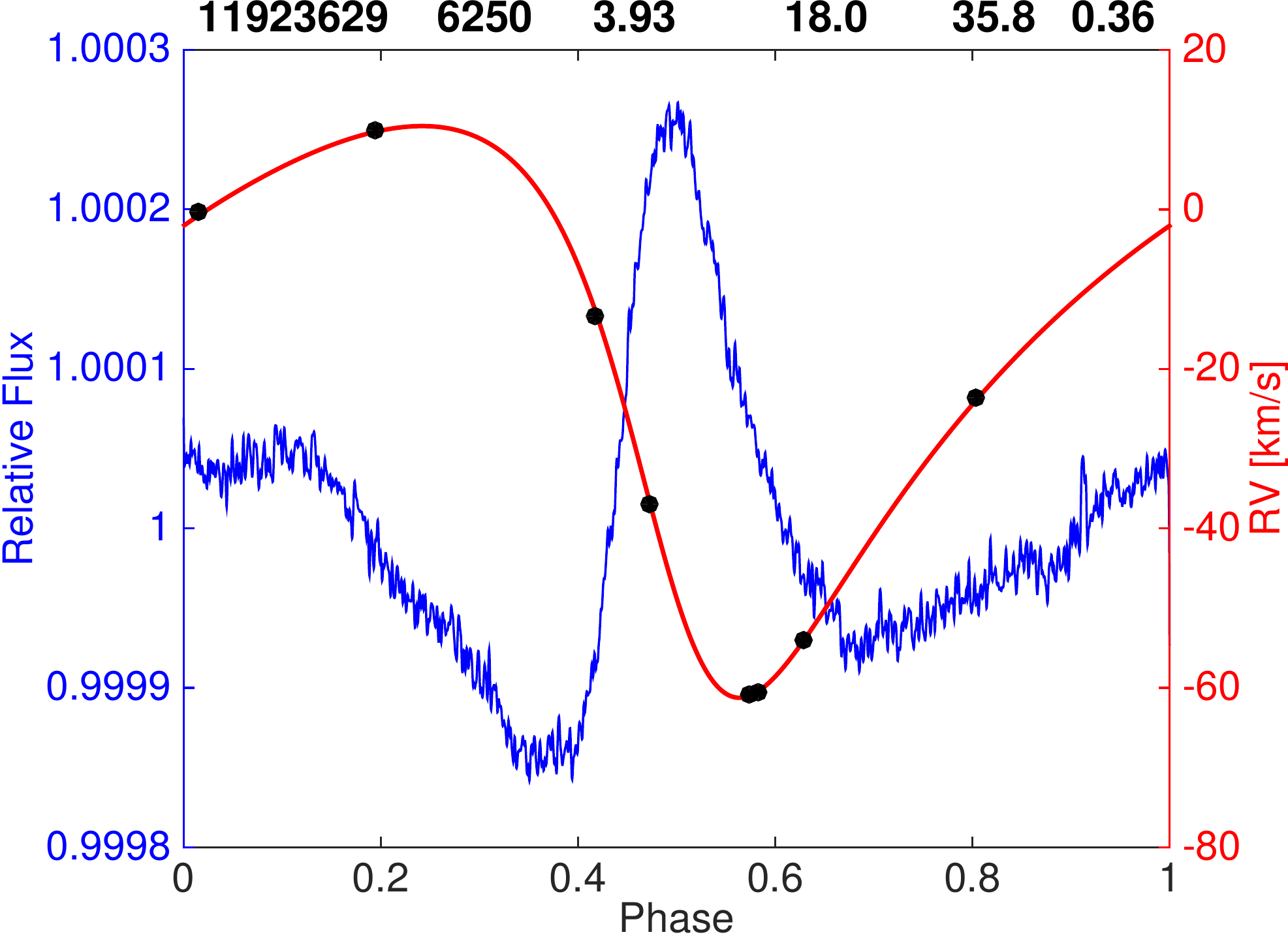} \\
\includegraphics[width=\figwidth]{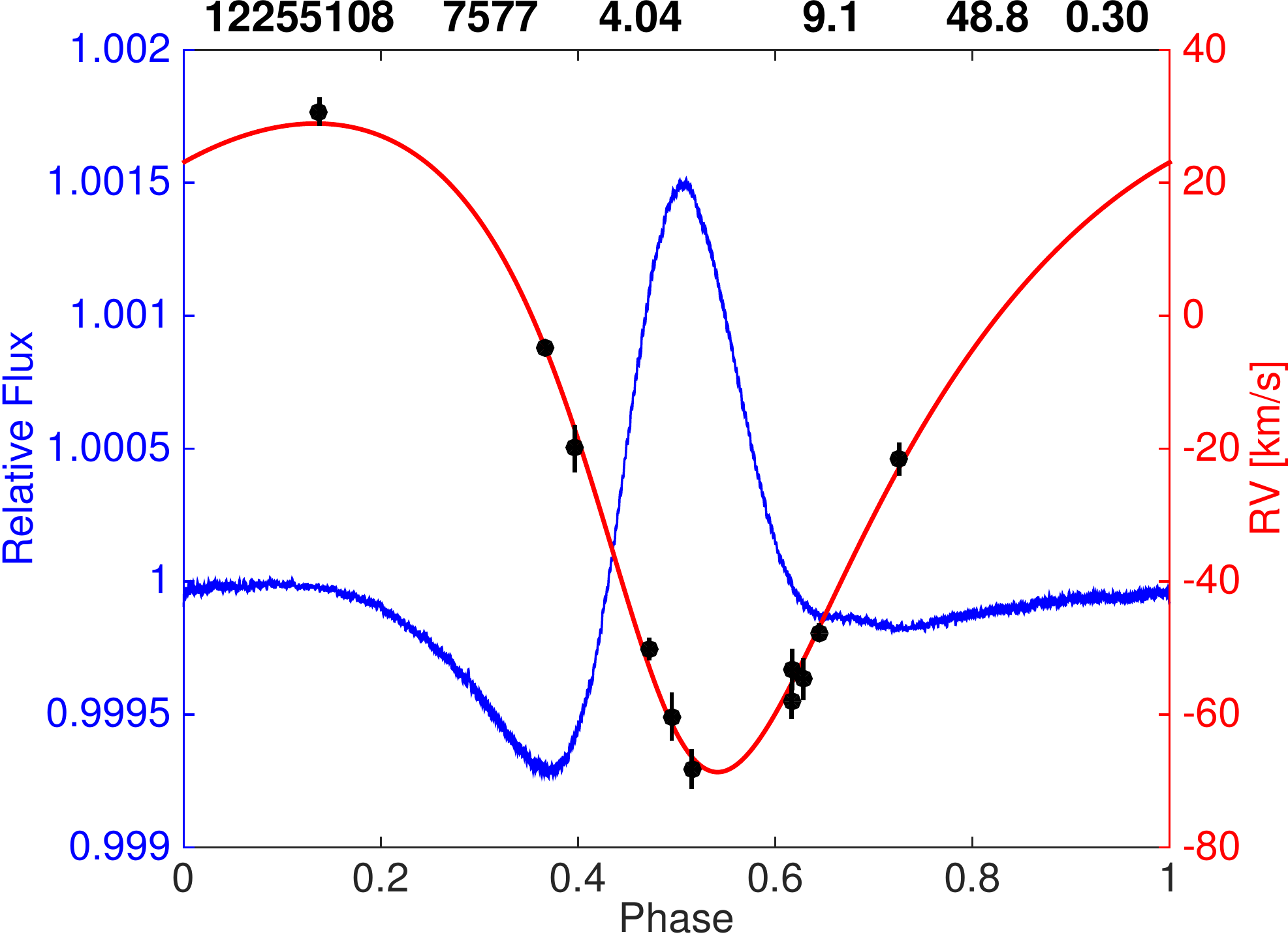}  
\end{tabular}
\begin{minipage}{60em}
Figure \thefigure: Same as Fig.~9 for KID 11649962, KID 11923629, and KID 12255108. In each panel the title lists, from left to right, the KIC ID, \teff\ (K), $\logg$, $P$ (d), $K$ (\kms), and $e$.
\end{minipage}
\label{fig:rvlc3}
\end{center}
\end{table}

%% file: figures_rvlc_const.tex

\def\figwidth{2.7in}

\stepcounter{figure} 

\begin{table}[h]
\begin{center}
\begin{tabular}{lcr}
\includegraphics[width=\figwidth]{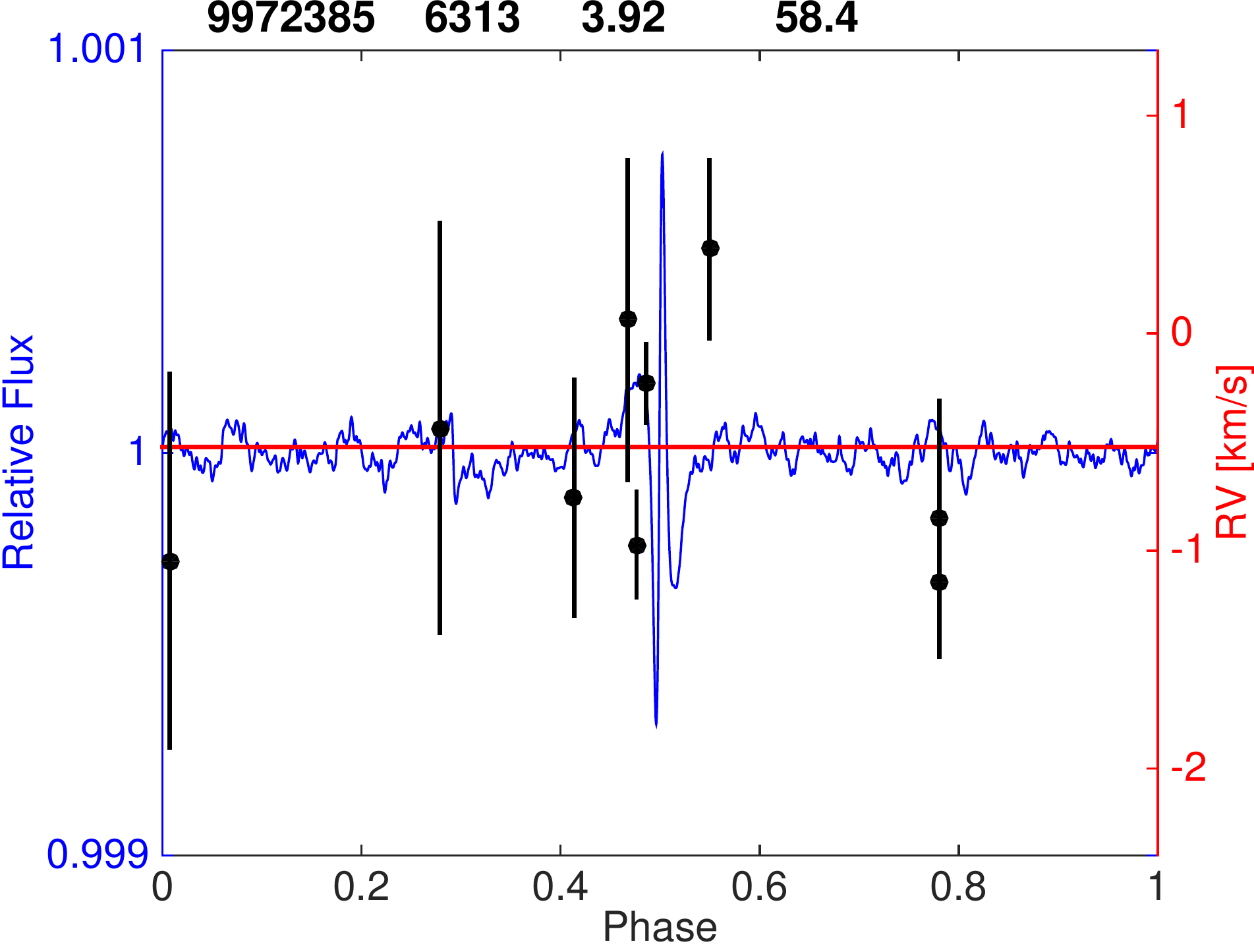}  \\
\includegraphics[width=\figwidth]{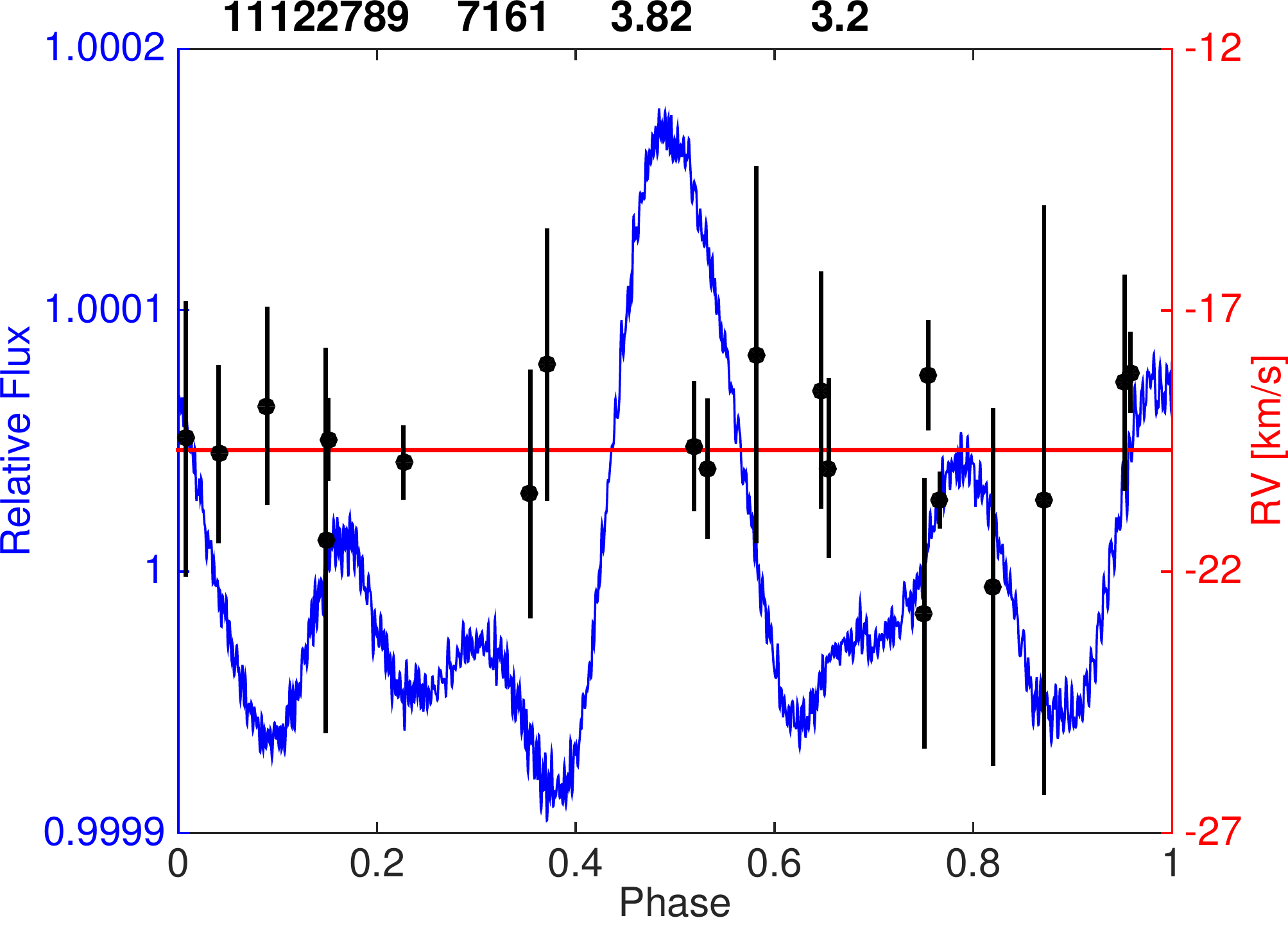} \\
\includegraphics[width=\figwidth]{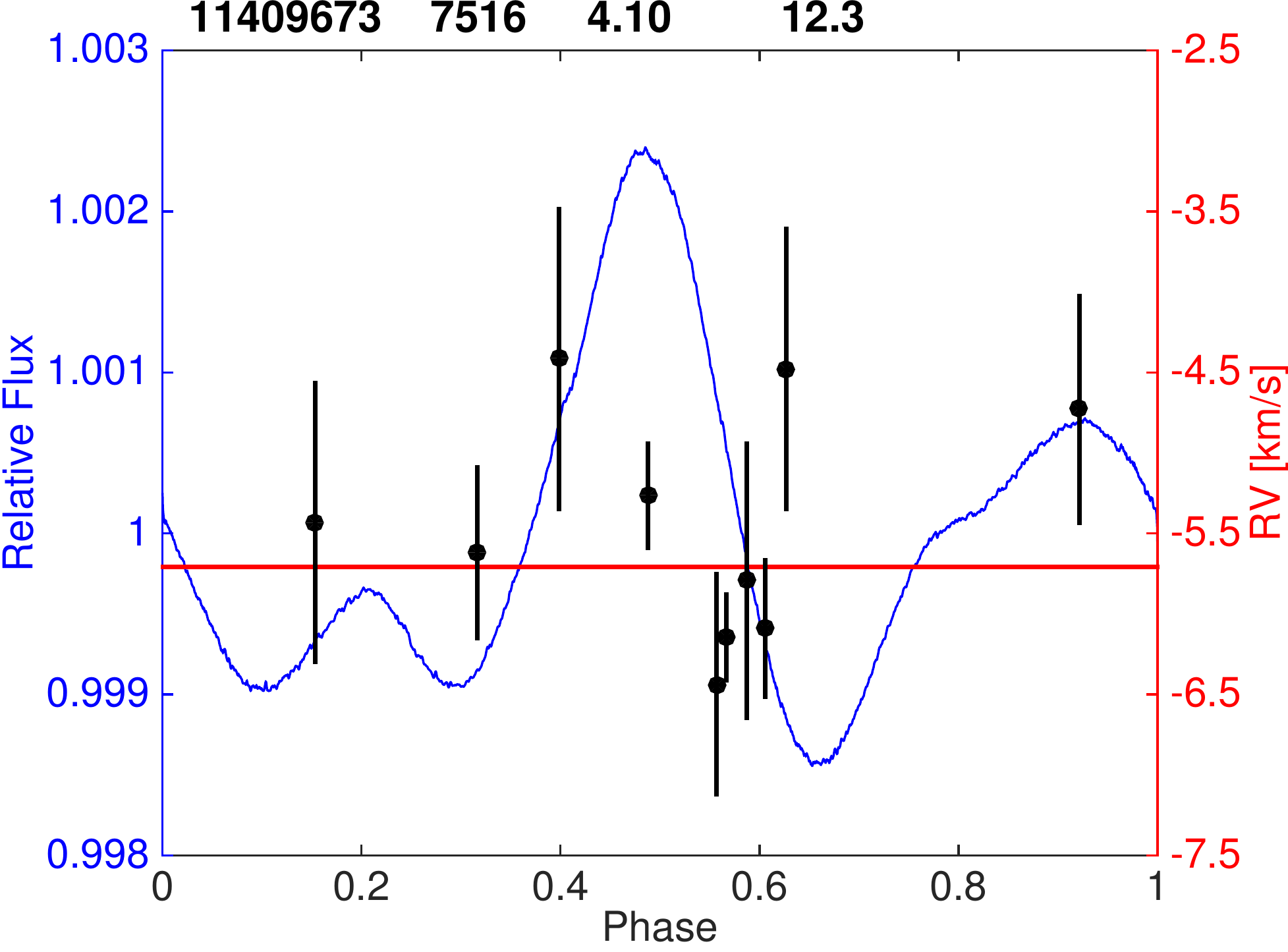}  
\end{tabular}
\begin{minipage}{60em}
Figure \thefigure: Same as Fig.~9 for the three RV non-variable stars, KID 9972385, KID 11122789, and KID 11409673. The red solid line marks the RVs weighted mean. The title of each panel lists the KIC ID, \teff\ (K), \logg, and $P$ (d).
\end{minipage}
\label{fig:rvlc3}
\end{center}
\end{table}

%% file: ms.bbl
\begin{thebibliography}{}


\bibitem[Abdul-Masih et al.(2016)]{abdul16} Abdul-Masih, M., Pr{\v s}a, A., Conroy, K., et al.\ 2016, \aj, 151, 101 

\bibitem[Beck et al.(2014)]{beck14} Beck, P.~G., Hambleton, K., Vos, J., et al.\ 2014, \aap, 564, A36 

\bibitem[Beers et al.(1990)]{beers90} Beers, T.~C., Flynn, K., \& Gebhardt, K.\ 1990, \aj, 100, 32 

\bibitem[Brown et al.(2011)]{brown11} Brown, T.~M., Latham, D.~W., Everett, M.~E., \& Esquerdo, G.~A.\ 2011, \aj, 142, 112 

\bibitem[Bryson et al.(2013)]{bryson13} Bryson, S.~T., Jenkins, J.~M., Gilliland, R.~L., et al.\ 2013, \pasp, 125, 889 

\bibitem[Burkart et al.(2012)]{burkart12} Burkart, J., Quataert, E., Arras, P., Weinberg, N.\ 2012, \mnras, 421, 983

\bibitem[Chubak et al.(2012)]{chubak12} Chubak, C., Marcy, G., Fischer, D.~A., et al.\ 2012, arXiv:1207.6212 

\bibitem[Coughlin et al.(2014)]{coughlin14} Coughlin, J.~L., Thompson, S.~E., Bryson, S.~T., et al.\ 2014, \aj, 147, 119 

\bibitem[Cowling(1941)]{cowling41} Cowling, T.~G.\ 1941, \mnras, 101, 367 

\bibitem[Duquennoy \& Mayor(1991)]{duquennoy91} Duquennoy, A., Mayor, M.\ 1991, \aap, 248, 485

\bibitem[Eastman et al.(2013)]{eastman13} Eastman, J., Gaudi, B.~S., \& Agol, E.\ 2013, \pasp, 125, 83 

\bibitem[Fuller \& Lai(2012)]{fuller12} Fuller, J., \& Lai,D.\ 2012, \mnras, 420, 3126 

\bibitem[Gregory(2005)]{gregory05} Gregory, P.~C.\ 2005, \apj, 631, 1198 

\bibitem[Hambleton et al.(2013)]{hambleton13} Hambleton, K.~M., Kurtz, D.~W., Pr{\v s}a, A., et al.\ 2013, \mnras, 434, 925 

\bibitem[Hambleton et al.(2016)]{hambleton16} Hambleton, K.~M., Kurtz, D.~W., Pr{\v s}a, A., et al.\ 2016, \mnras, accepted (arXiv:1608.02493)

\bibitem[Hareter et al.(2014)]{hareter14} Hareter, M., Papar{\'o}, M., Weiss, W., et al.\ 2014, \aap, 567, A124 

\bibitem[Hoaglin et al.(1983)]{hoaglin83} Hoaglin, D. C., Mosteller, F., \& Tukey, J.~W.~(ed.) 1983, Wiley Series in
Probability and Mathematical Statistics (New York: Wiley)

\bibitem[Holdsworth et al.(2016)]{holdsworth16} Holdsworth, D.~L., Kurtz, D.~W., Smalley, B., et al.\ 2016, arXiv:1607.03853 

\bibitem[Holman et al.(2006)]{holman06} Holman, M.~J., Winn, J.~N., Latham, D.~W., et al.\ 2006, \apj, 652, 1715 

\bibitem[Howard et al.(2009)]{howard09} Howard, A.~W., Johnson, J.~A., Marcy, G.~W., et al.\ 2009, \apj, 696, 75 

\bibitem[Howell et al.(2014)]{howell14} Howell, S.~B., Sobeck, C., Haas, M., et al.\ 2014, \pasp, 126, 398 

\bibitem[Huber et al.(2014)]{huber14} Huber, D., Silva Aguirre, V., Matthews, J.~M., et al.\ 2014, \apjs, 211, 2 

\bibitem[Hut(1981)]{hut81} Hut, P.\ 1981, \aap, 99, 126 

\bibitem[Kirk et al.(2016)]{kirk16} Kirk, B., Conroy, K., Pr{\v s}a, A., et al.\ 2016, \aj, 151, 68 

\bibitem[Kiseleva et al.(1994)]{kise94} Kiseleva, L.G., Eggleton, P.P., Anosova, J.P.\ 1994, \mnras, 267, 161 

\bibitem[Knutson et al.(2014)]{knutson14} Knutson, H.~A., Fulton, B.~J., Montet, B.~T., et al.\ 2014, \apj, 785, 126 

\bibitem[Kumar et al.(1995)]{kumar95} Kumar, P., Ao, C.O., Quataert, E., 1995, \apj, 449, 294

\bibitem[Kurtz(1982)]{kurtz82} Kurtz, D.~W.\ 1982, \mnras, 200, 807 
\bibitem[Lillo-Box et al.(2015)]{lillo15} Lillo-Box, J., Barrado, D., Mancini, L., et al.\ 2015, \aap, 576, A88 

\bibitem[Latham et al.(2002)]{latham02} Latham, D.~W., Stefanik, R.~P., Torres, G., et al.\ 2002, \aj, 124, 1144 

\bibitem[Maceroni et al.(2009)]{maceroni09} Maceroni, C., Montalb{\'a}n, J., Michel, E., et al.\ 2009, \aap, 508, 1375 

\bibitem[Meibom \& Mathieu(2005)]{meibom05} Meibom, S., \& Mathieu, R.~D.\ 2005, \apj, 620, 970 

\bibitem[Mazeh(2008)]{mazeh08} Mazeh, T.\ 2008, EAS Publications Series, 29, 1 

\bibitem[Murphy et al.(2014)]{murphy14} Murphy, S.~J., Bedding, T.~R., Shibahashi, H., Kurtz, D.~W., \& Kjeldsen, H.\ 2014, \mnras, 441, 2515 

\bibitem[Naoz(2016)]{naoz16} Naoz, S.\ 2016, arXiv:1601.07175 

\bibitem[Naoz \& Fabrycky(2014)]{naoz14} Naoz, S., \& Fabrycky, D.~C.\ 2014, \apj, 793, 137 

\bibitem[Ngo et al.(2015)]{ngo15} Ngo, H., Knutson, H.~A., Hinkley, S., et al.\ 2015, \apj, 800, 138 

\bibitem[Nidever et al.(2002)]{nidever02} Nidever, D.~L., Marcy, G.~W., Butler, R.~P., Fischer, D.~A., \& Vogt, S.~S.\ 2002, \apjs, 141, 503 

\bibitem[O'Leary \& Burkart(2014)]{oleary14} O'Leary R., \& Burkart, J.\ 2014, \mnras, 440, 3036 

\bibitem[Pejcha et al.(2013)]{pejcha13} Pejcha, O., Antognini, J., Shappee, B., Thompson, T.\ 2013, \mnras, 435, 943 

\bibitem[Piskorz et al.(2015)]{piskorz15} Piskorz, D., Knutson, H.~A., Ngo, H., et al.\ 2015, \apj, 814, 148 

\bibitem[Pr{\v s}a et al.(2011)]{prsa11} Pr{\v s}a, A., Batalha, N., Slawson, R.~W., et al.\ 2011, \aj, 141, 83 

\bibitem[Rauer et al.(2014)]{rauer14} Rauer, H., Catala, C., Aerts, C., et al.\ 2014, Experimental Astronomy, 38, 249 

\bibitem[Raghavan et al.(2010)]{raghavan10} Raghavan, D., McAlister, H., Henry, T., et al.\ 2010, \apjs, 190, 1

\bibitem[Ricker et al.(2014)]{ricker14} Ricker, G.~R., Winn, J.~N., Vanderspek, R., et al.\ 2014, \procspie, 9143, 914320 

\bibitem[Schmid et al.(2015)]{schmid15} Schmid, V.~S., Tkachenko, A., Aerts, C., et al.\ 2015, \aap, 584, A35 

\bibitem[Shibahashi \& Kurtz(2012)]{shibahashi12} Shibahashi, H., \& Kurtz, D.~W.\ 2012, \mnras, 422, 738 
\bibitem[Schlaufman \& Winn(2013)]{schlaufman13} Schlaufman, K., \& Winn, J.\ 2013, \apj, 772, 143

\bibitem[Shporer et al.(2009)]{shporer09} Shporer, A., Mazeh, T., Pont, F., et al.\ 2009, \apj, 694, 1559 

\bibitem[Shporer et al.(2014)]{shporer14} Shporer, A., O'Rourke, J.~G., Knutson, H.~A., et al.\ 2014, \apj, 788, 92 

\bibitem[Slawson et al.(2011)]{slawson11} Slawson, R.~W., Pr{\v s}a, A., Welsh, W.~F., et al.\ 2011, \aj, 142, 160 

\bibitem[Smullen \& Kobulnicky(2015)]{smullen15} Smullen, R.~A., \& Kobulnicky, H.~A.\ 2015, \apj, 808, 166 

\bibitem[Thompson et al.(2012)]{thompson12} Thompson, S.~E., Everett, M., Mullally, F., et al.\ 2012, \apj, 753, 86 

\bibitem[Tokovinin et al.(2006)]{tokovinin06} Tokovinin, A., Thomas, S., Sterzik, M., Udry, S.\ 2006, \aap, 450, 681

\bibitem[Vogt et al.(1994)]{vogt94} Vogt, S.~S., Allen, S.~L., Bigelow, B.~C., et al.\ 1994, \procspie, 2198, 362 

\bibitem[Welsh et al.(2011)]{welsh11} Welsh, W.~F., Orosz, J.~A., Aerts, C., et al.\ 2011, \apjs, 197, 4 

\bibitem[Zahn(1975)]{zahn75} Zahn, J.-P.\ 1975, \aap, 41, 329 

\bibitem[Zahn(1977)]{zahn77} Zahn, J.-P.\ 1977, \aap, 57, 383 

\end{thebibliography}
